\documentclass[usenatbib,fleqn]{mnras}
\usepackage{setspace}
\usepackage{epsf}
\usepackage{graphicx}
\usepackage{color}
\usepackage{txfonts}
\usepackage{tabularx}
\usepackage{here}
\usepackage{float}
\usepackage{threeparttable}
\usepackage{tabu}
\usepackage{dcolumn}
\usepackage{enumitem}
\usepackage[T1]{fontenc}
\usepackage{aecompl}
\usepackage{times}

\newcommand{\oi}{[O\,{\sc i}]}
\newcommand{\oii}{[O\,{\sc ii}]}
\newcommand{\oiii}{[O\,{\sc iii}]}

\newcommand{\NI}{[N\,{\sc i}]}
\newcommand{\nii}{[N\,{\sc ii}]}

\newcommand{\sii}{[S\,{\sc ii}]}
\newcommand{\siii}{[S\,{\sc iii}]}

\newcommand{\hei}{He\,{\sc i}}

\newcommand{\neii}{[Ne\,{\sc ii}]}

\newcommand{\arii}{[Ar\,{\sc ii}]}
\newcommand{\ariii}{[Ar\,{\sc iii}]}

\newcommand{\clii}{[Cl\,{\sc ii}]}
\newcommand{\cliii}{[Cl\,{\sc iii}]}

\newcommand{\feiii}{[Fe\,{\sc iii}]}

\newcommand{\ha}{H$\alpha$}
\newcommand{\hb}{H$\beta$}

\newcommand{\hi}{H\,{\sc i}}

\usepackage{color} 
\newcommand{\kms}{km s$^{-1}$}

\newcommand{\te}{$T_{\rm e}$}
\newcommand{\Ne}{$n_{\rm e}$}

\newcolumntype{C}{>{\centering\arraybackslash}X}
\newcolumntype{L}{>{\raggedright\arraybackslash}X}
\newcolumntype{R}{>{\raggedleft\arraybackslash}X}

\title[Fullerene-containing PN SaSt\,2-3]
{Physical properties of the fullerene C$_{60}$-containing planetary nebula
SaSt\,2-3
\thanks{Based on observations made with NAOJ Subaru Telescope under the programme IDs:
S13B-188S and S16A-227S (PI of both programme is M.~Otsuka)
and made with NOAO/WIYN telescope (programme ID: 2013A-0429, PI: M.~Meixner)}
}

\author[M.~Otsuka]
{
\begin{minipage}{1.0\linewidth}
Masaaki Otsuka$^{1,2}$\thanks{E-mail: otsuka@kusastro.kyoto-u.ac.jp}
\end{minipage}
\\
\\
 \begin{minipage}{1.0\linewidth}
$^{1}$Okayama Observatory, Kyoto University, Kamogata, Asakuchi, Okayama, 719-0232, Japan\\
  $^{2}$Academia Sinica, Institute of Astronomy and Astrophysics,
  11F Astronomy-Mathematics Building, NTU/AS campus, No. 1, Sec. 4,
  Roosevelt Rd., Taipei 10617, Taiwan, Republic of China
\end{minipage}
}

\begin{document}

\date{}

\pagerange{\pageref{firstpage}--\pageref{lastpage}} \pubyear{2018}

\maketitle

\label{firstpage}

\begin{abstract}
 We perform a detailed analysis of the
 fullerene C$_{60}$-containing planetary nebula (PN) SaSt\,2-3
 to investigate the physical properties of the central star (B0-1II)
 and nebula based on our own Subaru/HDS spectra and multiwavelength archival data.
 By assessing the stellar absorption,
 we derive the effective temperature, 
 surface gravity, and photospheric abundances. For the first
 time, we report time variability of the central star's radial velocity,
 strongly indicating a binary central star. Comparison between
 the derived elemental abundances
 and those predicted values by asymptotic giant branch (AGB)
 star nucleosynthesis models indicates that the progenitor
 is a star with initial mass of $\sim$1.25\,M$_{\sun}$ and metallicity
 $Z = 0.001$/$\alpha$-element/Cl-rich ([$\alpha$,Cl/Fe]$\sim$+0.3-0.4).
 We determine the distance (11.33\,kpc) to be
 consistent with the post-AGB evolution of 1.25\,M$_{\sun}$ initial mass
 stars with $Z = 0.001$. Using the photoionisation model, we fully reproduce
 the derived quantities by adopting a cylindrically shaped nebula. 
We derive the mass fraction of the C-atoms present in atomic gas, graphite grain, and C$_{60}$. 
The highest mass fraction of C$_{60}$ ($\sim$0.19\,\%) indicates 
that SaSt\,2-3 is the C$_{60}$-richest PN amongst Galactic PNe. 
From comparison of stellar/nebular properties with 
other C$_{60}$ PNe, we conclude that the C$_{60}$ formation depends on 
the central star's properties and its surrounding environment (e.g., binary disc), rather 
than the amount of C-atoms produced during the AGB phase.
\end{abstract}

\begin{keywords}
  ISM: planetary nebulae: individual (SaSt\,2-3)
  --- ISM: abundances --- ISM: dust, extinction
\end{keywords}

\section{Introduction}
\label{S-intro}

\setcounter{footnote}{2}

      \begin{figure*}
       \centering
       \includegraphics[clip,width=\textwidth]{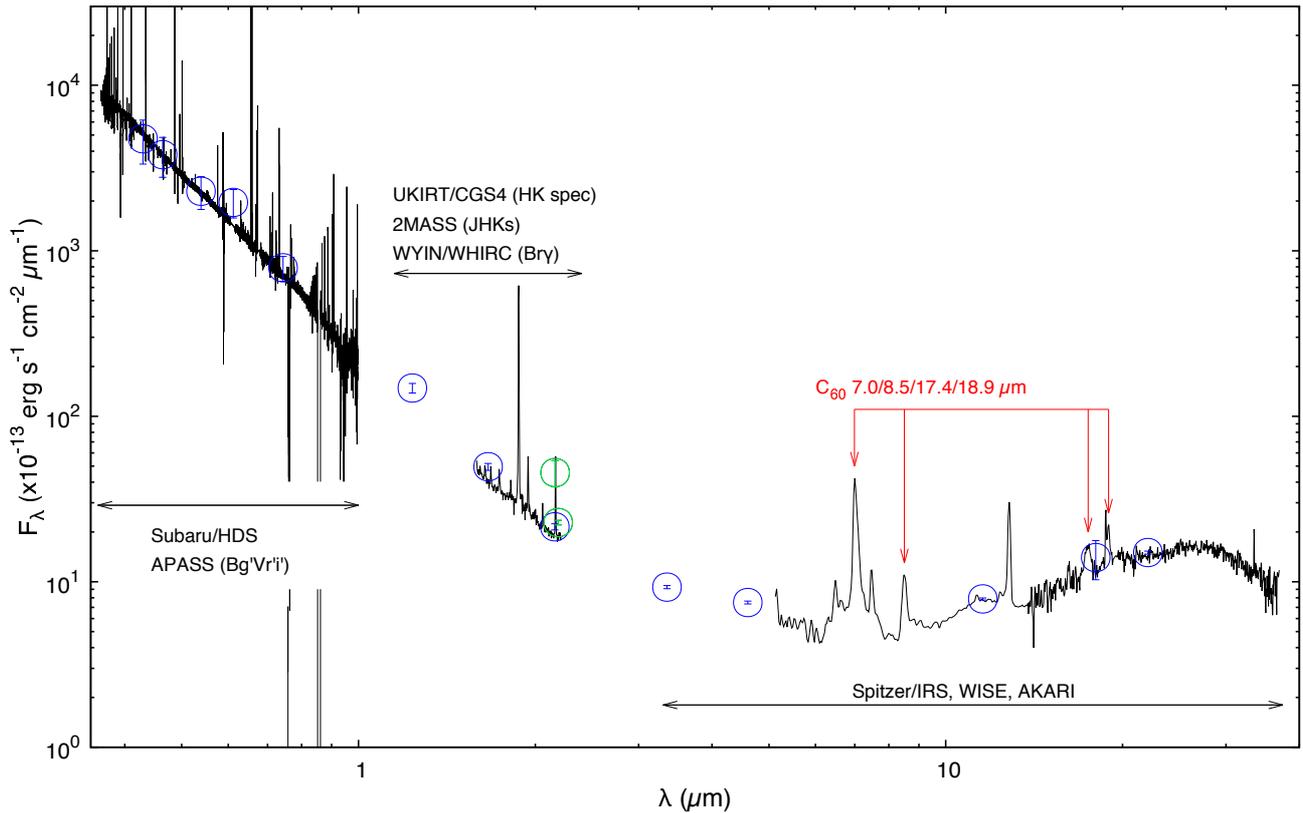}
      \caption{Panchromatic data of SaSt\,2-3 used in the present
       work. The green (WHIRC Br$\gamma$-on/off bands) and blue circles
       (the other bands) are the photometry data and the black
       lines are the spectra, respectively. The flux density of the
       Subaru/HDS spectrum is scaled to match the APASS $Bg'Vr'i'$-bands.
       The UKIRT 3.8 m/Cooled Grating Spectrometer 4 (CGS4) $HK$-band 
       spectrum (we downloaded the raw data of the spectrum presented in \citet{Lumsden:2001aa} 
       from the UKIRT archive data website and reduced them using {\sc IRAF}) 
       is scaled to match the 2MASS
       $Ks$-band. The \emph{Spitzer}/IRS spectrum is scaled to match
       the \emph{WISE}/W3 and W4 and \emph{AKARI} L18W bands.}
      \label{F-sedobs}
      \end{figure*}

 Mid-infrared (mid-IR) spectroscopic observations made by the  
 \emph{Spitzer}/Infrared Spectrograph
 \citep[IRS,][]{Houck:2004aa} have recently detected 
 fullerene C$_{60}$ and C$_{70}$ in a variety of space environments
 such as R Coronae Borealis stars \citep{Garcia-Hernandez_11_Are}, 
 reflection nebulae \citep{Sellgren_10_C60},
 young stellar objects \citep{Roberts_12_Detection},
 post asymptotic giant branch (AGB) stars
 \citep{Gielen_11_Silicate,Gielen_11_Carbonaceous}, 
 proto planetary nebula \citep[PNe;][]{Zhang_11_Detection}, and 
 PN \citep{Cami:2010aa,Garcia-Hernandez:2010aa,
 Garcia-Hernandez:2011aa,Garcia-Hernandez:2012aa,Otsuka:2013aa,
 Otsuka:2014aa,Otsuka:2016aa}.
  At the moment, PNe represent the largest fraction of fullerene
  detection; since the first detection of the mid-IR C$_{60}$ and C$_{70}$
 bands in the C-rich PN Tc\,1 by \citet{Cami:2010aa}, 24
 fullerene-containing PNe
 have been identified in the Milky
 Way and the Large and Small Magellanic Clouds (LMC and SMC,
 respectively).

 In general, C$_{60}$ PNe show very similar IR dust features and stellar/nebular 
 properties; their mid-IR spectra display broad $6-9$, 11, and 30\,{\micron} features in addition to 
 C$_{60}$ bands at 7.0, 8.5, 17.4, and 18.9\,{\micron}, 
 and they have cool central stars and low-excitation nebulae, indicating that their age after the AGB phase 
 is very young \citep[e.g.,][]{Otsuka:2014aa}. The excitation mechanisms 
 \citep[e.g.,][]{Bernard-Salas:2012aa} and the formation paths \citep[e.g.,][]{Berne_15_Top,Duley_12_FULLERENES} 
 are not well understood and are still a subject of debate. However, it remains unclear why these objects 
 exhibit the C$_{60}$ features - is the span of time during which spectral features of C$_{60}$ 
 are present a short-lived phase that all C-rich PNe go through, or are C$_{60}$ PNe distinct objects 
 in terms of their stellar/nebular properties and/or evolution? This is directly linked to the question 
 of how C$_{60}$ forms in evolved star environments. We would like to answer this fundamental 
 question by investigating the physical properties of C$_{60}$ PNe and comparing them with non-C$_{60}$ PNe.

 Amongst C$_{60}$ PNe, SaSt\,2-3 \citep[PN G232.0+05.7,][]{Acker:1992aa} firstly identified by \citet{Sanduleak:1972aa}
 is a particularly interesting object to that we should pay more attention. \citet{Otsuka:2014aa} discovered 
 C$_{60}$ bands in this PN for the first time. Surprisingly, the mid-IR C$_{60}$ band strengths in SaSt\,2-3 and Tc\,1 are 
 the strongest amongst all the fullerene-containing objects. This strongly indicates that the fullerene 
 formation in these two PNe is particularly efficient. Tc\,1 has been extensively studied since the discovery of C$_{60}$. 
 However, SaSt\,2-3 is not entirely understood due to the lack of available data for the central star and nebula 
 and its uncertain distance ($D$). The uncertain $D$ towards SaSt\,2-3 has led to different estimates of 
 the central star luminosity ($L_{\ast}$) and effective temperature ($T_{\rm eff}$); accordingly, this has led to inconsistencies in understanding the evolutionary status of this PN \citep{Gesicki:2007aa,Otsuka:2014aa}. 
 What we know from the prior studies is that this object has 
 low-metallicity \citep[$\epsilon({\rm S})$ = 5.48\footnote{$\epsilon$(X)
 equals to \mbox{12 + $\log_{10}$\,$n$(X)/$n$(H)}, where X is the target element and $n$(X)/$n$(H) is the number density
 ratio relative to hydrogen.},][]{Pereira:2007aa} and is (possibly) a Type\,IV PN \citep[i.e, halo population,][]{Pereira:2007aa}.

  If we obtain the UV to optical wavelength spectra of the
  central star as well as the nebula, we can resolve issues raised and verify conclusions from previous studies 
  of SaSt\,2-3; by so doing, we can hope to gain insights into the C$_{60}$ formation. 
  Fortunately, the UV-optical photometry data from the AAVSO Photometric All Sky Survey \citep[APASS,][]{Henden:2016aa} 
  can rigorously constrain $L_{\ast}$, and \citet{Frew:2016aa} improved its distance estimate 
  (\mbox{$D = 14.86~\pm~4.26$\,kpc}). Therefore, we perform a comprehensive analysis on our own
  high-dispersion spectra of SaSt\,2-3 taken using the 8.2 m Subaru telescope/high 
  dispersion spectrograph \citep[HDS,][]{Noguchi:2002aa} and archived multiwavelength data.

 We organise the next sections as follows.
 In \S\,\ref{S:obs}, we describe our
 HDS spectroscopy and near-IR imaging using the
 NOAO WIYN 3.5 m/WIYN High-Resolution Infrared Camera 
\citep[WHIRC,][]{Meixner:2010aa} and the reduction of this data.
 The \emph{Spitzer}/IRS observation and its data reduction are
 described in \citet{Otsuka:2014aa}. In Fig.\,\ref{F-sedobs},
 we plot all the data used in the present work.
 In \S\,\ref{S:nebula}, we perform
 plasma-diagnostics and derive ionic/elemental abundances. 
 In \S\,\ref{S:stellar},
 we derive photospheric elemental abundances, $T_{\rm eff}$, and
 surface gravity $\log\,g$ by fitting the stellar absorption using the theoretical stellar
 atmosphere code {\sc TLUSTY} \citep{Hubeny:1988aa}. 
 In \S\,\ref{S:AGB}, we compare the derived nebular and stellar
 elemental abundances with those values predicted by
 AGB nucleosynthesis models in order to infer the initial mass of the progenitor star.
 In \S\,\ref{S:cloudy}, we build
 the spectral energy distribution (SED) model using the photoionisation
 code {Cloudy} \citep[v.13.05,][]{Ferland:2013aa} to be consistent with all the derived
 quantities based on our determined $D$.
 In \S\,\ref{S-Discuss}, we discuss the origin and evolution of
 SaSt\,2-3 and the C$_{60}$ formation in PNe by comparison of the derived 
 nebular/stellar properties with other non-C$_{60}$ and C$_{60}$-containing PNe. 
 Finally, we summarise the present work.

 \section{Data Set and Reduction} \label{S:obs}

  \subsection{Subaru/HDS observation}

  We secured high-dispersion echelle spectra using the
  HDS located at one of the Nasmyth loci of the 8.2 m
  Subaru Telescope at the top of Mauna Kea in Hawai'i.
  We summarise our observations in
  Table\,\ref{T-HDS}.
  We selected the $2\times2$ on-chip binning pattern. We set the
  slit-width to be 1.2{\arcsec}. We used the blue
  cross disperser for the $3640-5390$\,{\AA} observation and the red one
  for the $4740-7490$\,{\AA} and $7190-9960$\,{\AA} observations,
  respectively. We utilised the atmospheric dispersion corrector (ADC)
  during the observations.
  In all the observations, we observed the standard
  star Hiltner\,600 for correcting echelle blaze functions
  and flux density simultaneously. In the
  $7190-9960$\,{\AA} observation, we observed the telluric standard stars
  HD\,61017 (B9III, $m_{\rm V} = 6.68$) and HD\,62217 (B9V, $m_{\rm V} =
  8.26$) at similar airmass.

  We reduced the data using {\sc IRAF}\footnote{{\sc IRAF}
  is distributed by the National Optical Astronomy Observatories,
  operated by the Association of Universities for Research in
  Astronomy (AURA), Inc., under a cooperative agreement with the
  National Science Foundation.} in a standard manner, including
  over-scan subtraction, scattered light
  subtraction between echelle orders, and telluric absorption removal.
  We adopted the atmospheric extinction correction function measured by 
  \citet{Buton:2013aa} at Mauna Kea. 
  We measured
  the actual spectral resolution ($R = \lambda/\delta{\lambda} = 32\,300-33\,500$, see Table\,\ref{T-HDS}) 
  using $>$300 Th-Ar comparison lines. The signal-to-noise ratio (S/N) for
  continuum reaches $\sim$40 at $\sim$3640\,{\AA} and
  $\sim$11 at $\sim$9950\,{\AA}. In $\sim$$3700-4800$\,{\AA} (this
  range is important in stellar absorption fittings), S/N is $>$70.
  We scaled both spectra to the average flux density in the overlapping
  regions ($4740-5390$\,{\AA} and $7190-7490$\,{\AA}), and
  we connected these scaled spectra into a single $3640-9960$\,{\AA}
  spectrum. The resultant spectrum is presented in Fig.\,\ref{F-sedobs}.

  \subsection{Flux measurements and interstellar extinction correction}
 \label{S-flux}

 We measure emission line fluxes by multiple Gaussian component
 fitting. Then, we correct
 these observed fluxes $F$($\lambda$) to obtain the interstellar
 extinction corrected fluxes $I$($\lambda$) using the following formula;
 
 \begin{equation}
  I(\lambda) = F(\lambda)~\cdot~10^{c({\rm H\beta})(1 + f(\lambda))},
   \label{eq-1}
 \end{equation}

 \noindent where $f$($\lambda$) is the interstellar
 extinction function at $\lambda$ computed by the reddening law of
 \citet{Cardelli:1989aa}. To verify $R_{V}$ accurately,
 $\sim$$2000-2500$\,{\AA} spectra/photometry data would be necessary
 because $f({\lambda})$ and $R_{V}$ are sensitive to this wavelength
 range. At the moment, there are no available spectra/photometry
 data in such wavelength range. Therefore, we adopt an average
 $R_{V} = 3.1$ in the Milky Way. Applying $R_{V} = 3.1$ to
 SaSt\,2-3 seems to be acceptable because \citet{Wegner:2003aa}
 reported \mbox{$R_{V} = 3.64~\pm~0.43$} towards HD\,60855 (B2Ve).
 HD\,60855 is a star in the direction closer to SaSt\,2-3 amongst
 stars whose $R_{V}$ has been measured.
 $c$({\hb}) is the reddening coefficient at {\hb}, corresponding
 to $\log_{10}$\,$I$({\hb})/$F$({\hb}).

 We determine $c$({\hb}) values by comparing the observed 
 Balmer and Paschen line ratios to {\hb} 
 with the theoretical ratios of \citet{Storey:1995aa} for the case with
 an electron temperature {\te} = 10$^{4}$\,K and an electron density {\Ne}
 = 2000\,cm$^{-3}$ under the Case B
 assumption. We calculate this {\Ne} using the {\oii}
 $F$(3726\,{\AA})/$F$(3729\,{\AA}) and the {\cliii}
 $F$(5517\,{\AA})/$F$(5537\,{\AA}) ratios. 
 For the $3640-5390$\,{\AA} spectrum, we obtain $c$({\hb}) =  \mbox{$0.20~\pm~0.01$}, which is the
 intensity-weight average amongst the
 H$\gamma$, H$\delta$, H$\epsilon$, and H$\eta$ to the
 {\hb} ratios. For the $4740-7490$\,{\AA} spectrum, we obtain $c$({\hb})
 =  \mbox{$0.40~\pm~0.01$} from the {\ha}/{\hb} ratio. For the
 $7190-9960$\,{\AA} spectrum, we determine $c$({\hb}) =  \mbox{$0.24~\pm~0.03$} from
 the Paschen H\,{\sc i}\,9014\,{\AA} (P10) to the {\hb} ratio. For all HDS spectra, we adopt the average 
 \mbox{$c$({\hb}) = $0.28~\pm~0.11$} amongst three HDS observations.

 \citet{Tylenda:1992aa} reported $c$({\hb}) = 1.11
 (observation date is unknown). We derive the average
 \mbox{$c$({\hb}) = $1.11~\pm~0.26$} using the ratio of
 $F$(H$\alpha$), $F$(H$\gamma$), and $F$(H$\delta$)
 to $F$({\hb}) measured from their
 spectrum\footnote{We downloaded this spectrum from
 HASH PN database. \url{http://202.189.117.101:8999/gpne/index.php}.}.
 Based on the $F$({\ha}) and $F$({\hb}) reported by
 \citet{Dopita:1997aa}, we obtain \mbox{$c$({\hb}) = $0.43~\pm~0.04$} (obs date: 1997 March).
 \citet{Pereira:2007aa} reported \mbox{$E(B-V) = 0.11~\pm~0.02$}, which
 corresponds to \mbox{$c$({\hb}) = $0.13 - 0.19$} (obs date: 2005 Feb).
 Using the line flux table of \citet{Pereira:2007aa}, we obtain
 the average \mbox{$c$({\hb}) = $0.41~\pm~0.23$} calculated from
 $F$(H$\alpha$), $F$(H$\gamma$), and $F$(P10) to $F$({\hb}).
 Using the archived ESO Faint Object Spectrograph and Camera (EFOSC) 
 spectrum taken on 2000
 April\footnote{The spectrum was taken by Acker et al
 (Programme ID: 64.H-0279(A)).}, we obtain a
 \mbox{$c$({\hb}) = $0.68~\pm~0.11$} measured from the $F$({\ha})/$F$({\hb})
 ratio. A time variation of $c$({\hb}) seen between 1992 and 2016 might
 be due to affect of stellar H\,{\sc i}
 absorption to corresponding nebular H\,{\sc i}
 and also orbital motion of the binary central star
 (\S\,\ref{S-binary}).

  We scale the \emph{Spitzer}/IRS spectrum to match the
  Wide-field Infrared Survey Explorer (\emph{WISE})
  W3/W4 band flux densities of
  \citet{Cutri:2014aa} and the L18W \emph{AKARI}/IRC
  mid-infrared all-sky survey of \citet{Ishihara:2010aa} (see \S\,\ref{S-phot}).
  For this scaled \emph{Spitzer}/IRS spectrum, we do not correct
  interstellar extinction because the interstellar extinction is negligibly
 small in the mid-IR wavelength. 
 It is common practice in nebular analyses to scale all line intensities in such a way that {\hb} 
 has a line flux of 100. To achieve this, we first normalise the line fluxes with respect the complex of
 the H\,{\sc i}\,7.46\,{\micron} ($n = 5-6$, $n$
 is the quantum number) and 7.50\,{\micron} ($n = 6-8$) lines.
 $F$(7.48/7.50\,{\micron}) is
  \mbox{$(4.87~\pm~0.30)\times10^{-15}$}\,erg\,s$^{-1}$\,cm$^{-2}$
  ($A(-B)$ means $A\times10^{-B}$ hereafter).
 According to \citet{Storey:1995aa} for the Case B assumption with
 {\te} = 10$^{4}$\,K and {\Ne} = 2000\,cm$^{-3}$,
 the ratio of H\,{\sc i} $I$(7.48/7.50\,{\micron})/$I$({\hb}) =
 3.102/100. Finally, we multiply all the normalised line fluxes by 3.102 
to express them relative to {\hb} with $I$({\hb}) = 100.

 In Appendix Table\,\ref{AT-line}, we list the identified emission lines
 in the Subaru/HDS and \emph{Spitzer}/IRS spectra. The first column is
 the laboratory wavelength in air. Here, $I$({\hb}) is 100. 
 The last column $\delta$\,$I(\lambda)$ corresponds to 1-$\sigma$.

  \begin{table}
   \centering
   \footnotesize
    \caption{HDS and WHIRC observation log for SaSt\,2-3.}
     \begin{tabular}{@{}l@{\hspace{4pt}}c@{\hspace{4pt}}
      c@{\hspace{4pt}}c@{\hspace{4pt}}c@{}}
   \hline
      Date       & $\lambda$ ({\AA})& $\lambda/\delta\,\lambda$ (ave.)
      &Exp. time &Condition/Seeing\\
    \hline
      2013/10/06&$3640-5390$&33\,500  &$2\times200$\,s, 2600\,s
		  &thin cloud, $\sim$0.7{\arcsec}\\
      2013/12/10&$4740-7490$&33\,300 &100, 500, 900\,s
		  &clear, $\sim$0.7{\arcsec}\\
      2016/02/01&$7190-9960$&32\,300  &180\,s, $4\times1600$\,s
		  &clear, $\sim$0.7-1.0{\arcsec}\\
   \hline
    Date       & Band & Pixel scale     &Exp. time &Condition/Seeing\\
    \hline
      2013/04/24&Br$\gamma$, Br$\gamma$45&0.1{\arcsec}$\times$0.1{\arcsec}
	         &5\,pts~$\times$~120\,s&clear,$\sim$0.6-0.7{\arcsec}\\
      \hline
     \end{tabular}
     \label{T-HDS}
     \end{table}

\subsection{NOAO/WHIRC near-IR imaging observation}

  We took the high-resolution images using NOAO WIYN 3.5 m/WHIRC.
  We summarise the observation log in
  Table\,\ref{T-HDS}. We took the two narrowband images using the
  Br$\gamma$ ($\lambda_{\rm c} = 2.162$\,{\micron}, effective band width ($W_{\rm eff}$)
  = 0.210\,{\micron}) and Br$\gamma$45
  ($\lambda_{\rm c} = 2.188$\,{\micron}, $W_{\rm eff} =
  0.245$\,{\micron}) filters\footnote{
  \url{https://www.noao.edu/kpno/manuals/whirc/filters.html}}.
  We selected a 5\,pts dithering pattern.
  We followed a standard
  manner for near-IR imaging data reductions using {\sc IRAF}, including
  background sky and dark current subtraction, bad pixel masking,
  flat-fielding, and distortion correction. Finally, we obtained
  a single averaged image for each band.

  For the flux calibration, we utilised the SED of the standard star
  2MASS07480394-1407155 based on its photometry between
  the two micron all sky survey
 \citep[2MASS,][]{Cutri:2003aa} $JHKs$ and \emph{WISE} bands W1/W2. The
  SED of this star
  can be well fitted with a single blackbody temperature of 3510\,K. Then,
  we derived the flux density in each band by taking each
  filter transmission curve into account.
  Next, we measured the respective count of the standard star in the 
  Br$\gamma$ and Br$\gamma$45 images. Thus, we obtained the conversion
  factor from the counts in ADU to flux density in
  \mbox{erg\,s$^{-1}$\,cm$^{-2}$\,{\micron}$^{-1}$}. The measured
  flux density in each band is listed in Appendix Table\,\ref{T-photo}
  and plotted in Fig.\,\ref{F-sedobs} (green circles).

  \subsection{Photometry data \label{S-phot}}

 To support the present work, we
 collected the data taken from APASS, 2MASS, \emph{WISE}, and
 \emph{AKARI}/Infrared Camera (IRC). In Table\,\ref{T-photo}, we list
 the observed and reddening corrected flux densities $F_{\lambda}$ and
 $I_{\lambda}$, respectively.
 We obtain $I_{\lambda}$ using Equation
 (\ref{eq-1}) and 
 the average \mbox{$c$(H$\beta$) = $0.28~\pm~0.11$} amongst three HDS observations (\S\,\ref{S-flux}).
 Due to negligibly small reddening effect, we do not correct $F_{\lambda}$
 in the longer wavelength than \emph{WISE} W1 band (3.35\,{\micron}).

  \section{Nebular line analysis} \label{S:nebula}

  \subsection{Systemic nebular radial velocity \label{S-radial}}

  We obtain the average heliocentric radial velocity
  of +166.6\,{\kms} measured from the identified
   128 nebular lines in the HDS spectrum (the standard deviation is 
   3.2\,{\kms} amongst all these lines and that of each radial velocity 
   is 0.52\,{\kms} in the average). The LSR radial
   velocity $v_{r}$(LSR) of +149.1\,{\kms}
   is much faster than $v_{r}$(LSR) in other Galactic PNe toward
   \mbox{$l$ $\sim$$220-240$$^{\circ}$}
   and \mbox{$b \lesssim \pm$10$^{\circ}$}
   \citep[$\lesssim+80$\,{\kms};][]{Quireza:2007aa}.
   $v_{r}$(LSR) of +105\,{\kms} in the C$_{60}$ PN M1-9
  \citep[PN G214.0+04.3,][]{Otsuka:2014aa} is the closest to
  SaSt\,2-3's $v_{r}$(LSR) \citep{Quireza:2007aa}.
  Peculiar velocity relative to Galactic rotation is calculated using 
 $v_{r}$(LSR) and $D$ in order to classify PNe into 
  Type\,I-IV \citep[i.e., thin/thick disc and halo; see e.g.,][]{Peimbert:1978aa}. 
  We discuss classification of SaSt\,2-3 in
  \S\,\ref{S-evolution}. We do not find a
  time-variation of the radial velocity measured by the nebular
  lines. We report the radial velocity
  measurements from the stellar absorption in \S\,\ref{S-binary}.

    \subsection{{\hb} flux of the entire nebula}
    \label{S-Hbflux}
    
    From the measured $F$(H\,{\sc i}\,7.48/7.50\,{\micron}) 
    and the theoretical $I$(H\,{\sc i}\,7.48/7.50\,{\micron})/$I$({\hb}) ratio = 3.102/100 (see \S\,\ref{S-flux}), 
    we obtain $I$({\hb}) of the entire nebula to be
  \mbox{$(1.57~\pm~0.28)(-12)$}\,erg\,s$^{-1}$\,cm$^{-2}$.  
  $F$({\ha}) and $F$({\hb}) using the 5{\arcsec} wide slit
  observation by \citet{Dopita:1997aa} and \mbox{$c$({\hb}) = $0.43~\pm~0.04$}
  (\S\,\ref{S-flux}) yields \mbox{$I$({\hb}) = 
  $(1.97~\pm~0.20)(-12)$}\,erg\,s$^{-1}$\,cm$^{-2}$, which is consistent with 
  ours. In the present work, we adopt our own calculated
  $I$({\hb}) for the entire nebula 
  because our $I$({\hb}) is based on
  interstellar extinction free and stellar H\,{\sc i} absorption effect
  less mid-IR H\,{\sc i} 7.48/7.50\,{\micron}.

  \subsection{Plasma diagnostics}

  \begin{figure}
   \includegraphics[width=\columnwidth]{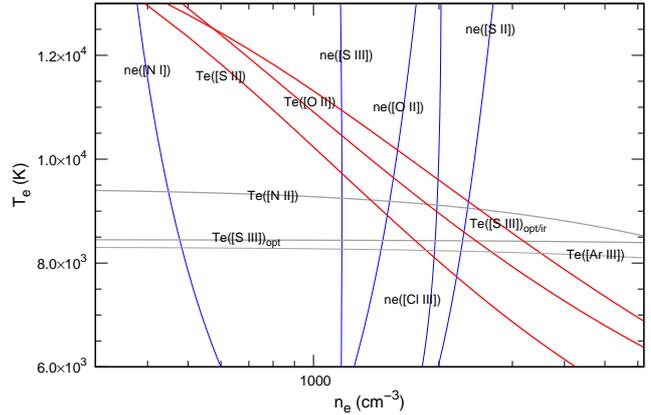}
   \vspace{-10pt}
   \caption{$n_{\rm e} - T_{\rm e}$ diagram of SaSt\,2-3. {\Ne},
   {\te}, and both {\Ne} and {\te} diagnostic curves
   are indicated by the blue, grey, and red lines, respectively.
   {\te}({\siii})$_{\rm opt}$ and {\te}({\siii})$_{\rm opt/ir}$
   curves are the results by the {\siii}\,$I$(9069\,{\AA})/$I$(6313\,{\AA}) and
   {\siii}\,$I$(9069\,{\AA})/$I$(18.71/33.47\,{\micron}) ratios,
   respectively. See also Table\,\ref{T-neTe}.
   \label{F-diagno}
   }
  \end{figure}

   \begin{table}
    \centering
    \caption{Summary of plasma diagnostics. Since 
   the lower limit {\NI} $I$(5198\,{\AA})/$I$(5200\,{\AA}) ratio is out of
   the range of the {\Ne}({\NI}) derivation, we give the upper
   limit {\Ne}($[$N\,{\sc i}$]$) = 1460\,cm$^{-3}$.
    \label{T-neTe}
    }
    \begin{tabularx}{\columnwidth}{@{}@{\extracolsep{\fill}}l@{\hspace{3pt}}
     D{p}{\pm}{-1}D{p}{\pm}{-1}@{}}
    \hline
    CEL {\Ne}-diagnostic line ratio& \multicolumn{1}{c}{Ratio}  &
    \multicolumn{1}{c}{Result (cm$^{-3}$)}   \\ 
    \hline
    $[$N\,{\sc i}$]$\,$I$(5198\,{\AA})/$I$(5200\,{\AA}) & 1.088~p~0.498
	& 460    \\ 
    {\sii}\,$I$(6717\,{\AA})/$I$(6731\,{\AA}) & 0.663~p~0.074 & 2680~p~1070 \\ 
    {\oii}\,$I$(3726\,{\AA})/$I$(3729\,{\AA}) & 1.440~p~0.163 & 1560~p~490 \\ 
    {\cliii}\,$I$(5517\,{\AA})/$I$(5537\,{\AA}) & 1.048~p~0.130 & 2180~p~1070 \\ 
    {\siii}\,$I$(18.71\,{\micron})/$I$(33.47\,{\micron}) & 1.314~p~0.334
	& 1190~p~570 \\ 
    \hline
    CEL {\te}-diagnostic line ratio   & \multicolumn{1}{c}{Ratio}   &
	    \multicolumn{1}{c}{Result (K)}   \\ 
    \hline
 
    {\nii}\,$I$(6548/83\,{\AA})/$I$(5755\,{\AA}) & 114.881~p~8.621
	& 9200~p~260 \\ 
    {\siii}\,$I$(9069\,{\AA})/$I$(6313\,{\AA}) & 14.443~p~2.558 & 8530~p~570 \\ 
    {\ariii}\,$I$(7135/7751\,{\AA})/$I$(8.99\,{\micron}) & 0.952~p~0.197 & 8250~p~790 \\ 
    \hline
    CEL {\Ne}~\&~{\te}-diagnostic line ratio   & \multicolumn{1}{c}{Ratio}   &
	    \multicolumn{1}{c}{Result (K)}   \\ 
    \hline
  {\sii}\,$I$(6717/31\,{\AA})/$I$(4069\,{\AA}) & 10.322~p~0.958 & 7220~p~430 \\ 
    {\oii}\,$I$(3726/3729\,{\AA})/$I$(7320/30\,{\AA}) & 26.725~p~2.103
	 & 9670~p~430 \\
      {\siii}\,$I$(9069\,{\AA})/$I$(18.71/33.47\,{\micron}) & 0.278~p~0.054
	 & 11\,040~p~1860 \\
      \hline
    RL {\te}-diagnostic line ratio   & \multicolumn{1}{c}{Ratio}   &
	    \multicolumn{1}{c}{Result (K)}   \\ 
    \hline
    $[I_{\lambda}(8194\,{\AA}) - I_{\lambda}(8169\,{\AA})]/I$(P11)
    & 0.022~p~0.004 & 7340~p~2610 \\ 
    {\hei}\,$I$(7281\,{\AA})/$I$(6678\,{\AA}) & 0.256~p~0.036 & 11\,830~p~1790 \\ 
   \hline
   \end{tabularx}
   \end{table}

We determine {\Ne} and {\te} using diagnostic line ratios listed in Table\,\ref{T-neTe}, 
with the resulting {\Ne}-{\te} diagnostic curves
for the collisionally excited lines (CELs)
shown in Fig.\,\ref{F-diagno}. The roughly vertical (blue) lines can
be used to determine {\Ne}; more horizontal (grey)
lines {\te}. Although the other diagnostic curves (red)
yield both {\Ne} and {\te}, we use them as {\te} indicators here.
Since the emission of each ion originates from regions
of different {\te} and {\Ne}, we need to determine
both parameters for each ion independently so that
we can determine accurate ionic abundances later on.
This involves several steps. First, we note that
{\te} $\sim$9000\,K from {\te}({\ariii}), 
{\te}({\siii})$_{\rm opt}$, and {\te}({\nii}) curves. 
Next, we adopt {\te} = 9000\,K
to solve each equation of population at $\ge 5$ multiple energy levels
for each {\Ne} sensitive ions; from this, we then calculate {\Ne}
from the corresponding diagnostic line ratios for
{\NI}, {\sii}, {\oii}, {\cliii}, and {\siii} (see Table\,\ref{T-neTe}). Note that the precise {\te} we assume 
here does not matter much, since these {\Ne} diagnostic line ratios are 
fairly insensitive to {\te}. With the {\Ne} values established, we then determine {\te}
by adopting the derived {\Ne} values corresponding to each ion. We adopt
{\Ne}({\oii}) for {\te}({\nii}) derivation. Our derived values are in agreement with 
those by \citet{Pereira:2007aa} who found \mbox{{\Ne}({\sii}) = $2100~\pm~600$\,cm$^{-3}$} 
and \mbox{{\te}({\nii}) = $9600~\pm~930$\,K}. 

We compute {\te}({\hei}) using 
singlet {\hei} lines. 
To calculate {\te}(PJ) from the Paschen continuum discontinuity 
by utilising the equation (7) of \citet{Fang:2011aa}, first 
we determine the He$^{+}$ abundance of \mbox{$1.09(-2)~\pm~2.28(-4)$} 
under the obtained {\te}({\hei}). Eventually, 
we utilise {\te}({\hei}) for both He$^{+}$ and C$^{2+}$ abundance 
calculations due to higher {\te}(PJ) uncertainty.

  \subsection{\label{S-abund} Ionic abundance derivations}

      \begin{table}
       \centering
       \caption{Adopting {\te} and {\Ne} for the CEL ionic abundance
       calculations.
       \label{T-tene}
       }
    \begin{tabularx}{\columnwidth}
     {@{\extracolsep{\fill}}lD{p}{\pm}{-1}D{p}{\pm}{-1}@{}}
    \hline
    Ion &\multicolumn{1}{c}{{\te} (K)}&\multicolumn{1}{c}{{\Ne} (cm$^{-3}$)}\\
    \hline
    N$^{0}$, O$^{0}$, S$^{+}$ &
    \multicolumn{1}{c}{{\te}({\sii})}&\multicolumn{1}{c}{{\Ne}({\sii})}\\
    N$^{+}$ &\multicolumn{1}{c}{{\te}({\nii})}
    &\multicolumn{1}{c}{{\Ne}({\oii})}\\
    O$^{+}$ &\multicolumn{1}{c}{{\te}({\oii})}
    &\multicolumn{1}{c}{{\Ne}({\oii})}\\
  O$^{2+}$, Ne$^{+}$&9270~p~1070&1690~p~820\\
  S$^{2+}$ &9790~p~1220&\multicolumn{1}{c}{{\Ne}({\siii})}\\
  Ar$^{2+}$ &\multicolumn{1}{c}{{\te}({\ariii})}&1690~p~820\\ 
  Cl$^{+}$, Fe$^{2+}$ &9440~p~350&\multicolumn{1}{c}{{\Ne}({\oii})}\\
  Cl$^{2+}$ &9270~p~1070&\multicolumn{1}{c}{{\Ne}({\cliii})}\\
    \hline
   \end{tabularx}
      \end{table}

 We calculate the CEL
 ionic abundances by solving an equation of population at multiple
 energy levels under the adopted {\te} and {\Ne} as listed in
 Table\,\ref{T-tene};
 {\te} = 9270\,K is the average
 value amongst two {\te}({\siii}) and {\te}({\ariii}), 
 {\te} = 9790\,K is the average
 value amongst two {\te}({\siii}), {\te} = 9440\,K
 is the average between {\te}({\nii}) and {\te}({\oii}), and {\Ne} =
 1690\,cm$^{-3}$ is the average between {\Ne}({\cliii}) and
 {\Ne}({\siii}). For the recombination line (RL) He$^{+}$ and C$^{2+}$,
 we adopt {\te}({\hei}) and
 {\Ne} = 10$^{4}$\,cm$^{-3}$.
 Our choice of the {\te}-{\Ne} pair of each ion depends on the
 potential (IP) of the targeting ion. Except
 for the CEL N$^{+}$, O$^{+,2+}$, and S$^{+}$
 which \citet{Pereira:2007aa} already measured,
 the first measurements of all the ionic abundances
 are done by us.
 We summarise the resultant CEL and RL ionic abundances in Appendix
 Table~\ref{T-ionic}. 
 We calculate each ionic abundance using each line intensity.
 Then, we adopt the weight-average value as the representative
 ionic abundance as listed in the last line of each ion.
 We give 1-$\sigma$ uncertainty of each ionic abundance,
 which accounts for the uncertainties of line fluxes
 (including $c$({\hb}) uncertainty), {\te}, and {\Ne}.

The He$^{+}$ abundance of 9.72(--3) in SaSt2-3 is ten times smaller than in evolved PNe 
(e.g., $T_{\rm eff} \gtrsim 50\,000$\,K). For instance, in the C$_{60}$ PN M\,1-20 
\citep[$T_{\rm eff} = 45\,880$\,K,][]{Otsuka:2014aa}, \citet{Wang:2007aa} 
find He$^{+}$ abundance of 9.50(--2). Moreover, the He$^{+}$ abundance is also 
significantly lower than in other Galactic C$_{60}$ PNe 
with  $T_{\rm eff} \lesssim 40\,000$\,K where He$^{+}$ abundances have 
been determined: 6.99(--2) in IC\,418 \citep{Hyung:1994aa},
6.57(--2) in M\,1-6 (Otsuka in prep), 
3.93(--2) in M\,1-11 \citep{Otsuka:2013aa}, 3.5(--2) 
in M\,1-12 \citep{Henry:2010aa}, 
and 6.0(--2) in Tc\,1 \citep{Pottasch:2011aa}.
Similar to the C$_{60}$ PN Lin\,49 in the SMC \citep{Otsuka:2016aa}, 
the low He$^{+}$ abundance is due to the smaller number of 
ionising photons for He$^{+}$ ($\geq 21$\,eV):  
using the spectra synthesised by {\sc TLUSTY} (with 
$L_{\ast} = 7000$\,L$_{\sun}$, $\log\,g$ = 3.11 cm\,s$^{-2}$, 
metallicity $Z = 1/10$\,Z$_{\sun}$, see below), 
we estimate the number of photon with energy $\geq 21$\,eV to 
be 8.3(+45)~s$^{-1}$ in a $T_{\rm eff} = 28\,100$\,K star like SaSt\,2-3 
(see \S\,\ref{S:stellar}) and 4.8(+46)~s$^{-1}$ in $T_{\rm eff}  
= 32\,000$\,K stars like M\,1-11 and M\,1-12. 
Thus, the majority of the He atoms in SaSt\,2-3 are in the neutral state.

 The higher multiplet C\,{\sc ii} lines are generally
 reliable because these lines are less
 affected by resonance fluorescence.
 However, the higher C$^{2+}$
 abundances from the C\,{\sc ii}\,3918.98/20.69\,{\AA} ($4s ^{2}S
 - 3p ^{2}P$) and 7231.32/36.42\,{\AA} ($3d ^{2}D - 3p ^{2}P$)
 are likely due to the enhancement by resonance
 from the 635.25/636.99\,{\AA}
 ($4s ^{2}S - 2p ^{2}P$) and the 687\,{\AA}
 ($3d ^{2}D - 2p ^{2}P$), respectively. Thus, we
 exclude the C$^{2+}$
 abundances from these C\,{\sc ii} lines and
 C\,{\sc ii}\,6451.95\,{\AA}\footnote{Because the C$^{2+}$
 abundance from this line
 is about three time larger than that from the C\,{\sc ii}\,4267\,{\AA}.
 The C\,{\sc ii}\,4267\,{\AA} is the most reliable RL C$^{2+}$
 indicator.} in the representative RL
 C$^{2+}$ determination.

  Our N$^{+}$ and O$^{+,2+}$ are comparable
  with \citet{Pereira:2007aa}, who calculated
  N$^{+}$ = 2.42(--5),
  O$^{+}$ = 1.87(--4), O$^{2+}$ = 1.22(--6),
  and S$^{+}$ = 3.0(--7)
  (they note that
  their derived ionic abundances has $\pm$30\,$\%$
  uncertainty) under {\te}({\nii}) = 9600\,K and {\Ne}({\sii})
  = 2100\,cm$^{-3}$. The discrepancy between their
  and our S$^{+}$ (5.89(--7)) is caused by {\te}
  selection;
    if we adopt {\te} = 9600\,K and {\Ne} = 2100\,cm$^{-3}$, we obtain
  S$^{+}$ = 3.48(--7).

  \subsection{\label{S-abund2} Elemental abundance derivations using the ICFs}

 \begin{table}
  \centering
  \caption{Nebular elemental abundances using the ICFs. The last column
  (PM07) is the $\epsilon$(X) value derived
  by \citet{Pereira:2007aa}. \label{T-e}}
 \begin{tabular}{@{}cc@{\hspace{7pt}}c@{\hspace{7pt}}c@{\hspace{7pt}}c@{}}
 \hline
 X&\multicolumn{1}{c}{$n$(X)/$n$(H)}&\multicolumn{1}{c}{$\epsilon$(X)}
 &\multicolumn{1}{c}{$\epsilon$(X) $-$ $\epsilon$(X$_{\sun}$)}&\multicolumn{1}{c}{$\epsilon$(X)}\\
&\multicolumn{1}{c}{(Ours)} &\multicolumn{1}{c}{(Ours)}
  &\multicolumn{1}{c}{(Ours)}
      &\multicolumn{1}{c}{(PM07)}\\
 \hline
He& 5.58(--2) -- 1.26(--1)    & 10.75 -- 11.10     & \phantom{..}--0.15 -- +0.20    &$\cdots$\\ 
C & 1.61(--3) $\pm$ 4.61(--4) & 9.21 $\pm$ 0.12  & +0.82 $\pm$ 0.13 &$\cdots$  \\ 
N & 2.95(--5) $\pm$ 4.09(--6) & 7.47 $\pm$ 0.06  & --0.36 $\pm$ 0.13 &7.38 $\pm$ 0.14\\ 
O & 1.30(--4) $\pm$ 1.10(--5) & 8.11 $\pm$ 0.04  & --0.58 $\pm$ 0.06 &8.27 $\pm$ 0.14\\ 
Ne & 2.91(--5) $\pm$ 2.85(--6) & 7.46 $\pm$ 0.04 & --0.41 $\pm$ 0.11 &$\cdots$\\ 
S & 1.26(--6) $\pm$ 8.80(--8) & 6.10 $\pm$ 0.03  & --1.09 $\pm$ 0.05 &5.48 $\pm$ 0.14\\ 
Cl & 3.68(--8) $\pm$ 5.51(--9) & 4.57 $\pm$ 0.07 & --0.69 $\pm$ 0.09 &$\cdots$\\ 
Ar & 4.62(--7) $\pm$ 1.43(--7) & 5.66 $\pm$ 0.13 & --0.89 $\pm$ 0.16 &$\cdots$\\ 
Fe & 1.94(--7) $\pm$ 2.74(--8) & 5.29 $\pm$ 0.06 & --2.18 $\pm$ 0.07 &$\cdots$\\ 
\hline
\end{tabular}
 \end{table}

  To obtain the elemental abundances using the derived ionic
  abundances, we introduce the ionisation correction factors
  \citep[ICFs, see e.g.,][for details]{Delgado-Inglada:2014ab}.
  Here, the number density ratio of the element X with respect to the
  hydrogen, $n$(X)/$n$(H) is equal to
  \mbox{ICF(X)$~\cdot~\sum_{\rm m = 1}$\,$n$(X$^{\rm m+}$)/
  $n$(H$^{+}$)}. The ICFs have been empirically determined 
  based on the fraction of observed ion number densities with 
  similar ionisation potentials to the target element, and have also 
  been determined based on the fractions of the ions calculated 
  by photoionisation models. Since SaSt\,2-3 is very low-excitation PN, 
  the ICFs (He in particular) adopted for more highly excited 
  PNe do not work well. Therefore, we need a special treatment for SaSt\,2-3. 
  Thus, in addition to the ICFs established by photoionisation grid models 
  of \citet{Delgado-Inglada:2014ab}, we refer to ICFs 
  used in Lin\,49 by \citet{Otsuka:2016aa}.

 In Lin\,49, the C$^{2+}$/C ratio is similar
 to the Ar$^{2+}$/Ar ratio. 
  For SaSt\,2-3, we adopt equation (A6) of \citet{Otsuka:2016aa}
  for C and Ar. Both ICF(C) and
  ICF(Ar) are \mbox{2.7~$\cdot$~(Cl/Cl$^{2+}$)}
  =  \mbox{$5.12~\pm~1.41$}. 
  For He derivation, we adopt two ICF(He) 
  calculated using the equation (42) of 
  \citet[][11.57]{Peimbert:1969aa} and from the ratio of 
  \mbox{Ar/Ar$^{2+}$} (5.12). 
  ICF(N) = \mbox{$1.04~\pm~0.13$}
   is from \citet{Delgado-Inglada:2014ab}.
   ICF(Fe) =
   \mbox{$1.31~\pm~0.16$} is from \citet{Delgado-Inglada:2014aa}
   based on the observation results. ICF of the other elements is unity. 
   We verify whether the adopted ICFs here are proper by comparing with
   {\sc Cloudy} photoionisation model (\S\,\ref{S:cloudy}).

   In the second and third columns of Table\,\ref{T-e}, we present the resultant
  elemental abundances with 1-$\sigma$ uncertainty, 
  except for $\epsilon$(He), where we adopt its range.
  The two columns are the relative abundance
  to the solar value by \citet{Lodders:2010aa} and the $\epsilon$(X) by
  \citet{Pereira:2007aa}. 
  Our $\epsilon$(N) and $\epsilon$(O) are consistent
   with \citet{Pereira:2007aa}. As explained in
   \S\,\ref{S-abund}, $\epsilon$(S) discrepancy between
   theirs and ours is attributed to the S$^{+}$ abundance. 
   By the {\sc Cloudy} model under $D$ = 6\,kpc,
   \citet{Otsuka:2014aa} derived $\epsilon$(N/O/Ne/S/Ar) = 7.49, 8.23,
   7.68, 6.17, and 5.93 based on the optical spectrum of
   \citet{Pereira:2007aa} and the \emph{Spitzer}/IRS spectrum. 
   \citet{Otsuka:2014aa} estimated an expected CEL $\epsilon$(C) = 8.72
   using a [C/H]$-$[C/Ar] relation established amongst 115 Galactic PNe.
  \citet{Delgado-Inglada:2014aa} reported that
  the RL C$^{2+}$ to the CEL C$^{2+}$ ratio in
  IC\,418 is 2.4. Applying this value to SaSt\,2-3,
  we obtain an expected CEL $\epsilon$(C) =
  \mbox{$8.83~\pm~0.12$}, which is consistent with \citet{Otsuka:2014aa}.
  We attempt to obtain more plausible expected CEL $\epsilon$(C)
  using the stellar $\epsilon$(C) and $\epsilon$(O) in \S\,\ref{S-stellar}. 
  The [Ne/H] is comparable with the [O/H] because Ne together with O
  had been synthesised in the He-rich intershell during the AGB phase.
  The Ne enhancement would be due to the increase of $^{22}$Ne.

  The [S,Cl,Ar/H] abundances are low, and if these represent the stellar 
  abundances, then SaSt\,2-3 is the lowest metallicity object 
  amongst the Galactic C$_{60}$ PNe, and we infer 
  $Z$ $\sim$0.1\,Z$_{\sun}$ from the average 
  [S,Cl,Ar/H]. While most 
  Ar is probably in the gas phase in this object, S could be 
  incorporated into dust grains (e.g. MgS, suggested to 
  be a candidate for the carrier of the broad 30\,{\micron} feature that 
  is observed in the C$_{60}$ PNe). Fe is even more depleted, 
  but it is unlikely that this represents the initial abundance 
  given the other elemental abundances. Rather, a fraction of the 
  Fe will be incorporated into dust grains. 
  We discuss further the elemental abundances in \S\,\ref{S:AGB}.

  \section{Stellar absorption analysis } \label{S:stellar}

   \begin{figure*}
    \renewcommand{\thetable}{A\arabic{table}}
 \centering
  \includegraphics[width=0.85\textwidth,clip,viewport=0 138 424 254]{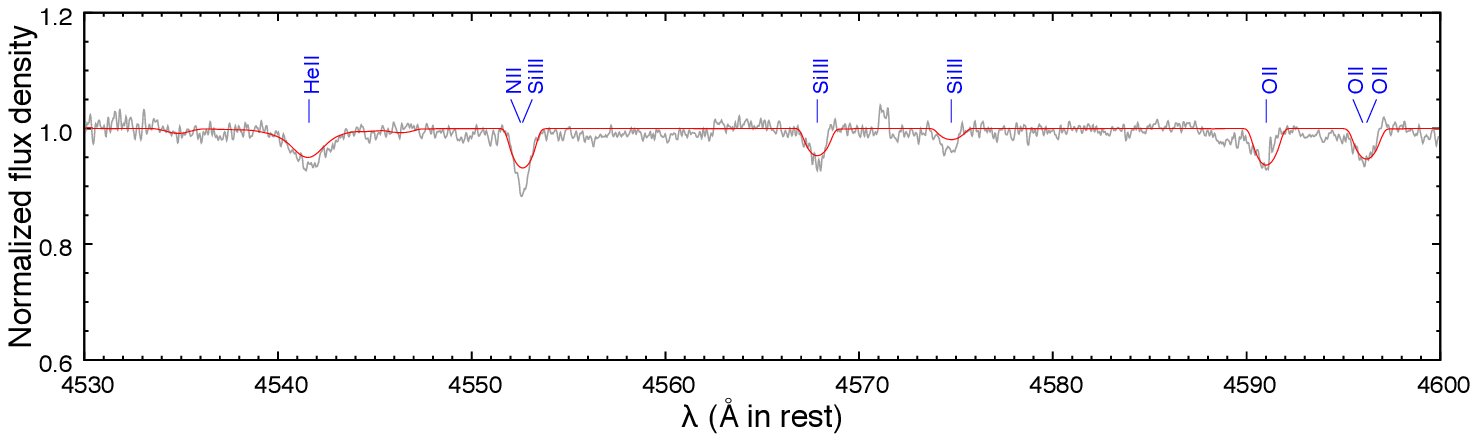}\\
  \includegraphics[width=0.85\textwidth,clip,viewport=0 138 424 254]{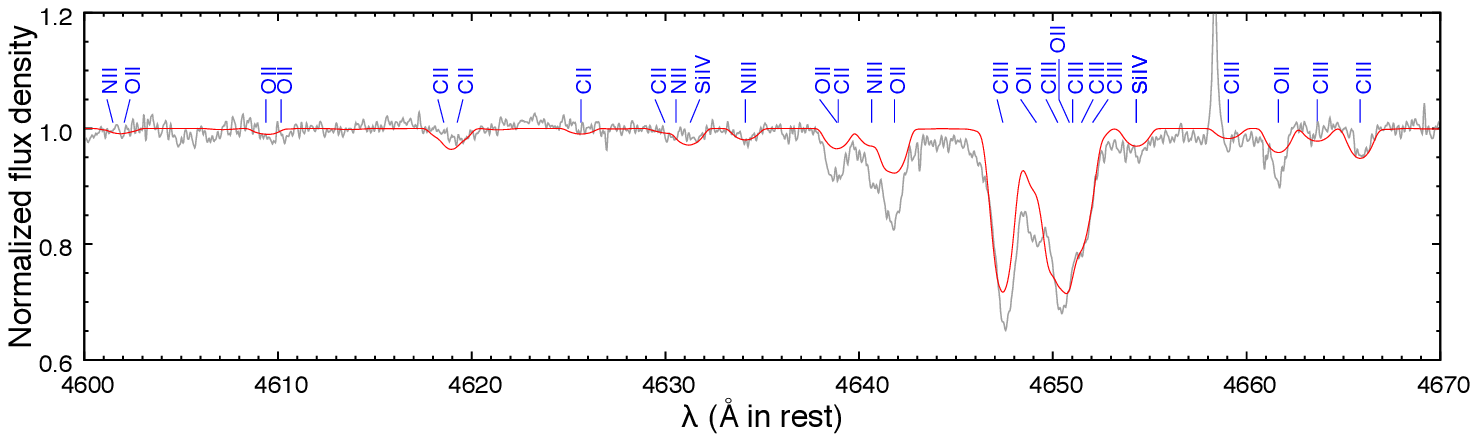}\\
  \includegraphics[width=0.85\textwidth,clip,viewport=0 127 424 254]{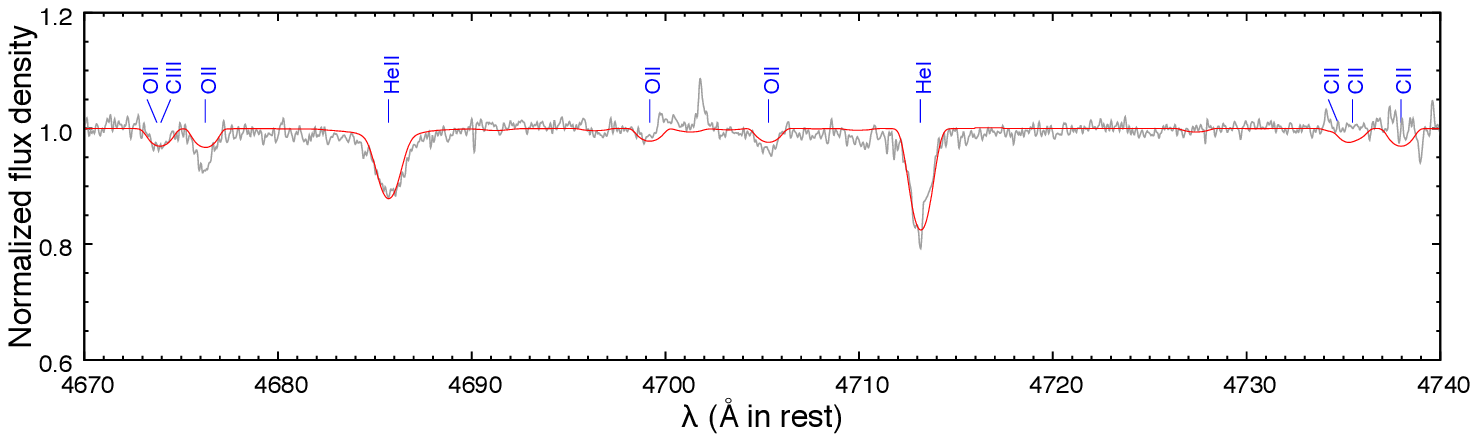}
 \caption{Comparison between the observed HDS (grey line)
      and the {\sc TLUSTY} synthetic spectrum (red line) in the range
    between 4530 and 4740\,{\AA}.
    The absorption lines (except for H\,{\sc i})
    with the model predicted $EW \ge 10$\,m{\AA} are
    indicated by the blue lines.
    The input parameters are listed in Table\,\ref{T-stellar}.
      \label{F-synspec}
      }
   \end{figure*}

  \subsection{Stellar parameter derivations \label{S-stellar}}

  We perform stellar absorption analysis of the HDS spectra
  taken on 2013 Oct 6 and Dec 10 using the non-local thermodynamic
  equilibrium (non-LTE) stellar atmosphere modelling code
  {\sc TLUSTY}. We detect strong Si\,{\sc iii,iv} and He\,{\sc ii}
  absorption lines. From our {\sc TLUSTY} modelling,
  $T_{\rm eff}$ of the central star is 28\,100\,K
  (Table\,\ref{T-stellar}), which is cooler than
  $T_{\rm eff} \ge 30\,000$\,K in the O-type stars. Thus,
  we classify the stellar spectrum of
  SaSt\,2-3 into early B-type giant B0-1II rather than O-type. 
  Thus, we use comprehensive grid of 1540 metal line-blanketed, non-LTE,
  plane-parallel,
  hydrostatic model atmospheres of
  B-type stars BSTAR2006\footnote{
  \url{http://tlusty.oca.eu/Tlusty2002/tlusty-frames-BS06.html}}
  by \citet{Lanz:2007aa}.

  We find the average nebular [Cl,S,Ar/H] of \mbox{$-0.89~\pm~0.12$}
  (\S\,\ref{S-abund2}). Assuming that
  the metallicity $Z$ of the central star and the nebula is roughly the
  same as seen in the case of IC\,418 by \citet{Morisset:2009aa},
  we adopt the $Z = 0.1$\,Z$_{\sun}$ model grid from BSTAR2006. All of
  the {\it initial} abundances in this model grid
  are set to $\epsilon$(He) = 11.00 and [X/H] = --1 except for He.
  Based on the $Z = 0.1$\,Z$_{\sun}$ model grid, we
  vary $\epsilon$(X) to yield each equivalent
  width ($EW$) of element X to compare with each $EW$(X) measured from
  the observed HDS spectra. Throughout our {\sc
  TLUSTY} synthesis analysis, we do not set [He,C,N,O,Si/H] = --1
  and we do not adopt the derived nebular
  He,C,N,O/H] as the stellar photospheric ones. Based on the
  measured $EW$ of the identified
  9 He\,{\sc i,ii}, 4 C\,{\sc iii,iv}, 2 N\,{\sc ii,iii}, 13 O\,{\sc
  ii}, and 5 Si\,{\sc iii,iv} absorption, we derive the 
  photospheric He/C/N/O/Si abundances, microturbulent velocity
  ($v_{\rm t}$), rotational velocity ($v\,\sin(i)$; $i$ is the angle
  between the rotation axis and the line of sight),
  $T_{\rm eff}$, and $\log$\,$g$ of the central star.
  These absorption
  lines are lesser affected by the nearby nebular lines and absorption
  lines of the other elements.
  As we report later, the central wavelength of the stellar
  absorption lines changes between observing dates whereas those
  of the nebular lines remain constant.
  Before analysis, we convert heliocentric wavelength frame of the HDS
  spectrum into rest frame using the radial velocity
  determined by the He\,{\sc
  ii}\,4686\,{\AA} for the 2013 Oct 6 data (\mbox{$154.8~\pm~2.3$\,{\kms}})
  and the He\,{\sc ii}\,5411\,{\AA} for the 2013 Dec 10 data 
  (\mbox{$182.3~\pm~4.0$\,{\kms}}).

  First, we set the basic parameters characterising the
  stellar atmosphere, i.e., $Z$, $v_{\rm t}$,
  $T_{\rm eff}$, and $\log$\,$g$. By setting
  $T_{\rm eff} = 28\,000$\,K and $\log\,g = 3.10$\,cm\,s$^{-2}$,
  we investigate $\epsilon$(O) versus selected 8 O\,{\sc ii} lines' $EW$
  to determine $v_{\rm t}$ using {\sc SYNFIT}.
  For each absorption, we set instrumental line broadening determined
  by measuring Th-Ar line widths. Since $v_{\rm t} \geq 10$\,{\kms}
  gives minimisation of the scatter in $\epsilon$(O) versus $EW$,
  we adopt $v_{\rm t} = 10$\,{\kms}. As a reference, \citet{Morisset:2009aa} 
  adopted $v_{\rm t} = 10$\,{\kms} for IC\,418.

  We determine $T_{\rm eff}$ and $\log\,g$
  using the $T_{\rm eff}$-$\log\,g$ curves generated by
  the model atmosphere with $v_{\rm t}
  = 10$\,{\kms}, $Z$ = 0.1\,Z$_{\sun}$, and $\epsilon$(He)
  = 10.90. The $T_{\rm eff}$-$\log\,g$ curves are 
  generated by the following process; 
  for a fixed $T_{\rm eff}$, we vary $\log\,g$ from $2.90-3.30$\,cm\,s$^{-2}$ 
  in a constant 0.01\,cm\,s$^{-2}$ step to find the best 
  fit value for each observed 
  He\,{\sc i,ii}'s EW. We test the range of $T_{\rm eff}$ 
  from $27\,500-28\,500$\,K (200\,K step).

  Based on the determined $v_{\rm t}$, $T_{\rm eff}$, and $\log$\,$g$,
  we further constrain $\epsilon$(He) and calculate $\epsilon$(C/N/O/Si)
  abundances by comparing the
  observed and model predicted $EW$s of each line by {\sc
  SYNABUND}\footnote{
  {\sc SYNABUND} is a code
  developed by Prof. I.~Hubeny in order to calculate $EWs$
  under {\sc TLUSTY} stellar model atmosphere.
  }. We
  summarise the result in Table\,\ref{T-stellar}. In Appendix
  Table\,\ref{T-stellar2}, we list the elemental abundances using
  each line. We adopt the average value as the representative
  abundance as listed in the last line of each element.
  The uncertainty of elemental abundances includes errors
  from the measured $EW$s, $T_{\rm eff}$, $\log$\,$g$, and
  the uncertainty when we adopt the model atmosphere with the
  [Z/H] = $-0.90$ or $-1.10$ and when we assume the uncertainty of
  $v_{\rm t}$ of 2\,{\kms}. We determine $v\,\sin(i)$ by line-profile
  fittings of the selected He\,{\sc i,ii} and H\,{\sc i}
  in $4000-4700$\,{\AA} using {\sc SYNFIT}\footnote{
  {\sc SYNFIT} is a code
  developed by Prof. I.~Hubeny in order to synthesise
  line-profiles under {\sc TLUSTY} stellar model atmosphere. 
  }

    \begin{table}
     \centering
     \caption{Summary of the set and derived parameters of the central
     star. \label{T-stellar}
     }
       \begin{tabularx}{\columnwidth}{@{\extracolsep{\fill}}
 cD{p}{\pm}{-1}cD{p}{\pm}{-1}@{}}
 \hline
 Parameter &\multicolumn{1}{c}{Value}&Parameter &\multicolumn{1}{c}{Value}\\
 \hline
       $T_{\rm eff}$ (K) &28\,100 ~p~ 300&$\epsilon$(He)&10.99 ~p~ 0.09\\
       $\log\,g$     (cm\,s$^{-2}$)&3.11 ~p~ 0.05&$\epsilon$(C) &8.56 ~p~ 0.10\\
       $v_{\rm t}$   ({\kms}) &10 ~p~ 2
    &$\epsilon$(N) &7.26 ~p~ 0.16\\
       $v\,\sin(i)$   (\kms)&56 ~p~ 4      &$\epsilon$(O) &8.10 ~p~ 0.17\\
                        &                &$\epsilon$(Si) &6.81 ~p~ 0.10\\
       \hline
      \end{tabularx}
    \end{table}

  In Fig.\,\ref{F-synspec}, we show the
  synthetic stellar spectrum generated using
  {\sc SYNSPEC}\footnote{\url{http://nova.astro.umd.edu/Synspec49/synspec.html}}.
  We identify the absorption lines (except for H\,{\sc i})
  with the model predicted
  $EW \ge 10$\,m{\AA} by the blue
  lines. The synthetic spectrum in $3700-4750$\,{\AA} is presented
  in Appendix Fig.\,\ref{F-synspecA}. Stellar Ne, S, Ar, Mg, Ca, and Ti
  ($\alpha$-elements) and Ni, Fe, and Zn are not derived in
  optical HDS spectra of SaSt\,2-3.
  These abundances are not small and also they are
  very important in characterising the spectrum of the central star and
  its radiation hardness (in particular, X-ray to UV wavelength). We
  know that the central star radiation is suppressed by the metal line-blanket
  effect and also its is very important in subsequent {\sc Cloudy} modelling.
  Thus, it is worth simulating these elements, too. 
We adopt the nebular Ne, S, Cl, and Ar abundances due to
  no detection of stellar absorption of these elements. We adopt
      $\epsilon$(Fe) = 6.38 ([Fe/H] = $-1.1$, see \S\,\ref{S-abund2}).
      Based on the
      discussion in \S\,\ref{S:AGB}, for the other elements up to Fe
      except for $\alpha$-elements Mg, Ca, and Ti, we adopt the
      predicted values by the AGB nucleosynthesis model of
      initially 1.25\,M$_{\sun}$ and $Z = 0.001$ stars by
      \citet{Fishlock:2014aa}.  For Mg, Ca, and Ti, we adopt
      $\epsilon$(Mg) = 6.80,
      $\epsilon$(Ca) = 5.43, and
      $\epsilon$(Ti) = 4.05, respectively (i.e., [Mg,Ca,Ti/H] = $-0.7$).

  The stellar $\epsilon$(He/N/O)
  is in agreement with the nebular
  $\epsilon$(He/N/O) within their uncertainties. 
  The stellar C/O ratio (\mbox{$2.93~\pm~1.33$}) indicates that
  SaSt\,2-3 is definitely a C-rich PN. Based on the consistency between the
  stellar and the nebular elemental abundances, we obtain an 
  expected CEL $\epsilon$(C) = \mbox{$8.58~\pm~0.20$} using the
  CEL $\epsilon$(O) and the stellar C/O ratio.

  \subsection{Time variation of line-profile and radial velocity;
  Evidence of a binary central star }
 \label{S-binary}

  Our important discovery is that the central wavelength of the 
  stellar absorption lines varies from date to date whereas there is no
  wavelength shift of the nebular emission lines.

\begin{figure}
    \centering
   \includegraphics[width=0.8\columnwidth,clip]{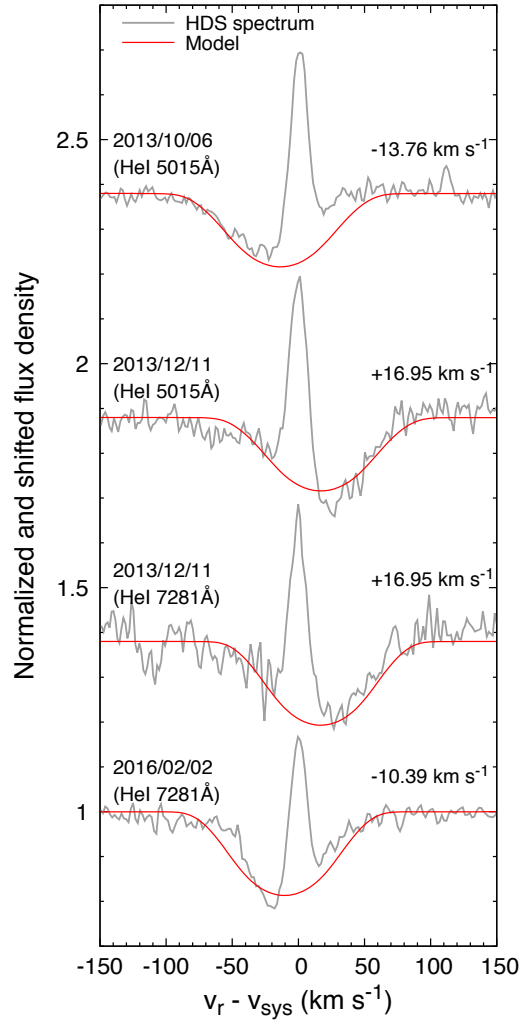}
    \vspace{-5pt}
  \caption{Time-variation of the singlet {\hei} line-profiles taken
   in three nights. The systemic radial velocity ($v_{\rm sys}$) is
    +166.6\,{\kms} (\S\,\ref{S-radial}). The synthetic spectrum
    is overplotted.
    The heliocentric radial velocity $v_{r}$ with respect
   to $v_{\rm sys}$ is indicated (see Table\,\ref{T-rv}).}
 \label{lineprof}
   \end{figure}

  We compute the
  heliocentric radial velocities $v_{r}$ of the central
  star via Fourier cross
  correlation between
  the observed spectra and the synthetic {\sc TLUSTY} spectrum using
  {\sc FXCOR} in {\sc IRAF}. {FXCOR} calculates the velocity
  shift between two different spectra in the selected wavelength
  regions\footnote{
  The wavelength ranges we set are as follows; for the 2013 Oct 6 spectrum,
  $3911-3932$, $4002-4036$,
  $4082-4096$, $4112-4182$,
  $4302-4330$, $4357-4442$, and $4533-4730$\,{\AA}. For
  the 2013 Dec 10, $5014-5023$, $4920-4934$, $5043-5056$,
  $5403-5420$, and  $5589-5599$\,{\AA}. For the 2016 Feb 1, $7280-7293$\,{\AA}.
  }. 
  Here, we select good S/N regions. 
  In Table\,\ref{T-rv}, we list $v_{r}$ and $v_{r} -
  v_{sys}$, where $v_{\rm sys}$ is the systemic radial velocity measured
  from the 128 nebular emission lines (+166.6\,{\kms}, see
  \S\,\ref{S-radial}). In Fig.\,\ref{lineprof}, we show the singlet He\,{\sc
  i}\,5015/7281\,{\AA} absorption and the {\sc TLUSTY}
  synthetic spectrum as the guide.

  We interpret that the radial velocity time-variation is caused by orbital
  motion in a binary system.
  \citet{Mendez:1986aa}
  reported the photometric and radial velocity variations
  of the CSPN of C$_{60}$ PN IC\,418. They measured the radial velocities using
  the stellar C\,{\sc iii}\,5695\,{\AA} and C\,{\sc
  iv}\,5801/11\,{\AA}. The systemic radial velocity was derived
  using the nebular {\nii}\,5755\,{\AA} line.
  Later, \citet{Mendez:1989aa}
  concluded that the central star is not likely to
  be a binary because the orbital motion alone (if present) would not be
  enough to explain the observed variations. We note that
  the C\,{\sc iii}\,5695\,{\AA} and C\,{\sc iv}\,5801/11\,{\AA}
  lines are good indicators of the stellar activity (e.g., wind
  velocity) and these lines would be unlikely to give more accurate
  radial velocity of the central star. Thus, as far as we know,
  this would be the firm detection case of the binary central stars
  amongst all the C$_{60}$ PNe. Since we have only three periods of the
  binary motion, we do not determine any parameters of the binary
  central star yet.

  We expected near-IR excess from the binary circumstellar
  disc from \citet{Otsuka:2016aa} who
  detected near-IR excess in most of the SMC C$_{60}$ PNe
  and discussed possible links between near-IR excess, disc, and fullerene
  formation; since the ejected material from the central star can be
  stably harboured for a long time, even smaller molecules could aggregate
  into much larger molecules. However, in SaSt\,2-3, we do not
  find near-IR excess in the observed SED (Fig.\,\ref{F-sedobs}).
  No near-IR excess might
  mean a possibility of a nearly edge-on disc rather an inclined disc.

      \begin{table}
       \centering
       \caption{Heliocentric radial velocity $v_{r}$ of the central
       star. The systemic radial velocity $v_{sys}$
       is +166.6\,{\kms} (\S\,\ref{S-radial}).}
       \label{T-rv}
      \begin{tabularx}{\columnwidth}{@{\extracolsep{\fill}}lD{.}{.}{-1}cc@{}}
       \hline
       Obs Date &\multicolumn{1}{c}{JD (-- 2456000.0)} &$v_{r}$
       ({\kms})&$v_{r} - v_{sys}$ ({\kms})\\
       \hline
       2013/10/06 &573.096 &+152.8 $\pm$ 0.6&--13.76\\
       2013/12/10 &637.917 &+183.6 $\pm$ 0.4&+16.95\\
       2016/02/01 &1420.831 &+156.2 $\pm$ 0.6&--10.39\\
       \hline
      \end{tabularx}
      \end{table}

 \section{Comparison with AGB model predictions}  \label{S:AGB}

     \begin{table}
    \centering
    \caption{Comparison between the derived abundances
    and the predicted values by the AGB nucleosynthesis models of
    \citet{Fishlock:2014aa} for
    1.25\,M$_{\sun}$ stars with $Z = 0.001$
    and \citet{Karakas:2010aa} for 1.5\,M$_{\sun}$ stars with $Z =
    0.004$. The CEL $\epsilon$(C) is the predicted value
     by our analysis (\S\,\ref{S-stellar}).
    }
   \begin{tabularx}{\columnwidth}{@{\extracolsep{\fill}}ccD{p}{\pm}{-1}rr@{}}
 \hline
   X &\multicolumn{1}{c}{Nebular}  &\multicolumn{1}{c}{Stellar}
   &1.25\,M$_{\sun}$&1.50\,M$_{\sun}$\\
   &&    &$Z = 0.001$&$Z = 0.004$\\
   \hline
 He         &10.75 -- 11.10 &10.99 ~p~ 0.09&11.01&10.97\\
 C(RL)      &9.21 $\pm$ 0.12&8.55 ~p~ 0.10 &8.56 &8.46 \\
 C(CEL)     &8.58 $\pm$ 0.20&\multicolumn{1}{c}{$\cdots$}\\
 N          &7.47 $\pm$ 0.06 &7.25 ~p~ 0.16  &7.26 &7.65\\
 O          &8.11 $\pm$ 0.04 &8.10 ~p~ 0.17  &7.68 &8.23\\
 Ne         &7.46 $\pm$ 0.04 &\multicolumn{1}{c}{$\cdots$}&7.37 &7.42 \\
 Si         &\multicolumn{1}{c}{$\cdots$}&6.81 ~p~ 0.10   &6.39 &6.85 \\
 S          &6.10 $\pm$ 0.03 &\multicolumn{1}{c}{$\cdots$}&6.00 &6.70 \\
 Cl         &4.57 $\pm$ 0.07 &\multicolumn{1}{c}{$\cdots$}&4.08 &$\cdots$ \\
 Ar         &5.66 $\pm$ 0.13 &\multicolumn{1}{c}{$\cdots$}&5.28 &$\cdots$ \\
 Fe         &5.29 $\pm$ 0.06 &\multicolumn{1}{c}{$\cdots$}&6.38 &6.80 \\
\hline
  \end{tabularx}
    \label{T-AGB}
   \end{table}

   In Table\,\ref{T-AGB}, we compile the derived abundances.
 The nebular CEL $\epsilon$(C) is an expected value by our analysis
 (\S\,\ref{S-stellar}). As the comparisons, we list the AGB
 nucleosynthesis model predictions by \citet{Fishlock:2014aa} for 
 initially 1.25\,M$_{\sun}$ stars with $Z = 0.001$ and
 \citet{Karakas:2010aa} for initially 1.50\,M$_{\sun}$ stars
 with $Z = 0.004$. Note that \citet{Fishlock:2014aa} and
 \citet{Karakas:2010aa} set the initial [X/H] to be
 --1.1 and --0.7, respectively. We calculate reduced
 chi-squared values ($\chi_{\nu}^{2}$, $\nu$ is degree of freedom) between the nebular
 $\epsilon$(X) and the AGB model predicted values for each
 of $1.00-3.25$\,M$_{\sun}$ star with $Z = 0.001$ (9 models in total).
 We use $\chi_{\nu}^{2}$ as the guide to find out which AGB model's
 predicted abundances is the closest to the derived abundances.
 The aim of this analysis is to infer the initial mass of the
 progenitor. 
 We should note that these AGB grid models do not aim to explain the observed elemental abundances of 
 SaSt\,2-3. We exclude Fe in $\chi_{\nu}^{2}$ evaluation.
 We adopt the nebular $\epsilon$(X) values. For $\epsilon$(He) and $\epsilon$(C), we 
 adopt \mbox{$10.96~\pm~0.17$} (intermediate value, \mbox{$9.10(-2)~\pm~3.52(-3)$}) 
and \mbox{$8.58~\pm~0.20$} (an expected nebular CEL C value, \mbox{$3.80(-4)~\pm~1.76(-4)$}), 
 respectively.

 Since the reduced-$\chi^{2}$ for the 1.25\,M$_{\sun}$ model marks the minimum
 ($\chi_{7}^{2}$ = 14 (= 99/($8-1$)) in 8 elements), 
 this model is the closet to the
 derived $\epsilon$(X). $\chi_{4}^{2}$ is 17 (= 66/($5-1$)) limited to
 $\epsilon$(He/C/N/O/Ne). Next, we compare the AGB models
 for the same mass
 stars with $Z = 0.004$ because these models could account for the
 derived abundances except for S (no predictions for Cl and Ar, however).
 The model for 1.5\,M$_{\sun}$ initial mass
 stars with $Z = 0.004$ gives the closest fit
 to the observation ($\chi_{5}^{2}$ = 360 (= 1800/($6-1$))in 6 elements).
 Limited to $\epsilon$(He/C/N/O/Ne), $\chi_{4}^{2}$ is 7 (= 28/($5-1$)).

The B-type central star indicates 
that SaSt\,2-3 is an extremely young PN and just finished the AGB phase. The presence of H absorption 
lines (Fig.\,\ref{F-synspec}) indicates that SaSt\,2-3 did not experience very late thermal pulse evolution, 
so this PN is probably in the course of H-burning post-AGB evolution. According 
to the H-burning post-AGB evolution model of \citet{Vassiliadis:1994ab}, stars with initially 
 1.5\,M$_{\sun}$ and $Z = 0.004$ would evolve into hot stars 
 with the core mass ($M_{\ast}$) of 0.64\,M$_{\sun}$. 
 $L_{\ast}$ of such stars is $\sim$7380\,L$_{\sun}$ when 
 $T_{\rm eff}$ is $\sim$28\,100\,K in $\sim$1050 years 
 after the AGB phase. Whereas, we infer that 
 1.25\,M$_{\sun}$ stars with $Z = 0.001$ would
 evolve into stars with $M_{\ast}$ of 0.649\,M$_{\sun}$;  
 their $L_{\ast}$ and $T_{\rm eff}$ is 
 $\sim$7765\,L$_{\sun}$ and $\sim$28\,100\,K, respectively 
in $\sim$1770\,yrs after the AGB-phase based on the models of 
 \citet{Fishlock:2014aa} and \citet{Vassiliadis:1994ab}. 
 The main difference in the post-AGB evolution of 
1.50\,M$_{\sun}$/$Z = 0.004$ stars and 1.25\,M$_{\sun}$/$Z = 
 0.001$ stars is evolutionary timescale.

 Through these discussions, we summarise as follows.
 The $Z = 0.001$ model gives the closest values to the derived elemental
 abundances, although there is
 systematically $\sim$0.3--0.4 dex discrepancy of
 $\epsilon$(O,Cl,Ar). The $Z = 0.004$ model shows excellent
 fit to the derived nebular and stellar $\epsilon$(He/C/N/O/Si).
 Considering initial settings of [X/H] in the models, 
 we conclude that
 the progenitor of SaSt\,2-3 would be a $\sim$1.25\,M$_{\sun}$ star with
 initially $Z$$\sim$0.001 ([Fe/H]$\sim$--1.1) and [$\alpha$,Cl/Fe]
 $\sim$+0.3--0.4. This is consistent or comparable with the Galaxy chemical
 evolution model of \citet{Kobayashi:2011aa}; in the Galactic thick
 disc, the predicted [Si/Fe], [S/Fe], [Cl/Fe], and [Ar/Fe] are
 $\sim$+0.6, $\sim$+0.4,
 $\sim$--0.3, and $\sim$+0.3 in [Fe/H] $<-1$, respectively.

 \section{Photoionisation model}  \label{S:cloudy}

    In the previous sections, we characterised the central star
    and dusty nebula. In this section, we build the photoionisation
    model using {\sc Cloudy} and {\sc TLUSTY} to be
    consistent with all the derived quantities, AGB nucleosynthesis model,
    and post-AGB evolution model. Below, we explain how to set
    each parameter in the model, and then we show the result.

    \subsection{Modelling approach}

  \subsubsection{Distance}
  \label{S-distance}

    \begin{figure}
     \centering
     \includegraphics[width=\columnwidth]{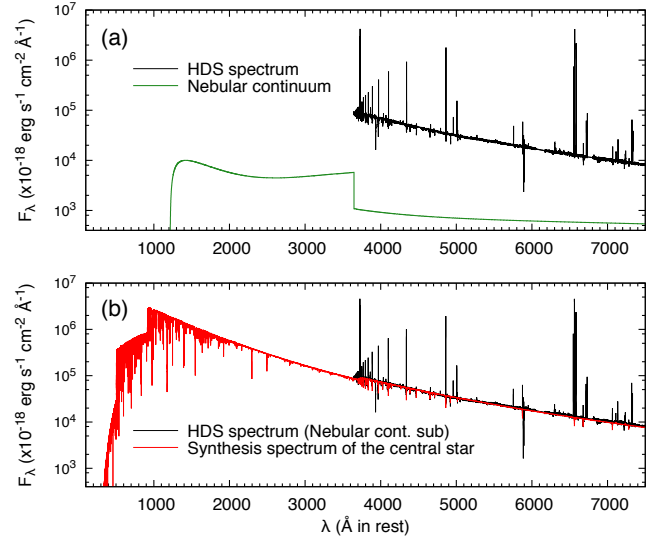}
     \vspace{-5pt}
    \caption{({\bf Upper panel}) The de-reddened HDS spectrum scaled up
    to the flux density at the APASS $Bg'Vr'i'$ bands and the
     calculated nebular continuum by {\sc NEBCONT}.
     ({\bf Lower panel}) The synthetic spectrum of the central star by
     {\sc TLUSTY} (\S\,\ref{S-stellar}) scaled to match
     the residual spectrum produced by subtracting the nebular continuum
     from the HDS spectrum.
     }
    \label{nebcont}
    \end{figure}

    Since the distance $D$ is an important parameter, we estimate
    it by our own method as explained below. 
    We first extract the stellar spectrum from the observed HDS
     spectrum because the observed spectrum is the sum of the
     nebular emission lines and continuum and the central star's
     continuum. For this purpose, we scale the HDS spectrum flux density to
     match the APASS $Bg'Vr'i'$ bands. Then, we
     subtract the theoretically calculated nebular continuum from the
     scaled HDS spectrum.
     We utilise the {\sc NEBCONT} code in the {\sc Dispo} package of
     {\sc STARLINK} v.2015A
     \footnote{\url{http://starlink.eao.hawaii.edu/starlink}}
     to generate the nebular continuum under adopting
     $I$({\hb}) = \mbox{1.57(--12)\,erg~s$^{-1}$~cm$^{-2}$}
     (\S\,\ref{S-Hbflux}), {\te} = 10$^{4}$\,K, {\Ne} =
     2000\,cm$^{-3}$ (Table\,\ref{T-neTe}), and $n$(He$^{+}$)/$n$(H$^{+}$)
     = 1.09(--2)
     (Table\,\ref{T-ionic}). In Fig.\,\ref{nebcont}(a), we show the scaled
     HDS spectrum and the synthetic nebular continuum.
     Fig.\,\ref{nebcont}(b) displays the {\sc TLUSTY} synthetic spectrum of the
     central star (in the case of $T_{\rm eff} = 28\,100$\,K) scaled to
     match the residual spectrum generated by subtracting the nebular
     continuum from the HDS spectrum.

     Next, by integrating the scaled central star's synthetic
     spectra in $T_{\rm eff} = 27\,800 - 28\,400$\,K
     (Table\,\ref{T-stellar}) by
     our {\sc TLUSTY} analysis (\S\,\ref{S-stellar}) in over the wavelength,
     we obtain $L_{\ast}$ as a function of $D$ and $T_{\rm eff}$
     for this $T_{\rm eff}$ range;

 \begin{equation}
 L_{\ast} =
  \left(48.56\cdot\left(T_{\rm eff}/10^{4}\right) - 75.94\right)
  ~\cdot~D_{\rm kpc}^{2}\,{\rm L}_{\sun}.
 \label{F-Lumi}
 \end{equation}

 Assuming that the progenitor is an initially   
 1.25\,M$_{\sun}$/$Z = 0.001$ star and its luminosity is currently
 7765\,L$_{\sun}$ (\S\,\ref{S:AGB}), we obtain $D = 9.90-12.76$\,kpc.
 If we assume an initially 1.5\,M$_{\sun}$ progenitor
 star with $Z = 0.004$, $D$ is $9.66 - 12.44$\,kpc.

 In our {\sc Cloudy} model, we adopt $D = 11.33$\,kpc, which is the
 intermediate value of $D$ when we assume that the central star evolved
 from a star with initially 1.25\,M$_{\sun}$ and $Z = 0.001$.
 Adopting our measured Galactocentric distance of 17.35\,kpc, the
 predicted $\epsilon$(O/Ne/Cl/S/Ar) from the Galaxy
 $\epsilon$(O/Ne/Cl/S/Ar) gradient established amongst Galactic PN
 nebular abundances by \citet{Henry:2004aa} 
 are \mbox{$8.33~\pm~0.21$}, \mbox{$7.61~\pm~0.35$},
 \mbox{$6.22~\pm~0.25$}, \mbox{$4.67~\pm~0.34$},
 and \mbox{$6.06~\pm~0.25$}, respectively.
 These values are in line with the derived nebular abundances
 (Table\,\ref{T-AGB}).
 Our adopted $D = 11.33$\,kpc is in agreement with
 \citet{Frew:2016aa}, who reported 
 \mbox{$14.86~\pm~4.26$}\,kpc. Our derived $D$ is also comparable with the value
 (\mbox{$14.31~\pm~8.54$}\,kpc) determined from the parallax measured
 using \emph{Gaia} DR2 
 \citep[\mbox{$\sigma_{\pi} = 0.0699~\pm~0.0417$\,mas;}][]{Gaia-Collaboration:2018aa}.
 Thus, we simultaneously justify our
 estimated $D$ and nebular $\epsilon$(O/Ne/Cl/S/Ar).

    \subsubsection{Central star}

    As input to {\sc Cloudy}, we use the {\sc TLUSTY} synthetic spectrum of 
    the central star, adopting the parameters from Table\,\ref{T-stellar}. 
    In our iterations here, we only vary $T_{\rm eff}$ in the range of 
    $27\,800-28\,500$\,K and $L_{\ast}$ in the range of $7300-8300$\,L$_{\sun}$.

    \subsubsection{Nebula geometry and boundary condition \label{S-nebgeo}}

   \begin{figure}
    \centering
   \includegraphics[clip,width=\columnwidth]{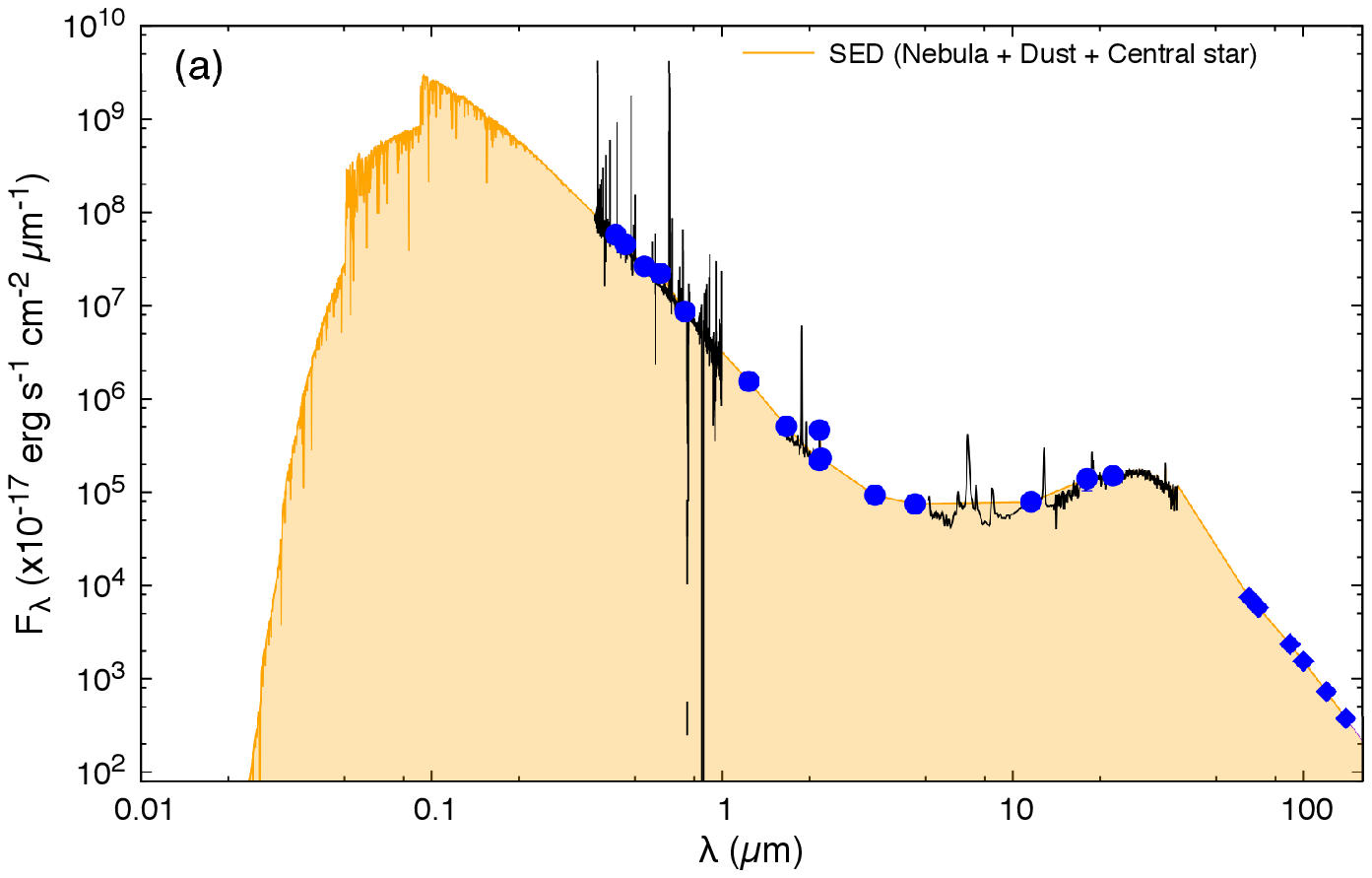}\\
   \includegraphics[clip,width=\columnwidth]{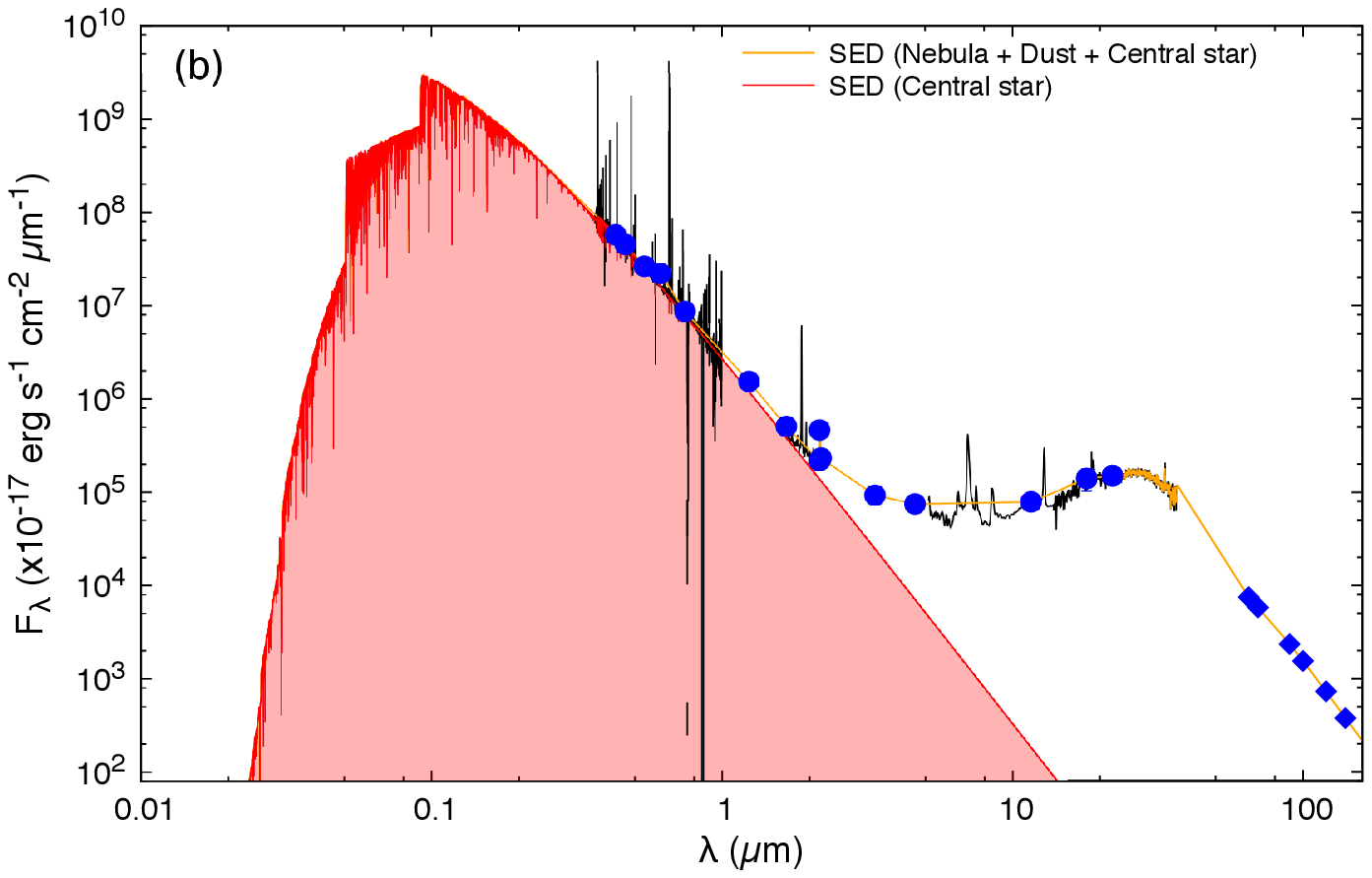}
     \vspace{-10pt}
   \caption{({\bf Upper panel}) SED based on the observed data
    (black lines and blue filled circles) and its
    interpolated curve
    (orange line). The far-IR flux density at 65, 90, 100, 120,
    and 140\,{\micron} (the blue diamonds)
   is an expected value. The
   integrated flux density between $\sim$5(--3) to 140\,{\micron}
   (indicated by the orange region) is $\sim$8215\,L$_{\sun}$
   in $D = 11.33$\,kpc.
   ({\bf Lower panel}) SED based on the observed data and its
   interpolated curve, and the synthetic spectrum of the central
    star. The integrated flux density of the central star
    within the same wavelength range
    (indicated by the red region) is $\sim$7765\,L$_{\sun}$
    in $D = 11.33$\,kpc. See text in details.
   }
   \label{F-SED0}
   \end{figure}

   \begin{figure}
   \centering
   \begin{tabular}{@{}c@{\hspace{5pt}}c@{}}
   \includegraphics[width=0.33\columnwidth,clip]{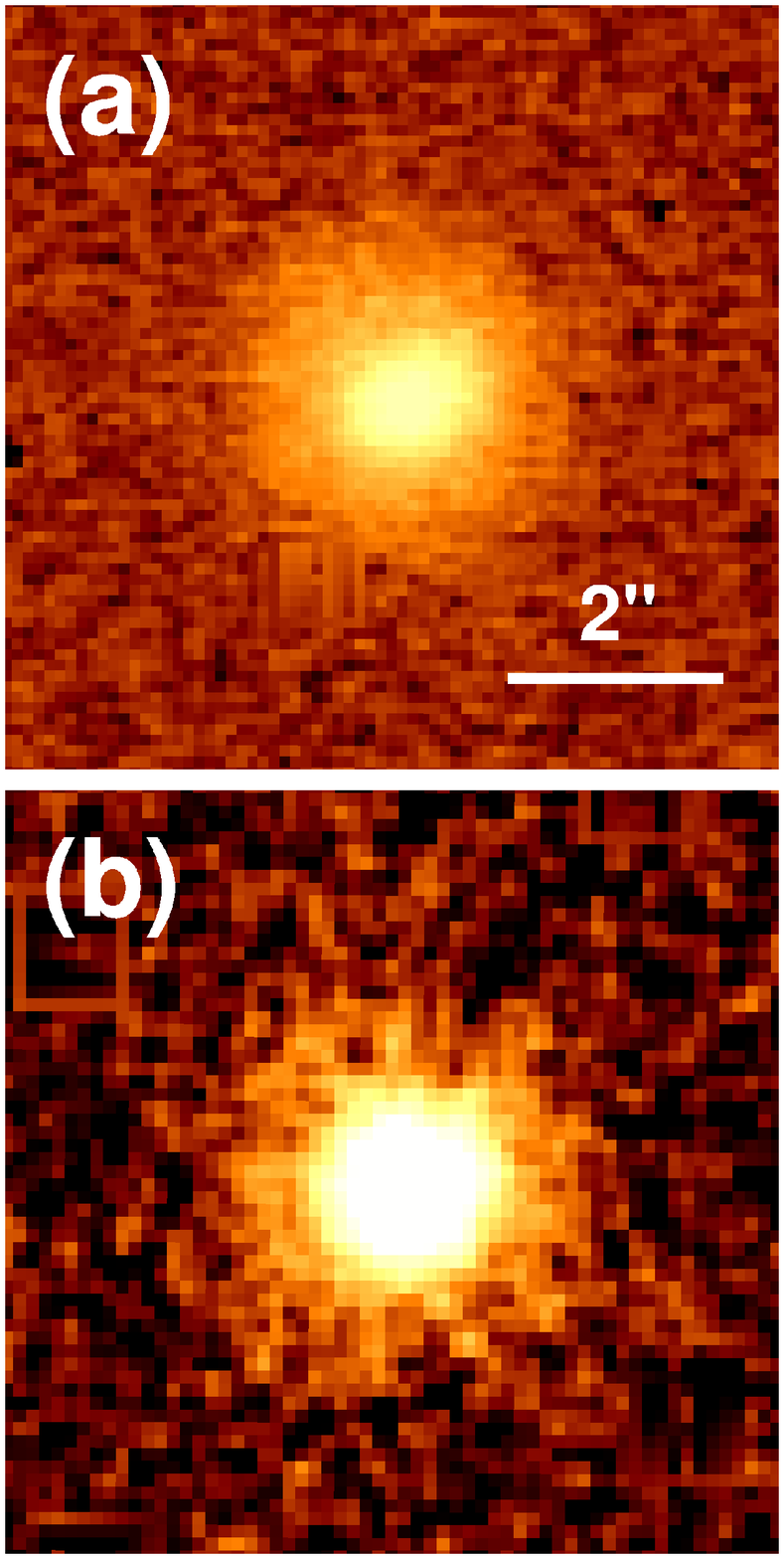}&
    \includegraphics[width=0.66\columnwidth,clip]{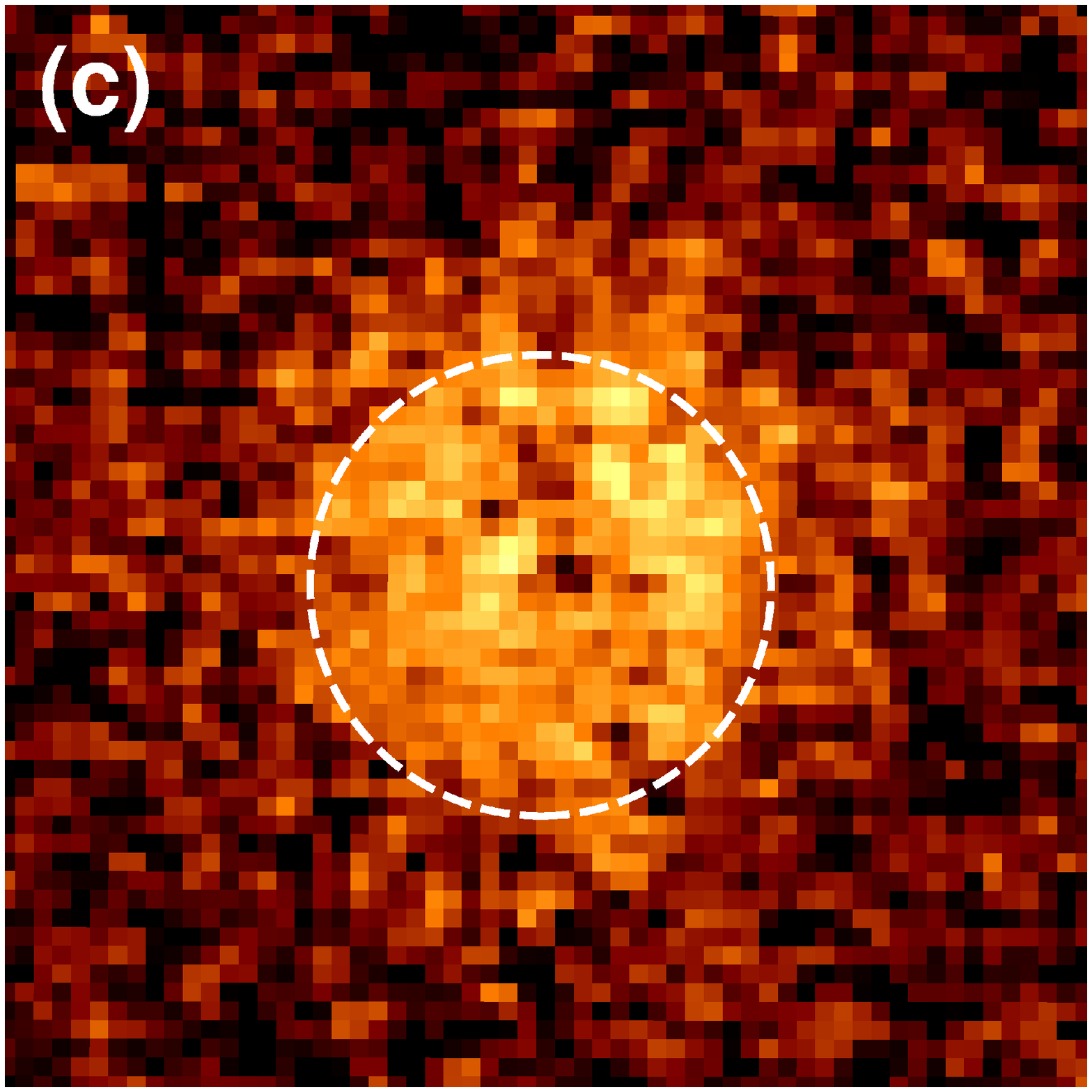}
   \end{tabular}
     \vspace{-5pt}
   \caption{({\bf LEFT two panels}) The Br\,$\gamma$ and Br\,$\gamma$45
   images of SaSt\,2-3. 
    ({\bf RIGHT panel}) The Br\,$\gamma$ minus Br\,$\gamma$45
    image. The radius of the dashed circle is 1.2{\arcsec}.  
    North is up and east is left in these images.}
   \label{image}
   \end{figure}

      \begin{figure}
      \centering
       \includegraphics[width=0.9\columnwidth,
       viewport= 44 0 287 212,clip]{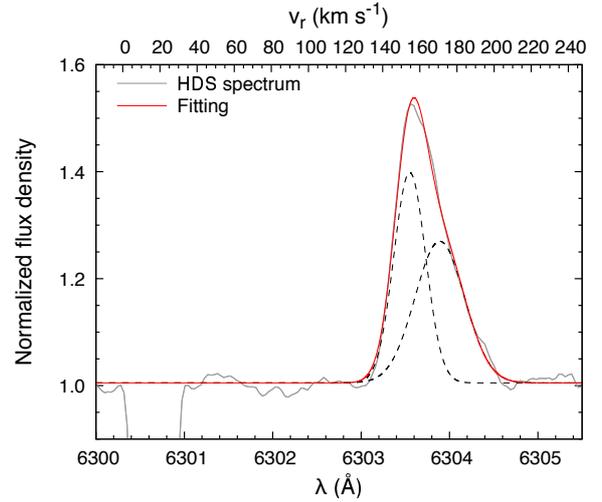}
        \vspace{-5pt}
       \caption{Fitting for the {\oi}\,6300\,{\AA} line. This line can
       be fitted with two Gaussian components (dashed lines). The red line
       is the sum of these components.}
    \label{F-O1}
      \end{figure}

    We plot the observed data and its interpolated curve in Fig.\,\ref{F-SED0}. The
    continuum spectrum in the wavelength $\lesssim 0.36$\,{\micron}
    corresponds to the
    sum of
    (1) the nebular continuum (green line in Fig.\,\ref{nebcont}(a)) and
    (2) the synthetic spectrum of the central star (red line
    in Fig.\,\ref{nebcont}(b)).
    The far-IR flux density
    at 65, 90, 100, 120, and 140\,{\micron} is an expected value
    obtained by fitting for the $15-40$\,{\micron}
    \emph{Spitzer}/IRS spectrum.
    
    We use Equations (2) and (3) \citet{Otsuka:2014aa}, who fitted the \emph{Spitzer}/IRS
    spectra of Galactic C$_{60}$ PNe in $15-40$\,{\micron} with the synthetic absorption
    coefficient $Q_{\lambda}$ value. For SaSt\,2-3, we set the minimum dust temperature = 20\,K and 
    adopt $p = q = 2$ and $\alpha = 0$ which are the same values used in \citet{Otsuka:2014aa}.
    From the fitting, we derive the maximum dust temperature of \mbox{$136.2~\pm~0.4$\,K}, 
    the expected $F_{\nu}$ at 65, 90, 100, 120, and 140\,{\micron} are 105.3, 63.5, 51.6, 35.0, 
    and 24.6\,mJy, respectively.

    From integrating this SED (i.e., the orange region in Fig.\,\ref{F-SED0}(a)), 
    we find a total luminosity of $\sim$8215\,L$_{\sun}$. 
    The luminosity of each component is
    \begin{description}[leftmargin=4em]
     \item[Central star:] $\sim$7765\,L$_{\sun}$,
     \item[Nebular continuum + dust continuum:] $\sim$392\,L$_{\sun}$,
     \item[Nebular emission line:] $\sim$58\,L$_{\sun}$.
    \end{description}
    Here, the luminosity of nebular emission line is
    the sum of all the detected emission lines in the
    HDS and Spitzer spectra. We obtain $L_{\ast}$ of
    $\sim$7765\,L$_{\sun}$ by integrating flux density of the central star's SED
    within the same wavelength range,
    indicated by the red region in Fig.\,\ref{F-SED0}(b).
    Thus, we find that only $\sim$6\, percent of the central
    star's radiation (= (392+58)/7765) seems to be absorbed by the nebula.

    To check whether nebula boundary is determined 
    by the power of the stellar radiation (i.e., ionisation bound) or material 
    distribution (i.e., material bound), 
    we tested both the ionisation boundary model and the material
    boundary model. The WHIRC Br$\gamma$ and Br$\gamma$45
    images in Fig.\,\ref{image} display the central
    bright region and the compact nebula.
    From the Br$\gamma$ -- Br$\gamma$45 image
    (Fig.\,\ref{image}(c)), 
    we measure the radius of the ionised nebula extended up to be 1.2{\arcsec}.

    The material boundary model 
    (the model calculation is stopped at the outer radius of 1.2{\arcsec}) 
    gives a good fit except for underestimates of
    the {\oi} and {\NI} line fluxes. However, when we adopt an open
    geometry such as a cylinder in the ionisation boundary model 
    (the model calculation is stopped when $T_{\rm e}$ is dropped down 
    to $\sim$4000\,K where most of the ionised species are not emitted), 
    we explained well the balance between the input
    energy from the central star and the output energy from the nebula plus dust
    (i.e., $L_{\ast} \gg L_{\rm neb+dust}$) and the observed {\oi} and
    {\NI} line fluxes. Although the WHIRC images do not
    clearly show a cylinder or bipolar nebula, it
    is plausible judging from the {\oi}\,6300\,{\AA} line-profile.
    Fig.\,\ref{F-O1} shows the {\oi}\,6300\,{\AA} line-profile fitting
    by two Gaussian components with $v_{r}$ = +154.6 and
    +170.6\,{\kms} at the peak intensity of each component.
    The {\oi} lines emitted from the most outer
    part of the nebula show blue-shifted asymmetry.
    Such asymmetric profiles are seen in e.g., PN Wray\,16-423
    \citep{Otsuka:2015ab}; 
    Wray16-423 has a bright cylindrical
    structure surrounded by an elliptically extended nebula shell.

    From these discussions, we adopt the cylinder geometry with the
    height = 0.8{\arcsec}. We determined this scale height
    through a small grid model, and we found that the cylinder
    height $\ge 0.8${\arcsec} is necessary.
    We adopt ionisation bounded condition,
    assuming the ionisation front radius of
    1.2{\arcsec}.

   \subsubsection{Elemental abundances and hydrogen density}

    We adopt the nebular value of $\epsilon$(N/O/Ne/S/Cl/Ar/Fe)
    (Table\,\ref{T-AGB}) as the initial value and then refine
    via model iterations within 0.2 dex of the input values
    so that the best-fit abundances would reproduce the observed
    emission line intensities.
    We adopt the nebular $\epsilon$(He) = 10.96 as the first guess, 
    and vary it in range from 10.75 to 11.10. 
    We keep an expected CEL $\epsilon$(C) = 8.58 
    (c.f. stellar $\epsilon$(C) = 8.55) 
    and stellar $\epsilon$(Si) = 6.81 through the model
    iterations. 
    For $\alpha$-elements Mg, Ca, and Ti (not derived, though),
    we fix the [Mg,Ca,Ti/Fe] =
    +0.3, where [Fe/H] is $-1.1$ (\S\,\ref{S:AGB}). We adopt a constant
    hydrogen number density ($n_{\rm H}$) radial profile.
    We first guess that $n_{\rm H}$
    is equal to {\Ne}; we adopt the average {\Ne} amongst the
    measured {\Ne} except for {\Ne}([N\,{\sc i}])
    (Table\,\,\ref{T-neTe}), then we vary $n_{\rm H}$ to get the best fit.

    \subsubsection{Dust grains}

    We assume that the underlying continuum is due to graphite grains based 
    on the fact that SaSt\,2-3 shows the spectral signature of
    carbon-rich species.
    
    We use the optical data of \citet{Martin:1991aa} for randomly
    oriented graphite spheres, and we assume the "$1/3-2/3$" approximation
    \citep{Draine:1993aa}. 
    We adopt the grain radius
    $a = 0.05-0.25$\,{\micron} and $a^{-3.5}$ size distribution. If we
    set the smallest $a = 0.005$\,{\micron}, the maximum grain
    temperature is over the sublimation temperature of 1750\,K. Thus,
    we set the smallest $a = 0.05$\,{\micron}. We resolve the size
    distribution into 20 bins (the smallest is $0.05-0.054$\,{\micron}
    and the largest is $\sim$$0.23-0.25$\,{\micron}). 
    We do not attempt to fit the broad $6-9$\,{\micron} and 11\,{\micron}
    features because the carriers of these features are not determined
    yet and also these profiles are
    different from typical band profile of the
    polycyclic aromatic hydrocarbons (PAHs) in the same
    wavelengths.

    \subsection{Modelling results}
    \label{S-modelR}

\begin{figure}
     \centering
    \includegraphics[width=\columnwidth,clip]{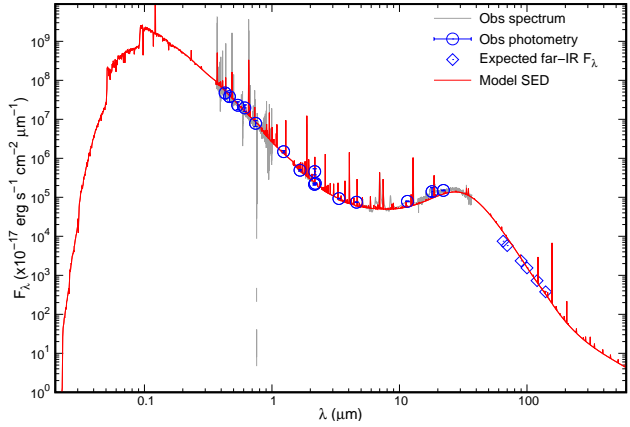}
    \vspace{-5pt}
     \caption{Comparison between the observed SED
     (blue circles and grey lines) and the SED predicted by the best fit
    model (red line). The spectral resolution ($R$) of the model SED is a
    constant 1000.
   \label{F-SED}
   }
    \end{figure}

\begin{figure}
     \centering
    \includegraphics[width=\columnwidth,clip]{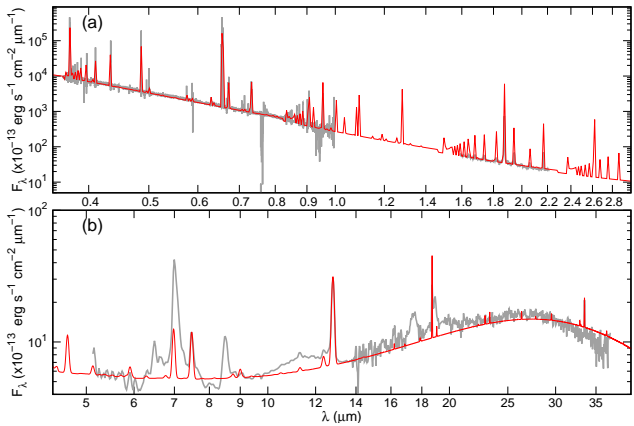}
     \vspace{-5pt}
     \caption{Comparison between the observed SED
     and the SED predicted by the best fit
   model in $0.35-3.0$\,{\micron} (upper panel) and in
    $4.4-40$\,{\micron} (lower panel). The legends in both panels are
     the same used in Fig.\,\ref{F-SED}. $R$ of the model
     SED in $0.35-3.0$\,{\micron}
    is a constant 1000. $R$ is a constant 90
     in $4.4-14$\,{\micron} (low-resolution module) and 570
     in $14-40$\,{\micron} (high-resolution one), which correspond
     to the \emph{Spitzer}/IRS resolution.
   \label{F-SED2}
   }
    \end{figure}

    The input parameters of the best fitting
    result and the derived quantities are summarised in
    Table\,\ref{T:model}.
    In total, we varied 12 parameters within a given range;
    $L_{\ast}$, $\epsilon$(He/N/O/Ne/S/Cl/Ar/Fe), inner radius
    ($r_{\rm in}$), $n_{\rm H}$, and grain abundance 
    until $\chi_{\nu}^{2}$ calculated from $I$({\hb}), 
    76 emission line fluxes, 25 broadband fluxes, 4 mid-IR flux
    densities, and ionisation bound radius (i.e., outer radius
    $r_{\rm out}$).
    Since there is no observed far-IR data,
    we stop the model calculation at the ionisation front, where
    {\te} is dropped down to $\sim$4000\,K. 
    To evaluate the goodness of the model fitting, we refer to
    $\chi_{\nu}^{2}$. 
    For the {\oi} and {\NI} lines, we adopt 30\,percent relative uncertainty because these lines are mostly from the PDRs. 
    For the higher order Balmer lines H\,{\sc i} (B24 - B14), we set 10\,percent relative uncertainty by considering into account the uncertainty 
    of these lines largely affected by the stellar absorption. 
    The reduced-$\chi^{2}$ value in the best model is 12. The relatively large reduced-$\chi^{2}$ value even in the best fitting 
    would be due to the uncertainty of the atomic data which we cannot control. 
    Therefore, we conclude that our best fitting result reproduces observations very well.
    The predicted line fluxes,
    broadband fluxes, and 
    flux densities are compiled in Appendix Table\,\ref{T-Cloudy}. 
    For references, we list {\it expected} $F_{\nu}$ at 65, 90, 100, 120, and 140\,$\mu$m obtained by 
    fitting for the $15 - 40$\,{\micron} \emph{Spitzer}/IRS spectrum (\S\,\ref{S-nebgeo}). 
    In Figs.\,\ref{F-SED} and \ref{F-SED2}, we compare the model SED with the observed one.

 \begin{table}
  \centering
  \caption{\label{T:model} The best-fit model parameters of SaSt\,2-3.}
  \begin{tabularx}{\columnwidth}{@{\extracolsep{\fill}}ll@{}}
   \hline
   {Central star}      &\multicolumn1c{Value}\\
   \hline
   $L_{\ast}$ / $T_{\rm eff}$ / $\log\,g$ / $D$ &7400\,L$_{\sun}$ / 28\,170\,K /
       3.11\,cm s$^{-2}$ / 11.33\,kpc\\
   $M_{V}$ / $R_{\ast}$ /$m_{\ast}$ &--2.10 / 3.606\,R$_{\sun}$ / 0.611\,M$_{\sun}$\\
   \hline
   {Nebula}      &\multicolumn1c{Value}\\
   \hline
   Geometry      &Cylinder with height = 0.8{\arcsec} (8600\,AU)\\
   Radius        &$r_{\rm in}$:0.006{\arcsec} (63\,AU),
                 $r_{\rm out}$:1.25{\arcsec} (14\,162\,AU)\\ 
   $\epsilon$(X)   &He:10.83, C:8.58, N:7.46, O:8.27,  Ne:7.46, \\
                   &Mg:6.80, Si:6.84, S:6.10, Cl:4.51, Ar:5.65, \\
                   &Ca:5.43, Ti:4.05, Fe:5.41, \\
                   &Others: \citet{Fishlock:2014aa}\\
   $n_{\rm H}$     &3098\,cm$^{-3}$\\ 
   $\log$\,$I$({\hb})      &--11.804\,erg\,s$^{-1}$\,cm$^{-2}$\\
   $m_{g}$         &6.13(--2)\,M$_{\sun}$ \\
   \hline
   {Dust}      &\multicolumn1c{Value}\\
   \hline
   Grain size    &$0.05-0.25$\,{\micron}\\
   $T_{d}$ / $m_{d}$  / DGR ($m_{d}$/$m_{g}$) &$66-909$\,K / 2.08(--5)\,M$_{\sun}$ / 3.39(--4) \\
   \hline
  \end{tabularx}
\begin{description}
 \item[Note --] Nebular $\epsilon$(He/N/O/Ne/S/Cl/Ar/Fe) abundances derived by 
empirical method (\S\,\ref{S-abund2}) are $10.75-11.10$/7.47/8.11/7.46/6.10/4.57/5.66/5.29, 
respectively.
 \end{description}
 \end{table}

     \begin{table}
       \centering
      \caption{Ionic abundance fraction predicted by our model
      and comparison between the predicted ICFs (ICF$_{\rm model}$)
      and the empirically determined ICFs (ICF$_{\rm emp}$).
      \label{T-ICF}}
       \begin{tabularx}{\columnwidth}{@{\extracolsep{\fill}}c
 D{.}{.}{-1}cD{.}{.}{-1}D{.}{.}{-1}
       @{\hspace{-2pt}}cc@{}}
   \hline 
   X  &\multicolumn{1}{c}{X$^{0}$} &X$^{+}$
      &\multicolumn{1}{c}{X$^{2+}$}
      &\multicolumn{1}{c}{X$^{3+}$}&ICF$_{\rm model}$ &ICF$_{\rm emp}$\\
   \hline
   He &0.845   &0.155   &<0.001 &              &6.70&\multicolumn1c{$5.12 - 11.57$}\\ 
   C  &<0.001&0.883   &0.116    &<0.001        &8.61&$5.12~\pm~1.41$\\
   N  &0.002   &0.948   &0.050    &<0.001      &1.05&$1.04~\pm~0.13$\\
   O  &0.010   &0.982   &0.008    &<0.001      &1.01&1.00\\
   Ne &0.012   &0.988   &<0.001 &<0.001        &1.01&1.00\\
   S  &<0.001&0.244   &0.756    &<0.001        &1.00&1.00\\
   Cl &<0.001&0.383   &0.617    &<0.001        &1.00&1.00\\
   Ar &0.003   &0.867   &0.131    &<0.001        &7.66&$5.12~\pm~1.41$\\
   Fe &<0.001&0.062   &0.923    &0.015   &1.08&$1.31~\pm~0.16$\\
      \hline
     \end{tabularx}
       \end{table}

\begin{figure}
      \centering
     \includegraphics[width=0.9\columnwidth,clip]{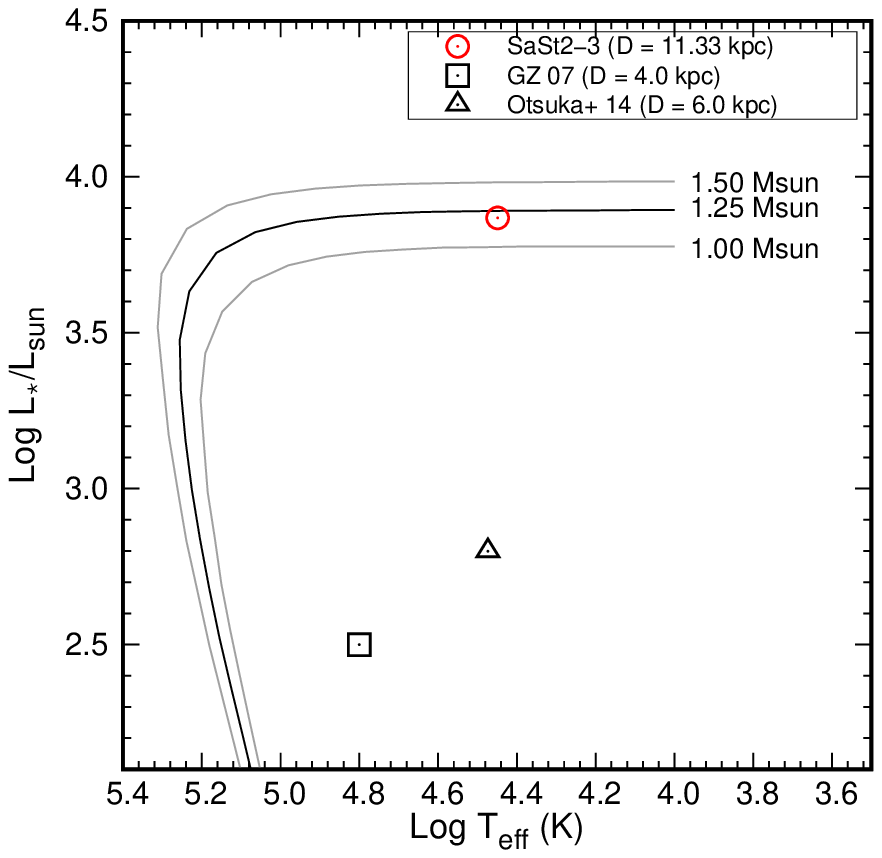}
     \vspace{-5pt}
      \caption{The location of the central star
      on the post-AGB evolutionary tracks for initially $Z = 0.001$ and
     1.0, 1.25, and 1.50\,M$_{\sun}$ \citep{Vassiliadis:1994ab}. We
      generate this 1.25\,M$_{\sun}$ track by linear interpolation between
     the 1.00 and 1.50\,M$_{\sun}$ tracks of
      \citet{Vassiliadis:1994ab}. For comparisons, we plot $T_{\rm eff}$
      and $L_{\ast}$ of the central star derived by
      \citet{Gesicki:2007aa} (GZ07, square; $T_{\rm eff} = 63\,095$\,K and $L_{\ast} = 316$\,L$_{\sun}$ under $D = 4$\,kpc) 
     and \citet{Otsuka:2014aa} (Otsuka+ 14, triangle; $T_{\rm eff} = 29\,750$\,K and $L_{\ast} = 630$\,L$_{\sun}$ under $D =6$\,kpc).
     }
      \label{F-HR}
     \end{figure}

    In Fig.\,\ref{F-HR}, we show the location of the CSPN predicted by
     our {\sc Cloudy} model on the post-AGB evolutionary tracks
     for initially $Z = 0.001$ and
     1.0, 1.25, and 1.50\,M$_{\sun}$ \citep{Vassiliadis:1994ab}. We
      generate this 1.25\,M$_{\sun}$ track by linear interpolation between
     the 1.00 and 1.50\,M$_{\sun}$ tracks of
      \citep{Vassiliadis:1994ab}. Our {\sc Cloudy} model predicts 
      $L_{\ast} = 7400$\,L$_{\sun}$, $T_{\rm eff} = 28\,170$\,K,
      and $m_{\ast} = 0.611$\,M$_{\sun}$. $L_{\ast}$ and $m_{\ast}$
      are justly close to predicted values
      ($\sim$7765\,L$_{\sun}$ and 0.649\,M$_{\sun}$) based on
      the models of initially 1.25\,M$_{\sun}$ stars with $Z = 0.001$
      \citep{Vassiliadis:1994ab,Fishlock:2014aa}. 
      In Fig.\,\ref{F-HR}, we plot the model results
      of \citet{Gesicki:2007aa} and
      \citet{Otsuka:2014aa} as well. These two models
      show a large discrepancy from the predicted post-AGB evolution
      track, but our model completely improves this.

    Our model succeeds in reproducing the derived
    $\epsilon$(X), the volume
    average {\te} (9150\,K, while 8930\,K in the observation), and $I$({\hb}).
    In Table\,\ref{T-ICF}, we present the fraction of each ion in each
    element. Except for C and Ar, the model predicted ICF is well consistent
    with the empirically determined ICF. 
    

    The gas mass $m_{\rm g}$ is the sum of the ionised and neutral
    atomic/molecular gas species. Note that we stopped the model
    calculation at the ionisation front. Our $m_{\rm g}$
    is $\sim$18\,percent of the ejected
    mass at the last thermal pulse (TP) of 1.25\,M$_{\sun}$ initial mass
    stars with $Z = 0.001$
    \citep[0.334\,M$_{\sun}$,][]{Fishlock:2014aa}.
    If we increase the emitting volume by adopting a closed-geometry
    such as a spherical nebula, the situation is slightly improved
    (we trace $\sim$35\,percent of the ejected mass) but the fitting
    model becomes worse as we explained. According to
    \citet{Fishlock:2014aa}, such stars experienced the
    superwind phase in the final few TPs during which the mass-loss
    rate reaches a plateau of $\sim$10$^{-5}$\,M$_{\sun}$\,yr$^{-1}$.
    One might think that our underestimated $m_{\rm g}$
    might be caused by excluding
    the neutral gas and molecular gas regions.
    However, it is unlikely that SaSt\,2-3 has
    the molecular gas rich envelope because
    the molecular hydrogen H$_{2}$ lines in $K$-band are
    not detected \citep{Lumsden:2001aa}. Indeed, we confirm this fact
    by analysis of the UKIRT CGS4 $HK$-band spectrum of SaSt\,2-3
    (Fig.\,\ref{F-sedobs}). Since our $m_{\rm g}$ is
    greater than the ejected mass 2.7(--3)\,M$_{\sun}$ at the last TP of
    initially 1.00\,M$_{\sun}$ stars with $Z = 0.001$, we can conclude
    that the progenitor should be a $> 1.00$\,M$_{\sun}$ initial mass star.
    \citet{Fishlock:2014aa} predicts that 1.00\,M$_{\sun}$ stars lost
    the majority of its stellar envelope before it reaches the superwind
    phase. The 1.25\,M$_{\sun}$ progenitor that we infer 
    might have experienced such mass loss.
    If this is true, non-detection of the H$_{2}$ lines
    might be because thin circumstellar envelope does not shield UV radiation
    from the CSPN and H$_{2}$ is dissociated. As an other
    explanation for the estimated small $m_{\rm g}$,
    the ejected mass during the AGB phase might be
    efficiently transported to the stellar surface of a companion star.
    Our small $m_{\rm g}$ would be largely improved by taking
    cold gas/dust components that can be traced by far-IR
    observation.

As presented in Fig.\,\ref{F-SED2}(b), our {\sc Cloudy} model predicts an emission line around 7\,{\micron}. 
This line is the complex of the {\arii}\,6.99\,{\micron} and H\,{\sc
    i}\,6.95/7.09\,{\micron} lines. In the \emph{Spitzer}/IRS spectrum,
    we measure the total line-flux of the
    {\arii}\,6.99\,{\micron}, H\,{\sc i}\,6.95/7.09\,{\micron},
    and C$_{60}$\,7.0\,{\micron} to be \mbox{$28.67~\pm~3.23$}, where $I$({\hb}) =
    100. The model predicts
    $I$({\arii}\,6.99\,{\micron} + H\,{\sc
    i}\,6.95/7.09\,{\micron})/$I$({\hb}) = 4.173 ($I$({\hb}) =
    100). The contribution of the atomic line complex to the
    C$_{60}$\,7.0\,{\micron} band (14.6\,$\%$) is not as significant
    as \citet{Otsuka:2014aa} expected (30.3\,$\%$).
    We estimate $I$(C$_{60}$ 7.0\,{\micron})/$I$({\hb}) to be
    \mbox{$24.5~\pm~2.8$}.

\section{Discussions}    
 \label{S-Discuss}

\begin{table*}
\renewcommand{\arraystretch}{0.9}
\caption{Nebular and central star's properties of C$_{60}$-containing PNe 
IC\,418, Tc\,1 and SaSt\,2-3 and non C$_{60}$-containing C-rich PNe, 
IC\,2165 and Me\,2-1. $\epsilon$(He) in SaSt\,2-3 listed in this table 
is the intermediate value of $\epsilon$(He) derived by empirical method. 
M$_{\rm ini.}$ is the initial mass of the progenitor 
inferred from the plot of $L_{\ast}$ and $T_{\rm eff}$ on the post-AGB evolutionary tracks (Fig.\,\ref{F-HR2}) 
based on \citet{Vassiliadis:1994ab}. 
}
\centering
 \begin{tabularx}{\textwidth}{@{\extracolsep{\fill}}l@{\hspace{5pt}}c@{\hspace{5pt}}
c@{\hspace{5pt}}c@{\hspace{5pt}}c@{\hspace{5pt}}c@{\hspace{5pt}}c@{\hspace{5pt}}
c@{\hspace{5pt}}c@{\hspace{5pt}}c@{\hspace{5pt}}c@{\hspace{5pt}}c@{\hspace{5pt}}
c@{\hspace{5pt}}c@{\hspace{5pt}}c@{}}
\hline
Nebula    &$\epsilon$(He)&$\epsilon$(C)&$\epsilon$(N)&$\epsilon$(O)&$\epsilon$(Ne)
&$\epsilon$(S)&$\epsilon$(Cl)&$\epsilon$(Ar)&$\epsilon$(Fe)&$Z$&$\log\,L_{\ast}/L_{\sun}$
&$T_{\rm eff}$ (K)&$\log\,g$ (cm\,s$^{-2}$)&$M_{\rm ini.}$ (M$_{\sun}$)\\
\hline
IC\,418   &11.08 & 8.90 &8.00 &8.60 &8.00 &6.65 &5.00&6.20&4.60&0.008&3.88&36\,700&3.55&$\sim$1.8\\
Tc\,1     &10.92 & 8.56 &7.56 &8.41 &7.80 &6.45 &4.97&6.08&5.19&0.004&3.85&32\,000&3.30&$\sim$1.5\\
SaSt\,2-3 &10.96 & 8.58 &7.47 &8.11 &7.46 &6.10 &4.57&5.66&5.29&0.001&3.87&28\,170&3.11&$\sim$1.3\\
\hline    
IC\,2165  &11.05 & 8.62 &8.07 &8.53 &7.73 &6.26 &$\cdots$&6.00&$\cdots$    &0.004&3.73     &181\,000&7.0&$\sim$2.1\\
Me\,2-1   &11.00 & 8.85 &7.71 &8.72 &7.97&6.96  &$\cdots$&6.20&$\cdots$    &0.008&3.56     &170\,000&7.0&$\sim$1.8\\

\hline
\end{tabularx}
\begin{description}
 \item[Note --] We estimated $L_{\sun}$ of Tc\,1 using $D = 3.0$\,kpc \citep[cf. 2.67\,kpc,][]{Frew:2013aa}, 
a theoretical model spectrum of \citet{Lanz:2007aa} for O-type stars with $T_{\rm eff} = 32\,000$\,K, 
$\log\,g = 3.3$\,cm\,s$^{-2}$ \citep{Mendez:1992aa}, and $Z = 0.008$ to match with 
the interstellar extinction corrected \emph{HST}/STIS spectrum of the central star \citep{Khan:2018aa}.
We estimated $T_{\rm eff}$ of IC\,2165 and Me\,2-1 
based the energy balance method of \citet{Dopita:1991aa}.  
$L_{\ast}$ is estimated using theoretical model spectra of 
\citet{Rauch:2003aa} with the derived $T_{\rm eff}$ and assumed 
$\log\,g = 7.0$\,cm\,s$^{-2}$ to match with the dereddened \emph{HST}/WFPC2 F547M/F555W ($V$-band) magnitude 
of the central stars measured by \citet{Wolff:2000aa}, and 
$D$ of \citet{Frew:2013aa}.
 
 \end{description}
\label{T-pro}
\end{table*}

 \subsection{Evolution of SaSt\,2-3  \label{S-evolution}}

 Using the Galactic rotation velocity based on the
 distance scale of \citet{Cahn:1992aa},
 \citet{van-de-Steene:1995aa}, and 
 \citet{Zhang:1995aa}, 
 $v_{r}$(LSR) = 149.1\,{\kms}, and $D = 11.33$\,kpc,
 we obtain ${\Delta}V = 68.2-77.9$\,{\kms}. The height above the
 Galactic plane $|z|$ is 1.13\,kpc. These results are in agreement
 with the Type\,III PN classification of \citet{Quireza:2007aa}.
 ${\Delta}V$ and $|z|$ do
 not exceed 120\,{\kms} and 1.99\,kpc for the Type\,IV PN classification
 of \citet{Quireza:2007aa}, respectively. The metallicity of
 SaSt\,2-3 is much richer than typical halo PNe such as K\,648, BoBn\,1, and
 H\,4-1 showing [Ar/H] $\lesssim -2$
 \citep{Otsuka:2010aa,Otsuka:2015aa,Otsuka:2013ab}.
 Thus, we conclude that SaSt\,2-3 belongs to the 
 thick disc younger population and a Type\,III PN rather than a Type\,IV
 PN \citep{Pereira:2007aa}.
 Note that classification of PN type does not matter whether the central
 star is binary or not.

 SaSt\,2-3 would have evolved from a binary composed of a
 $\sim$1.25\,M$_{\sun}$ initial mass star and an companion
 star. However, any parameters on binary motion are unknown yet.
 According to the simulation using the \verb*|binary_c| \footnote{This
 code can simulate single star evolution by adopting a large binary separation
 and a small companion star. Here, we adopted an initial binary separation of
 1(+6)\,R$_{\sun}$ and the initial companion star mass of
 0.1\,M$_{\sun}$.} code by \citet{Izzard:2004aa},
 an initially 1.25\,M$_{\sun}$ single star with $Z = 0.001$ will
 enter the PN phase within 3.5\,Gyr after the progenitor was in the
 main-sequence. Perhaps, evolutionary time required to
 reach the AGB phase would be shortened by binary interaction (i.e.,
 $<3.47$\,Gyr).

  SaSt\,2-3 composes of a $\sim$0.61\,M$_{\sun}$ B-type cool
  central star with $T_{\rm eff} = 28\,100$\,K and a companion
  star. We find the Ca\,I absorption centred at 
  6616.62\,{\AA} and 6689.28\,{\AA} in heliocentric wavelength (6613.13 and 
  6685.6\,{\AA} in rest wavelength, respectively). 
  $v_{r}$ using these two absorption lines is
  165.5\,{\kms} and 152.1\,{\kms}, respectively. These $v_{r}$ is close to
  $v_{r}$ of the central star (183.6\,{\kms}, Table\,\ref{T-rv}). Thus,
  we assume that these lines could be originated from the envelope in the
  companion star. However, since
  there is only one spectrum covering $6600-6690$\,{\AA} we have
  (Table\,\ref{T-HDS}), 
  we do not yet find radial velocity variations of this absorption.
  We suppose that the companion star might be a F-K
  spectral type star in the main-sequence in terms of the initial mass
  by referring to \citet{De-Marco:2013aa}.

\subsection{Comparisons with non-C$_{60}$ and C$_{60}$-containing PNe}

Our study can fully characterise the physical properties of SaSt\,2-3. Thus, 
we are able to compare nebular elemental abundances and central star properties 
with those of other C$_{60}$ PNe, and we attempt to gain insights into the C$_{60}$ formation in PNe.
For this purpose, we select C$_{60}$ PNe Tc\,1 and IC\,418 because they were previously modelled 
using {\sc Cloudy}, and they have been extensively studied. In Table\,\ref{T-pro}, 
we compile their properties. Due to the lack of existing UV spectra, the CEL $\epsilon$(C) in SaSt\,2-3 is not 
determined yet. However, since we adopt an expected CEL $\epsilon$(C) 
(= (C/O)$_{\ast}$ $\times$ O$_{\rm CEL}$) from the stellar C/O ratio, 
the reliability of the discussion here is not compromised. 
We adopt the results of Tc\,1 and IC\,418 by \citet{Pottasch:2011aa} and 
\citet{Morisset:2009aa}, respectively. For Tc\,1, since 
\citet{Pottasch:2011aa} calculated $\epsilon$(Ar) by the sum of the Ar$^{+}$ 
and Ar$^{2+}$ abundances without subtracting C$_{60}$\,7.0\,{\micron} 
and H\,{\sc i} lines from the complex line at 7\,{\micron}, 
their calculated $\epsilon$(Ar) is certainly overestimated. 
Therefore, we compute $\epsilon$(Ar) from Ar$^{2+}$ (7.0(--7)) and ICF(Ar) = S/S$^{2+}$ 
(1.71). $\epsilon$(He) in Tc\,1 is predicted by their {\sc Cloudy} model.

\begin{figure}
      \centering
     \includegraphics[width=0.9\columnwidth,clip]{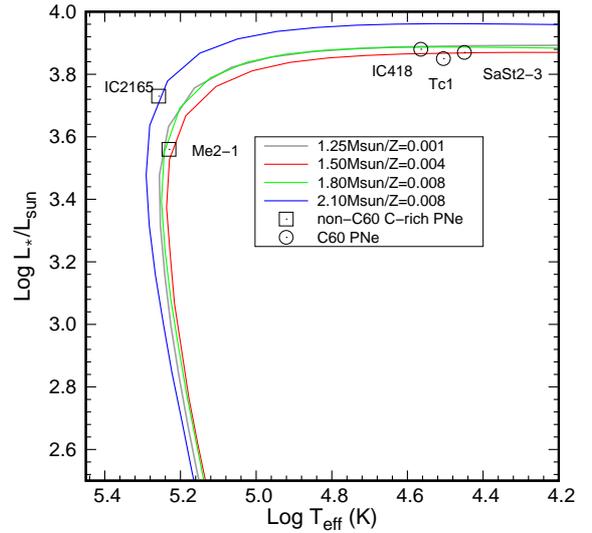}
     \vspace{-5pt}
      \caption{The location of C$_{60}$-containing PNe IC\,418, Tc\,1, and SaSt\,2-3 and non-C$_{60}$ C-rich PNe IC\,2165 and Me\,2-1 
      on the post-AGB evolutionary tracks based on \citet{Vassiliadis:1994ab}.
     }
      \label{F-HR2}
     \end{figure}

Comparisons with theoretical AGB nucleosynthesis models of \citet{Karakas:2010aa} 
indicate that the initial mass ($M_{\rm ini.}$) is 
$\sim$$1.90-2.10$\,M$_{\sun}$ for IC\,418 and $\sim$$1.50-1.90$\,M$_{\sun}$ for Tc\,1, 
respectively. From plots of $L_{\ast}$ and $T_{\rm eff}$ on the post-AGB 
evolutionary tracks based on \citet{Vassiliadis:1994ab} (Fig.\ref{F-HR2}), we have the same estimate 
for IC\,418 ($\sim$1.8\,M$_{\sun}$) and Tc\,1 ($\sim$1.5\,M$_{\sun}$). 
Based on our estimates for the initial mass and metallicity, 
the age of IC\,418 and Tc\,1 after the main sequence is $\sim$$2-3$\,Gyr. 
Thus, we attest that SaSt\,2-3 is the most metal-deficient and oldest Galactic C$_{60}$ PN.

Our findings in elemental abundances are as follows; (i) the values of $\epsilon$(C) in these 
C$_{60}$ PNe are not peculiar values that can be explained by the AGB nucleosynthesis models, 
(ii) despite that the C$_{60}$ band strength in SaSt\,2-3 and Tc\,1 is much 
stronger than that in IC\,418 \citep[see Table\,10 of][]{Otsuka:2014aa}, and 
$\epsilon$(C) in Tc\,1 and SaSt\,2-3 is smaller than that in IC\,418, and 
(iii) the C/O ratio (an indicator of the amount of the C-atoms unlocked in 
dust and molecules) in Tc\,1 is smaller than that in IC\,418. Supporting (i), 
non C$_{60}$-containing C-rich PNe IC\,2165 \citep{Miller:2018aa} 
and Me\,2-1 \citep{Pottasch:2010aa} (they are selected 
based on metallicity and initial mass) show similar abundances (including $\epsilon$(C)) 
to C$_{60}$ PNe. The difference between non-C$_{60}$ and C$_{60}$ PNe is $T_{\rm eff}$ only, supporting that 
the weak radiation field from the central star is in 
favour of the C$_{60}$ formation. The time during which C$_{60}$ 
is present might be a short-lived phase that C-rich PNe go through, 
but we do not yet have firm observational evidence of this. 
These findings would result in our conclusion that the C$_{60}$ formation does 
not largely depend on the amount of the C-atoms produced during the AGB 
phase, and Tc\,1 and SaSt\,2-3 efficiently produced the C$_{60}$ molecule 
by some mechanisms not present in IC\,418.

It is noticeable that Fe abundance is highly depleted in all C$_{60}$ PNe. 
The highly deficient Fe is possibly due to selective depletion in a binary disc 
\citep[e.g.,][reference therein]{Otsuka:2016aa}. If C$_{60}$ PNe have a disc around 
the central star, they can harbour mass-loss 
including AGB products for a long time and shield C$_{60}$ molecules from the 
intense central star's UV radiation. Then, 
most of the Fe-atoms might be tied up in dust grains (e.g., FeO) within a disc. 
Accordingly, an environment suitable for large carbon molecule formation 
might be created. This would be in the case of SaSt\,2-3. The central star of IC\,418 
would be not a binary \citep[e.g.,][]{Mendez:1989aa}. There are no reports on the binary 
central star of Tc\,1 so far.

If it is the case, is C$_{60}$ formation dependent on the central star's properties 
and its surrounding environment, such as a binary disc?  To answer this question, 
we need to calculate the mass of the C-atoms present in atomic gas, dust and C$_{60}$ 
by a fair means. From a more global perspective, the fraction of the C-atoms would 
be a critically important parameter to understand how much mass carbon PN progenitors 
had returned to their host galaxies. However, we find that the excitation diagram 
based on the observed C$_{60}$ 8.5/17.4/18.9\,{\micron} fluxes 
and the expected C$_{60}$ 7.0\,{\micron} flux ($I$(C$_{60}$ 7.0\,{\micron})/$I$({\hb}) = 
\mbox{$24.5~\pm~2.8$}, see \S\,\ref{S-modelR}) in SaSt\,2-3 indicates non-LTE conditions. 
The same situation exists in other C$_{60}$ PNe 
including Tc\,1 \citep{Cami:2010aa} and IC\,418 \citep{Otsuka:2014aa}. 
Therefore, we sought other ways not involving the excitation diagrams. 
Of these, the method proposed by \citet{Berne:2012aa} is seemingly suitable to 
our aim; this method requires the observed/modelled IR SED, all four mid-IR C$_{60}$ band fluxes, 
and dust mass. These three parameters are already determined in SaSt\,2-3 and IC\,418.

In Table\,\ref{T-Cmass}, we summarise the mass of the C-atoms present in atomic gas, dust, and C$_{60}$ 
in SaSt\,2-3, IC\,418, and Lin\,49 in the SMC (as a comparison of the possible binary C$_{60}$ PN). 
For SaSt\,2-3 and Lin\,49, we obtain $I$(IR) by integrating the atomic gas 
emission free SED (generated by the {\sc Cloudy} best model) in the range from 4 to 200\,{\micron}. 
Then, we calculate the total C$_{60}$ flux $I$(C$_{60}$) (Table\,\ref{AT-line}). 
Finally, assuming that the dust and C$_{60}$ are emitted in the same regions, we estimate the C$_{60}$ mass (= 
$m_{\rm d} \times I$(C$_{60}$)/$I$(IR)). The column density and total number of C$_{60}$ in SaSt\,2-3 are 
8.57(+12)\,cm$^{-2}$ and 7.53(+47), respectively. By applying the method to IC\,418 
(but using the combined Infrared Space Observatory (\emph{ISO})/Short Wavelength 
Spectrometer (SWS)/Long-Wave Spectrometer (LWS) 
and \emph{Spitzer} combined spectrum), we obtain their C$_{60}$ mass.

From this analysis, we quantitatively demonstrate that SaSt\,2-3 and Lin\,49 produced 
C$_{60}$ more efficiently than IC\,418. 
Thus, we conclude that the C$_{60}$ formation depends upon the central star's properties and its 
surrounding environment (e.g., a binary disc), rather than the amount of C-atoms.

\begin{table}
\centering
\caption{Mass of the C-atoms in each atomic gas, dust, and C$_{60}$. 
\label{T-Cmass}}
 \begin{tabularx}{\columnwidth}{@{\extracolsep{\fill}}lcccc@{}}
\hline
Nebula    &atomic C   &graphite grains &C$_{60}$ &Mass frac.  \\
          &($\times10^{-4}$\,M$_{\sun}$)&($\times10^{-5}$\,M$_{\sun}$)&($\times10^{-7}$\,M$_{\sun}$)&of C$_{60}$ (\%)\\
\hline
IC\,418       &3.99  &1.70  &2.26 &0.05\\
Lin\,49&3.00 &$4.15-4.29$ &$3.05-4.64$ &$0.09-0.14$\\
SaSt\,2-3     &2.17  &2.08  &4.53 &0.19\\
\hline
\end{tabularx} 
\begin{description}
 \item[Note --] Mass of C$_{60}$ is determined by (graphite grain mass for 
	    Lin\,49 and SaSt\,2-3) or 
	    (the total dust mass 7.76(--5)\,M${\sun}$ including graphite grain 
	    for IC\,418; \citealt{Gomez-Llanos:2018aa}) 
	    $\times~I$(C$_{60}$)/$I$(IR).
		 $I$(C$_{60}$) is the total flux of the  mid-IR C${60}$ bands measured by 
		 \citet[1.03(--10)\,erg\,s$^{-1}$\,cm$^{-2}$ for IC\,418 using 
	    the \emph{ISO}/SWS/LWS and \emph{Spitzer} combined spectrum]{Otsuka:2014aa}, 
	    \citet[9.26(--14)\,erg\,s$^{-1}$\,cm$^{-2}$ for Lin\,49]{Otsuka:2016aa}, 
	   and the present work for SaSt\,2-3 (9.43(--13)\,erg\,s$^{-1}$\,cm$^{-2}$). 
	    $I$(IR) in IC\,418, Lin\,49, and SaSt\,2-3 is 3.53(--8), 
	    1.26(--11) (two-shell model)/8.56(--12) (single shell model), 
	    and 4.33(--11) in erg\,s$^{-1}$\,cm$^{-2}$, 
	    respectively. 
     \end{description}
\end{table}

  \section{Summary}
  \label{S:summary}

  We have studied the fullerene-containing PN SaSt\,2-3
  in order to investigate its physical properties 
  and gain insights into the C$_{60}$ formation in PN progenitors.
  %
  We derived the nine and four elemental abundances
  from nebular line and stellar absorption analysis, respectively.
  The derived elemental abundances indicate that
  the progenitor is an
  initially $\sim$1.25\,M$_{\sun}$ star with $Z = 0.001$ and
  $\alpha$-element and Cl enhanced ([$\alpha$,Cl/Fe] $\sim$+0.3-0.4).
  The distance $D$ of 11.33\,kpc is determined by comparing the
  derived luminosity as
  a function of $D$ and $T_{\rm eff}$
  with the predicted luminosity by the post-AGB
  evolution model for the 1.25\,M$_{\sun}$ stars with $Z = 0.001$.
  SaSt\,2-3 is classified as part of the thick disc population with
  an early B-type central
  star with $T_{\rm eff} = 28\,100$\,K,
  $\log\,{g} = 3.11$\,cm\,s$^{-2}$, and the
  core-mass of $\sim$0.61\,M$_{\sun}$. We discovered
  the binary central star of SaSt\,2-3 from time-variation of
  the stellar radial velocity.
  Further observations are necessary to understand the binary system.
  We built the comprehensive photoionisation model.
  The calculated gas mass is much smaller than the AGB
  model prediction for the single 1.25\,M$_{\sun}$ stars
  with $Z = 0.001$. The lower gas mass could
  be due to the short duration time of the superwind phase
  or efficient mass-transfer into the stellar surface
  of the companion star during the AGB phase.
  %
  From the simple analysis, we quantitatively demonstrate that SaSt\,2-3 produced C$_{60}$ more efficiently 
  than other C$_{60}$ PNe. The C$_{60}$ formation would depend on the central star's properties and its surrounding environment. 
  There might be a link between the C$_{60}$ formation efficiency and the binary central star. 
  Spatially-resolved spectral maps of the atomic carbon, carbon dust, and C$_{60}$ are necessary to 
  identify the locations of C$_{60}$ and investigate the abundance distribution of these species within dusty nebula.
 We succeed to demonstrate what type of and how much mass of stars can produce
 how much C$_{60}$ molecules. We will further investigate stellar/nebular 
 properties and C$_{60}$ in order to find what parameters are of  
 critical importance in the C$_{60}$ formation.

\section*{Acknowledgements}

I am grateful to the anonymous referee for carefully reading and the
useful suggestions which greatly improved this article. I learned a
lot of things from his/her comments.
I was supported by the research fund 104-2811-M-001-138 and
104-2112-M-001-041-MY3 from the Ministry of Science and Technology
(MOST), R.O.C. I thank Dr. Akito Tajitsu for supporting
my Subaru HDS observations.
I sincerely thank Drs. Benjamin Sargent, Peter Scicluna, and Toshiya Ueta 
for critically reading the paper 
and giving suggestions. This work was partly based on archival
data obtained with the \emph{Spitzer} Space Telescope,
which is operated by the Jet Propulsion Laboratory, California
Institute of Technology under a contract with NASA. This research
is in part based on observations with
\emph{AKARI}, a JAXA project with the participation of ESA.
A portion of this work was based on
the use of the ASIAA clustering computing system.

\appendix

\section{Supporting results}

The following tables and figure support our works.

   \begin{table}
\renewcommand{\arraystretch}{0.9}
    \renewcommand{\thetable}{A\arabic{table}} 
  \footnotesize
  \centering
\caption{Identified emission lines in the HDS and IRS spectra. 
$f$($\lambda$) is reddening law of \citet{Cardelli:1989aa}, 
$F$($\lambda$) and $I$($\lambda$) are observed and interstellar extinction corrected line fluxes normalised 
to $F$({\hb}) and $I$({\hb}), respectively. 
 \label{AT-line}}
 \begin{tabularx}{\columnwidth}{@{\extracolsep{\fill}}D{.}{.}{-1}@{\hspace{-2pt}}
c@{\hspace{-2pt}}D{.}{.}{-1}@{\hspace{-2pt}}D{.}{.}{-1}@{\hspace{-2pt}}D{.}{.}{-1}@{\hspace{-2pt}}D{.}{.}{-1}@{\hspace{-2pt}}D{.}{.}{-1}@{}}
\hline
 \multicolumn{1}{c}{$\lambda_{\rm lab.}$ ({\AA})}
 &Line&\multicolumn{1}{r}{$f(\lambda)$}&\multicolumn{1}{r}{$F$($\lambda$)}&\multicolumn{1}{r}{$\delta$\,$F$($\lambda$)}
&\multicolumn{1}{r}{$I$($\lambda$)}&\multicolumn{1}{r}{$\delta$\,$I$($\lambda$)}
\\
\hline  
3656.66 & B37 & 0.336 & 0.015 & 0.010 & 0.019 & 0.013 \\ 
3657.27 & B36 & 0.336 & 0.062 & 0.019 & 0.077 & 0.024 \\ 
3657.92 & B35 & 0.336 & 0.048 & 0.013 & 0.060 & 0.017 \\ 
3658.64 & B34 & 0.336 & 0.076 & 0.012 & 0.094 & 0.017 \\ 
3659.42 & B32 & 0.336 & 0.131 & 0.015 & 0.162 & 0.023 \\ 
3660.28 & B33 & 0.335 & 0.162 & 0.014 & 0.201 & 0.024 \\ 
3661.22 & B31 & 0.335 & 0.265 & 0.021 & 0.330 & 0.038 \\ 
3662.26 & B30 & 0.335 & 0.324 & 0.028 & 0.402 & 0.048 \\ 
3663.40 & B29 & 0.335 & 0.377 & 0.032 & 0.469 & 0.055 \\ 
3664.68 & B28 & 0.334 & 0.762 & 0.052 & 0.945 & 0.101 \\ 
3666.09 & B27 & 0.334 & 0.676 & 0.049 & 0.839 & 0.092 \\ 
3667.68 & B26 & 0.334 & 0.623 & 0.039 & 0.773 & 0.080 \\ 
3669.46 & B25 & 0.334 & 0.581 & 0.032 & 0.721 & 0.071 \\ 
3671.48 & B24 & 0.333 & 0.593 & 0.029 & 0.735 & 0.070 \\ 
3673.76 & B23 & 0.333 & 0.733 & 0.030 & 0.909 & 0.083 \\ 
3676.36 & B22 & 0.332 & 0.739 & 0.033 & 0.916 & 0.085 \\ 
3679.35 & B21 & 0.332 & 0.844 & 0.031 & 1.046 & 0.094 \\ 
3682.81 & B20 & 0.331 & 0.897 & 0.034 & 1.111 & 0.100 \\ 
3686.83 & B19 & 0.330 & 1.121 & 0.031 & 1.388 & 0.119 \\ 
3691.55 & B18 & 0.329 & 1.184 & 0.023 & 1.465 & 0.122 \\ 
3697.15 & B17 & 0.328 & 1.298 & 0.027 & 1.605 & 0.134 \\ 
3703.85 & B16 & 0.327 & 1.778 & 0.031 & 2.196 & 0.181 \\ 
3709.65 & {\hei}        & 0.326 & 0.039 & 0.013 & 0.049 & 0.016 \\ 
3711.97 & B15 & 0.325 & 1.665 & 0.032 & 2.054 & 0.169 \\ 
3721.94 & B14 & 0.323 & 2.167 & 0.040 & 2.671 & 0.218 \\ 
3726.03 & {\oii}    & 0.322 & 107.494 & 1.338 & 132.411 & 10.617 \\ 
3728.81 & {\oii}    & 0.322 & 74.678 & 0.644 & 91.955 & 7.314 \\ 
3733.01 & {\hei}        & 0.321 & 0.071 & 0.010 & 0.088 & 0.014 \\ 
3734.37 & B13 & 0.321 & 2.375 & 0.036 & 2.922 & 0.235 \\ 
3750.15 & B12 & 0.317 & 3.125 & 0.027 & 3.836 & 0.301 \\ 
3770.63 & B11 & 0.313 & 4.076 & 0.033 & 4.991 & 0.386 \\ 
3797.90 & B10 & 0.307 & 5.102 & 0.035 & 6.224 & 0.472 \\ 
3819.60 & {\hei}        & 0.302 & 0.089 & 0.011 & 0.108 & 0.015 \\ 
3835.38 & B9 (H$\eta$)& 0.299 & 6.491 & 0.037 & 7.876 & 0.580 \\ 
3889.05 & B8 & 0.286 & 9.595 & 0.191 & 11.547 & 0.844 \\ 
3918.97 & C\,{\sc ii}       & 0.279 & 0.065 & 0.011 & 0.078 & 0.014 \\ 
3920.68 & C\,{\sc ii}       & 0.279 & 0.142 & 0.011 & 0.170 & 0.018 \\ 
3970.07 & B7 (H$\epsilon$)& 0.266 & 14.294 & 0.071 & 16.977 & 1.113 \\ 
4068.60 & {\sii}    & 0.239 & 0.808 & 0.035 & 0.944 & 0.069 \\ 
4101.73 & B6 (H$\delta$)& 0.230 & 23.318 & 0.106 & 27.051 & 1.531 \\ 
4267.00 & C\,{\sc ii}      & 0.180 & 0.210 & 0.030 & 0.236 & 0.035 \\ 
4340.46 & B5 (H$\gamma$)& 0.157 & 42.854 & 0.195 & 47.420 & 1.837 \\ 
4471.47 & {\hei}        & 0.115 & 0.569 & 0.013 & 0.613 & 0.022 \\ 
4571.10 & Mg\,{\sc i}]       & 0.084 & 0.106 & 0.019 & 0.112 & 0.020 \\ 
4658.05 & {\feiii}    & 0.058 & 0.391 & 0.015 & 0.406 & 0.017 \\ 
4701.53 & {\feiii}    & 0.045 & 0.093 & 0.008 & 0.096 & 0.009 \\ 
4754.69 & {\feiii}    & 0.030 & 0.077 & 0.013 & 0.079 & 0.014 \\ 
4769.43 & {\feiii}    & 0.025 & 0.025 & 0.008 & 0.025 & 0.008 \\ 
4861.33 & B4 (H$\beta$)& 0.000 & 100.000 & 0.251 & 100.000 & 0.251 \\ 
4881.00 & {\feiii}    & -0.005 & 0.131 & 0.009 & 0.130 & 0.009 \\ 
4888.87 & {\feiii}    & -0.007 & 0.021 & 0.007 & 0.021 & 0.007 \\ 
4921.93 & {\hei}        & -0.016 & 0.130 & 0.008 & 0.129 & 0.008 \\ 
4958.91 & {\oiii}    & -0.026 & 1.127 & 0.008 & 1.109 & 0.011 \\ 
5006.84 & {\oiii}     & -0.038 & 3.408 & 0.015 & 3.326 & 0.034 \\ 
5015.68 & {\hei}        & -0.040 & 0.385 & 0.010 & 0.375 & 0.011 \\ 
5055.98 & Si\,{\sc ii}       & -0.050 & 0.097 & 0.031 & 0.094 & 0.030 \\ 
5067.52 & Ni\,{\sc iii}    & -0.052 & 0.065 & 0.010 & 0.062 & 0.009 \\ 
5197.90 & {\NI}      & -0.082 & 0.059 & 0.022 & 0.056 & 0.021 \\ 
5200.26 & {\NI}     & -0.083 & 0.055 & 0.015 & 0.052 & 0.014 \\ 
5270.40 & {\feiii}    & -0.098 & 0.183 & 0.011 & 0.172 & 0.011 \\ 
5273.24 & Ne\,{\sc ii}      & -0.098 & 0.031 & 0.012 & 0.029 & 0.011 \\ 
5517.72 & {\cliii}    & -0.145 & 0.128 & 0.013 & 0.116 & 0.012 \\ 
5537.89 & {\cliii}    & -0.149 & 0.122 & 0.006 & 0.111 & 0.007 \\ 
5754.64 & {\nii}     & -0.185 & 1.489 & 0.017 & 1.322 & 0.062 \\ 
5875.60 & {\hei}        & -0.203 & 1.786 & 0.029 & 1.566 & 0.082 \\ 
5886.05 & C\,{\sc ii}      & -0.205 & 0.061 & 0.010 & 0.054 & 0.009 \\ 
5891.60 & C\,{\sc ii}       & -0.205 & 0.056 & 0.008 & 0.049 & 0.007 \\ 
5912.58 & C\,{\sc i}       & -0.208 & 0.058 & 0.005 & 0.051 & 0.005 \\ 
5931.78 & N\,{\sc ii}     & -0.211 & 0.068 & 0.008 & 0.059 & 0.007 \\ 
5950.71 & Fe\,{\sc iii}      & -0.214 & 0.040 & 0.009 & 0.035 & 0.008 \\ 
6300.30 & {\oi}       & -0.263 & 0.638 & 0.018 & 0.538 & 0.038 \\ 
6312.10 & {\siii}    & -0.264 & 0.212 & 0.013 & 0.179 & 0.016 \\ 
6347.03 & Ni\,{\sc ii}     & -0.269 & 0.070 & 0.010 & 0.059 & 0.009 \\ 
6363.78 & {\oi}       & -0.271 & 0.195 & 0.014 & 0.164 & 0.016 \\ 
\hline
\end{tabularx}
\end{table}

\setcounter{table}{0}
 \begin{table}
\renewcommand{\arraystretch}{0.9}
    \renewcommand{\thetable}{A\arabic{table}} 
  \footnotesize
  \centering
\caption{Continued.}
 \begin{tabularx}{\columnwidth}{@{\extracolsep{\fill}}D{.}{.}{-1}@{\hspace{-2pt}}
c@{\hspace{-2pt}}D{.}{.}{-1}@{\hspace{-2pt}}D{.}{.}{-1}@{\hspace{-2pt}}D{.}{.}{-1}@{\hspace{-2pt}}D{.}{.}{-1}@{\hspace{-2pt}}D{.}{.}{-1}@{}}
\hline
 \multicolumn{1}{c}{$\lambda_{\rm lab.}$ ({\AA})}
 &Line&\multicolumn{1}{r}{$f(\lambda)$}&\multicolumn{1}{r}{$F$($\lambda$)}&\multicolumn{1}{r}{$\delta$\,$F$($\lambda$)}
&\multicolumn{1}{r}{$I$($\lambda$)}&\multicolumn{1}{r}{$\delta$\,$I$($\lambda$)}
\\
\hline  
6379.58 & O\,{\sc ii}       & -0.273 & 0.098 & 0.023 & 0.082 & 0.020 \\ 
6461.95 & C\,{\sc ii}      & -0.284 & 0.074 & 0.009 & 0.061 & 0.008 \\ 
6548.04 & {\nii}     & -0.296 & 45.988 & 0.395 & 37.985 & 2.778 \\ 
6562.80 & B3 ({\ha})& -0.298 & 375.888 & 1.958 & 310.085 & 22.731 \\ 
6578.05 & C\,{\sc ii}       & -0.300 & 0.317 & 0.008 & 0.261 & 0.020 \\ 
6583.46 & {\nii}     & -0.300 & 138.238 & 1.323 & 113.836 & 8.471 \\ 
6678.15 & {\hei}        & -0.313 & 0.588 & 0.012 & 0.480 & 0.038 \\ 
6716.44 & {\sii}     & -0.318 & 4.770 & 0.047 & 3.884 & 0.306 \\ 
6730.81 & {\sii}     & -0.320 & 7.203 & 0.078 & 5.857 & 0.465 \\ 
7065.18 & {\hei}        & -0.364 & 0.494 & 0.011 & 0.391 & 0.036 \\ 
7135.80 & {\ariii}  & -0.374 & 0.740 & 0.009 & 0.581 & 0.054 \\ 
7231.32 & C\,{\sc ii}       & -0.387 & 0.352 & 0.018 & 0.274 & 0.024 \\ 
7236.42 & C\,{\sc ii}       & -0.387 & 0.741 & 0.034 & 0.576 & 0.042 \\ 
7281.35 & {\hei}        & -0.393 & 0.159 & 0.010 & 0.123 & 0.014 \\ 
7318.92 & {\oii}    & -0.398 & 1.465 & 0.061 & 1.132 & 0.120 \\ 
7319.99 & {\oii}    & -0.398 & 4.560 & 0.057 & 3.524 & 0.348 \\ 
7329.66 & {\oii}    & -0.400 & 2.382 & 0.026 & 1.839 & 0.182 \\ 
7330.73 & {\oii}    & -0.400 & 2.460 & 0.029 & 1.899 & 0.188 \\ 
7751.10 & {\ariii}    & -0.455 & 0.187 & 0.009 & 0.139 & 0.017 \\ 
7769.23 & Ca\,{\sc i} & -0.458 & 0.122 & 0.007 & 0.091 & 0.012 \\ 
7771.94 & O\,{\sc i}         & -0.458 & 0.084 & 0.006 & 0.063 & 0.008 \\ 
7874.65 & [Fe\,{\sc ii}]   & -0.471 & 0.032 & 0.004 & 0.023 & 0.004 \\ 
8240.19 & P45 & -0.515 & 0.067 & 0.008 & 0.048 & 0.008 \\ 
8241.88 & P44 & -0.516 & 0.090 & 0.008 & 0.064 & 0.010 \\ 
8243.69 & P43 & -0.516 & 0.055 & 0.006 & 0.039 & 0.007 \\ 
8245.64 & P42 & -0.516 & 0.023 & 0.005 & 0.016 & 0.004 \\ 
8247.73 & P41 & -0.516 & 0.071 & 0.006 & 0.051 & 0.008 \\ 
8249.97 & P40 & -0.517 & 0.037 & 0.003 & 0.027 & 0.004 \\ 
8252.40 & P39 & -0.517 & 0.077 & 0.008 & 0.055 & 0.009 \\ 
8255.02 & P38 & -0.517 & 0.093 & 0.008 & 0.066 & 0.010 \\ 
8257.85 & P37 & -0.517 & 0.093 & 0.007 & 0.066 & 0.010 \\ 
8260.93 & P36 & -0.518 & 0.103 & 0.010 & 0.074 & 0.012 \\ 
8264.28 & P35 & -0.518 & 0.107 & 0.009 & 0.076 & 0.012 \\ 
8267.94 & P34 & -0.519 & 0.114 & 0.008 & 0.082 & 0.012 \\ 
8271.93 & P33 & -0.519 & 0.118 & 0.012 & 0.084 & 0.014 \\ 
8276.31 & P32 & -0.520 & 0.120 & 0.011 & 0.086 & 0.014 \\ 
8281.12 & P31 & -0.520 & 0.114 & 0.010 & 0.081 & 0.013 \\ 
8286.43 & P30 & -0.521 & 0.135 & 0.011 & 0.096 & 0.015 \\ 
8292.31 & P29 & -0.521 & 0.190 & 0.012 & 0.135 & 0.019 \\ 
8298.83 & P28 & -0.522 & 0.164 & 0.010 & 0.117 & 0.017 \\ 
8306.11 & P27 & -0.523 & 0.207 & 0.012 & 0.148 & 0.021 \\ 
8333.78 & P24 & -0.526 & 0.307 & 0.017 & 0.218 & 0.031 \\ 
8340.80 & C\,{\sc ii}       & -0.527 & 0.018 & 0.005 & 0.013 & 0.004 \\ 
8345.47 & P23 & -0.527 & 0.296 & 0.017 & 0.211 & 0.030 \\ 
8359.00 & P22 & -0.529 & 0.385 & 0.021 & 0.274 & 0.038 \\ 
8374.48 & P21 & -0.531 & 0.366 & 0.020 & 0.260 & 0.037 \\ 
8392.40 & P20 & -0.533 & 0.382 & 0.020 & 0.271 & 0.038 \\ 
8413.32 & P19 & -0.535 & 0.449 & 0.021 & 0.318 & 0.044 \\ 
8502.48 & P16 & -0.544 & 0.668 & 0.031 & 0.470 & 0.067 \\ 
8665.02 & P13 & -0.560 & 1.274 & 0.057 & 0.887 & 0.128 \\ 
8750.47 & P12 & -0.568 & 1.558 & 0.069 & 1.079 & 0.158 \\ 
8776.83 & {\hei}        & -0.571 & 0.046 & 0.005 & 0.032 & 0.006 \\ 
8862.78 & P11 & -0.578 & 1.903 & 0.085 & 1.309 & 0.195 \\ 
9014.91 & P10 & -0.590 & 2.554 & 0.113 & 1.744 & 0.264 \\ 
9068.60 & {\siii} & -0.594 & 3.789 & 0.170 & 2.581 & 0.394 \\ 
9123.60 & {\clii}     & -0.598 & 0.128 & 0.010 & 0.087 & 0.014 \\ 
9545.97 & P8 & -0.626 & 4.546 & 0.202 & 3.033 & 0.486 \\ 
\hline 
\end{tabularx}
\end{table}

\setcounter{table}{0}
 \begin{table}
\renewcommand{\arraystretch}{0.9}
    \renewcommand{\thetable}{A\arabic{table}} 
  \footnotesize
  \centering
\caption{Continued.}
 \begin{tabularx}{\columnwidth}{@{\extracolsep{\fill}}D{.}{.}{-1}@{\hspace{-2pt}}
c@{\hspace{-2pt}}D{.}{.}{-1}@{\hspace{-2pt}}D{.}{.}{-1}@{\hspace{-2pt}}D{.}{.}{-1}@{\hspace{-2pt}}D{.}{.}{-1}@{\hspace{-2pt}}D{.}{.}{-1}@{}}
\hline
 \multicolumn{1}{c}{$\lambda_{\rm lab.}$ ({\AA})}
 &Line&\multicolumn{1}{r}{$f(\lambda)$}&\multicolumn{1}{r}{$F$($\lambda$)}&\multicolumn{1}{r}{$\delta$\,$F$($\lambda$)}
&\multicolumn{1}{r}{$I$($\lambda$)}&\multicolumn{1}{r}{$\delta$\,$I$($\lambda$)}
\\
\hline  
  70000.0  &C$_{60}$/{\arii}/H\,{\sc i}  &&    &&28.665 &3.226\\
  74578.2    &H\,{\sc i}&&&& 3.102 &0.226\\
  85000.0  &C$_{60}$ && & &7.003 &0.545\\
  89889.3  &{\ariii} && & &0.757 &0.145\\
 123800.0  &H\,{\sc i}&& & &1.029 &0.271\\
 128100.6  &{\neii}  && &  &20.607&1.555\\
  174000.0  &C$_{60}$/C$_{70}$ &&& &12.131 &1.654\\
           &C$_{60}$ &&& &9.399 &1.282\\
  187079.3  &{\siii}   && & &5.263 &0.724\\
  189000.0  &C$_{60}$/C$_{70}$&&& &<21.253&\\
            &C$_{60}$   &&& &<19.188&\\
 334719.0  &{\siii}    &&& &4.004 &0.857\\
\hline 
 \end{tabularx}
     \begin{description}
      \item[Note --] We measured the flux of the C$_{60,70}$\,18.9\,{\micron}
   by adopting FWHM of 0.347\,{\micron} measured in Tc\,1
   \citep{Otsuka:2014aa} because this line
   seems to be partially lacked due to spike noise. We
   estimated the expected solo intensity of the 
   C$_{60}$ 17.4/18.9\,{\micron} using
   the C$_{60}$/C$_{70}$ ratio at 17.4\,{\micron} = 3.44
   and the C$_{60}$/C$_{70}$ ratio at 18.9\,{\micron} = 9.29
   measured in Tc\,1 \citep{Cami:2010aa}. 
     \end{description}
  \end{table}

  \begin{table}
   \renewcommand{\thetable}{A\arabic{table}}
   \renewcommand{\arraystretch}{0.9}
  \centering
  \caption{Broadband flux density of SaSt\,2-3.}
   \label{T-photo}
  \begin{tabularx}{\columnwidth}{@{\extracolsep{\fill}}
   c@{\hspace{3pt}}c@{\hspace{-1pt}}
  D{p}{~\pm~}{-1}@{\hspace{-1pt}}D{p}{~\pm~}{-1}@{}}
 \hline
  Band &\multicolumn{1}{c}{$\lambda_{\rm c}$}
  &\multicolumn{1}{c}{$F_{\lambda}$ (reddened)}
  &\multicolumn{1}{c}{$I_{\lambda}$ (de-reddened)}\\ 
  & &\multicolumn{1}{c}{(erg s$^{-1}$ cm$^{-2}$ {\micron}$^{-1}$)}
   &\multicolumn{1}{c}{(erg s$^{-1}$ cm$^{-2}$ {\micron}$^{-1}$)}\\
 \hline
  $B$ & 0.4297\,\micron         & 2.24(-10) p 5.99(-12)  & 4.75(-10) p 1.41(-10) \\ 
  $g{'}$ & 0.4640\,\micron      & 1.93(-10) p 1.42(-12)  & 3.81(-10) p 1.03(-10) \\ 
  $V$ & 0.5394\,\micron         & 1.31(-10) p 3.25(-12)  & 2.29(-10) p 5.14(-11) \\ 
  $r{'}$ & 0.6122\,\micron      & 1.05(-10) p 2.61(-12)  & 1.96(-10) p 3.82(-11) \\ 
  $i{'}$ & 0.7440\,\micron      & 3.89(-11) p 3.29(-12)  & 7.89(-11) p 1.35(-11) \\ 
  $J$ & 1.235\,\micron          & 1.27(-11) p 3.03(-13)  & 1.48(-11) p 9.94(-13) \\ 
  $H$ & 1.662\,\micron          & 4.49(-12) p 1.24(-13)  & 4.95(-12) p 2.36(-13) \\ 
  $Ks$ & 2.159\,\micron         & 2.01(-12) p 6.86(-14)  & 2.15(-12) p 9.13(-14) \\ 
  Br$\gamma$ & 2.162\,\micron   & 4.29(-12) p 7.66(-13)  & 4.57(-12) p 8.25(-13) \\ 
  Br$\gamma$45 & 2.188\,\micron & 2.15(-12) p 4.14(-14)  & 2.28(-12) p 7.21(-14) \\ 
  W1 & 3.353\,\micron &                           & 9.28(-13) p 1.97(-14) \\ 
  W2 & 4.602\,\micron &                           & 7.49(-13) p 1.45(-14) \\ 
  W3 & 11.56\,\micron &                           & 7.87(-13) p 1.09(-14) \\ 
  L18W & 18.00\,\micron&                          & 1.40(-12) p 3.71(-13) \\ 
  W4 & 22.09\,\micron &                           & 1.50(-12) p 3.04(-14) \\ 
  \hline
 \end{tabularx}
  \end{table}

  \begin{table}
   \renewcommand{\thetable}{A\arabic{table}}
   \renewcommand{\arraystretch}{0.9}
  \centering
   \caption{Ionic abundances of SaSt\,2-3 using nebular emission lines.
   }
\begin{tabularx}{\columnwidth}{@{\extracolsep{\fill}}lcD{p}{\pm}{-1}D{p}{\pm}{-1}@{}}
 \hline
 Ion(X$^{\rm m+}$) &$\lambda_{\rm lab.}$
 &\multicolumn{1}{c}{$I$($\lambda$) ($I$({\hb}) = 100)}
 &\multicolumn{1}{c}{$n$(X$^{\rm m+}$)/$n$(H$^{+}$)}\\
 \hline
He$^{+}$ & 4471.47\,{\AA} & 0.613 ~p~ 0.022 & 1.11(-2) ~p~ 4.43(-4) \\ 
  & 4921.93\,{\AA} & 0.129 ~p~ 0.008 & 9.44(-3) ~p~ 6.10(-4) \\ 
  & 5015.68\,{\AA} & 0.375 ~p~ 0.011 & 1.20(-2) ~p~ 4.04(-4) \\ 
  & 5875.60\,{\AA} & 1.566 ~p~ 0.082 & 9.58(-3) ~p~ 5.33(-4) \\ 
  & 6678.15\,{\AA} & 0.480 ~p~ 0.038 & 1.19(-2) ~p~ 9.72(-4) \\ 
  & 7065.18\,{\AA} & 0.391 ~p~ 0.036 & 6.49(-3) ~p~ 6.06(-4) \\ 
  & 7281.35\,{\AA} & 0.123 ~p~ 0.014 & 1.06(-2) ~p~ 1.21(-3) \\ 
   &  &  & {\bf 1.09(-2)} ~p~ {\bf 2.28(-4)} \\ 
C$^{2+}$ & 4267.00\,{\AA} & 0.236 ~p~ 0.035 & 2.39(-4) ~p~ 3.58(-5) \\ 
 & 5891.60\,{\AA} & 0.049 ~p~ 0.007 & 3.35(-4) ~p~ 4.56(-5) \\ 
  & 6461.95\,{\AA} & 0.061 ~p~ 0.008 & 6.03(-4) ~p~ 8.07(-5) \\ 
  & 6578.05\,{\AA} & 0.261 ~p~ 0.020 & 3.16(-4) ~p~ 2.60(-5) \\ 
  & 7231.34\,{\AA} & 0.274 ~p~ 0.024 & 6.11(-4) ~p~ 8.85(-5) \\ 
  & 7236.42\,{\AA} & 0.576 ~p~ 0.042 & 6.68(-4) ~p~ 6.13(-5) \\ 
   &  &  & {\bf 3.14(-4)} ~p~ {\bf 2.49(-5)} \\ 
N$^{0}$ & 5197.90\,{\AA} & 0.056 ~p~ 0.021 & 7.60(-7) ~p~ 2.85(-7) \\ 
  & 5200.26\,{\AA} & 0.052 ~p~ 0.014 & 4.04(-7) ~p~ 1.35(-7) \\ 
   &  &  & {\bf 5.82(-7)} ~p~ {\bf 2.51(-7)} \\ 
N$^{+}$ & 5754.64\,{\AA} & 1.322 ~p~ 0.062 & 2.88(-5) ~p~ 5.08(-6) \\ 
  & 6548.04\,{\AA} & 37.985 ~p~ 2.778 & 2.83(-5) ~p~ 2.93(-6) \\ 
  & 6583.46\,{\AA} & 113.836 ~p~ 8.471 & 2.86(-5) ~p~ 3.00(-6) \\ 
   &  &  & {\bf 2.85(-5)} ~p~ {\bf 1.94(-6)} \\ 
O$^{0}$ & 6300.30\,{\AA} & 0.538 ~p~ 0.038 & 4.37(-6) ~p~ 1.25(-6) \\ 
  & 6363.78\,{\AA} & 0.164 ~p~ 0.016 & 4.16(-6) ~p~ 1.22(-6) \\ 
   &  &  & {\bf 4.26(-6)} ~p~ {\bf 8.72(-7)} \\ 
O$^{+}$ & 3726.03\,{\AA} & 132.411 ~p~ 10.617 & 1.29(-4) ~p~ 2.56(-5) \\ 
  & 3728.81\,{\AA} & 91.955 ~p~ 7.314 & 1.25(-4) ~p~ 1.38(-5) \\ 
  & 7318.92\,{\AA} & 1.132 ~p~ 0.120 & 1.38(-4) ~p~ 5.23(-5) \\ 
  & 7319.99\,{\AA} & 3.524 ~p~ 0.348 & 1.37(-4) ~p~ 5.16(-5) \\ 
  & 7329.66\,{\AA} & 1.839 ~p~ 0.182 & 1.35(-4) ~p~ 5.11(-5) \\ 
  & 7330.73\,{\AA} & 1.899 ~p~ 0.188 & 1.41(-4) ~p~ 5.31(-5) \\ 
   &  &  & {\bf 1.28(-4)} ~p~ {\bf 1.10(-5)} \\ 
O$^{2+}$ & 4958.91\,{\AA} & 1.109 ~p~ 0.011 & 1.68(-6) ~p~ 6.82(-7) \\ 
  & 5006.84\,{\AA} & 3.326 ~p~ 0.034 & 1.75(-6) ~p~ 7.08(-7) \\ 
   &  &  & {\bf 1.72(-6)} ~p~ {\bf 4.91(-7)} \\ 
Ne$^{+}$ & 12.81\,{\micron} & 20.607 ~p~ 1.555 & 2.91(-5) ~p~ 2.85(-6) \\ 
S$^{+}$ & 4068.60\,{\AA} & 0.944 ~p~ 0.069 & 6.22(-7) ~p~ 1.08(-7) \\ 
  & 6716.44\,{\AA} & 3.884 ~p~ 0.306 & 6.47(-7) ~p~ 6.72(-8) \\ 
  & 6730.81\,{\AA} & 5.857 ~p~ 0.465 & 6.38(-7) ~p~ 7.27(-8) \\ 
   &  &  & {\bf 6.39(-7)} ~p~ {\bf 4.49(-8)} \\ 
S$^{2+}$ & 18.71\,{\micron} & 5.263 ~p~ 0.724 & 6.08(-7) ~p~ 9.80(-8) \\ 
  & 33.47\,{\micron} & 4.004 ~p~ 0.857 & 5.93(-7) ~p~ 1.63(-7) \\ 
  & 6312.10\,{\AA} & 0.179 ~p~ 0.016 & 5.84(-7) ~p~ 3.00(-7) \\ 
  & 9068.60\,{\AA} & 2.581 ~p~ 0.394 & 7.28(-7) ~p~ 2.17(-7) \\ 
    &  &  & {\bf 6.18(-7)} ~p~ {\bf 7.57(-8)} \\ 
Cl$^{+}$ & 9123.60\,{\AA} & 0.087 ~p~ 0.014 & 1.74(-8) ~p~ 3.20(-9) \\ 
Cl$^{2+}$ & 5517.72\,{\AA} & 0.116 ~p~ 0.012 & 2.12(-8) ~p~ 8.45(-9) \\ 
  & 5537.89\,{\AA} & 0.111 ~p~ 0.007 & 1.87(-8) ~p~ 5.29(-9) \\ 
   &  &  & {\bf 1.94(-8)} ~p~ {\bf 4.48(-9)} \\ 
Ar$^{2+}$   & 8.99\,{\micron} & 0.757 ~p~ 0.145 & 8.79(-8) ~p~ 1.74(-8) \\
   & 7135.80\,{\AA} & 0.581 ~p~ 0.054 & 9.30(-8) ~p~ 2.58(-8) \\ 
  & 7751.10\,{\AA} & 0.139 ~p~ 0.017 & 9.29(-8) ~p~ 2.68(-8) \\ 
   &  &  & {\bf 9.02(-8)} ~p~ {\bf 1.27(-8)} \\ 
Fe$^{2}$ & 4658.05\,{\AA} & 0.406 ~p~ 0.017 & 1.93(-7) ~p~ 2.49(-8) \\ 
  & 4701.53\,{\AA} & 0.096 ~p~ 0.009 & 1.45(-7) ~p~ 2.40(-8) \\ 
  & 4754.69\,{\AA} & 0.079 ~p~ 0.014 & 2.00(-7) ~p~ 4.24(-8) \\ 
  & 4769.43\,{\AA} & 0.025 ~p~ 0.008 & 1.10(-7) ~p~ 3.71(-8) \\ 
  & 4881.00\,{\AA} & 0.130 ~p~ 0.009 & 1.09(-7) ~p~ 1.99(-8) \\ 
  & 5270.40\,{\AA} & 0.172 ~p~ 0.011 & 1.61(-7) ~p~ 2.06(-8) \\ 
    &  &  & {\bf 1.48(-7)} ~p~ {\bf 1.03(-8)} \\ 
 \hline
\end{tabularx}
\label{T-ionic}
  \end{table}

  \begin{table*}
   \renewcommand{\thetable}{A\arabic{table}}
   \renewcommand{\arraystretch}{0.9}
  \centering
   \caption{The measurements of the equivalent width ($EW$) of the stellar
     absorption and the derived elemental abundances. \label{T-stellar2}}
\begin{tabularx}{\textwidth}{@{\extracolsep{\fill}}lcD{p}{\pm}{-1}D{p}{\pm}{-1}D{p}{\pm}{-1}@{}}
\hline
 Line &$\lambda_{\rm lab.}$ ({\AA})& \multicolumn{1}{c}{$EW$ (m{\AA})}
 & \multicolumn{1}{c}{$n$(X)/$n$(H)}        & \multicolumn{1}{c}{$\epsilon$(X)}\\ 
\hline
{\hei}  & 3652.90 & 29.24 ~p~ 1.87  & 8.92(-2) ~p~ 2.77(-2)  & 10.95 ~p~ 0.14 \\ 
     & 3867.45 & 104.20 ~p~ 1.30 & 9.30(-2) ~p~ 1.41(-2)  & 10.97 ~p~ 0.07 \\ 
     & 4009.24 & 188.70 ~p~ 1.56 & 7.85(-2) ~p~ 9.76(-3)  & 10.89 ~p~ 0.05 \\ 
     & 4120.78 & 184.40 ~p~ 2.10 & 1.20(-1) ~p~ 2.12(-2)  & 11.08 ~p~ 0.08 \\ 
     & 4143.68 & 266.70 ~p~ 1.65 & 1.13(-1) ~p~ 1.47(-2)  & 11.05 ~p~ 0.06 \\ 
     & 4168.99 & 39.69 ~p~ 1.48  & 8.60(-2) ~p~ 1.96(-2)  & 10.93 ~p~ 0.10 \\ 
He\,{\sc ii} & 4541.62 & 129.90 ~p~ 2.34 & 1.01(-1) ~p~ 2.23(-2)  & 11.01 ~p~ 0.10 \\ 
     & 4685.71 & 210.00 ~p~ 1.96 & 9.85(-2) ~p~ 2.12(-2)  & 10.99 ~p~ 0.09 \\ 
     & 5411.52 & 136.90 ~p~ 2.21 & 9.85(-2) ~p~ 2.27(-2)  & 10.99 ~p~ 0.10 \\ 
     &         &               & {\bf 9.76(-2)} ~p~ {\bf 1.93(-3)}  &
   {\bf 10.99} ~p~ {\bf 0.09} \\ 
C\,{\sc iii} & 4156.56 & 65.84 ~p~ 1.68  & 3.83(-4) ~p~ 8.75(-5)  & 8.58 ~p~ 0.10 \\ 
     & 4162.96 & 75.54 ~p~ 1.79  & 3.39(-4) ~p~ 7.42(-5)  & 8.53 ~p~ 0.10 \\ 
     & 4665.92 & 67.63 ~p~ 1.49  & 3.47(-4) ~p~ 7.07(-5)  & 8.54 ~p~ 0.09 \\ 
C\,{\sc iv} & 5801.31 & 44.60 ~p~ 1.60  & 3.96(-4) ~p~ 1.12(-4)  & 8.60 ~p~ 0.12 \\ 
     &         &               & {\bf 3.66(-4)} ~p~ {\bf 8.60(-5)}  &
   {\bf 8.56} ~p~ {\bf 0.10} \\ 
N\,{\sc ii}  & 3995.00 & 15.58 ~p~ 1.45  & 1.68(-5) ~p~ 7.90(-6)  & 7.23 ~p~ 0.20 \\ 
N\,{\sc iii} & 4634.12 & 26.21 ~p~ 1.47  & 1.91(-5) ~p~ 5.61(-6)  & 7.28 ~p~ 0.13 \\ 
     &         &               & {\bf 1.85(-5)} ~p~ {\bf 6.76(-6)}  &
   {\bf 7.25} ~p~ {\bf 0.16} \\ 
O\,{\sc ii}  & 3882.17 & 38.17 ~p~ 1.34 & 8.77(-5) ~p~ 3.08(-5) & 7.94 ~p~ 0.15 \\ 
     & 3954.21 & 37.06 ~p~ 0.93 & 1.66(-4) ~p~ 3.19(-5) & 8.22 ~p~ 0.08 \\ 
     & 3982.72 & 26.86 ~p~ 1.89 & 8.70(-5) ~p~ 5.19(-5) & 7.94 ~p~ 0.26 \\ 
     & 4092.89 & 22.34 ~p~ 1.23 & 6.36(-5) ~p~ 4.62(-5) & 7.80 ~p~ 0.32 \\ 
     & 4119.22 & 54.49 ~p~ 1.67 & 1.38(-4) ~p~ 2.93(-5) & 8.14 ~p~ 0.09 \\ 
     & 4132.79 & 24.40 ~p~ 1.15 & 1.99(-4) ~p~ 4.91(-5) & 8.30 ~p~ 0.11 \\ 
     & 4189.75 & 30.55 ~p~ 1.22 & 2.24(-4) ~p~ 4.86(-5) & 8.35 ~p~ 0.09 \\ 
     & 4366.77 & 53.74 ~p~ 1.13 & 1.07(-4) ~p~ 1.92(-5) & 8.07 ~p~ 0.08 \\ 
     & 4590.83 & 63.35 ~p~ 2.44 & 8.57(-5) ~p~ 2.01(-5) & 7.93 ~p~ 0.10 \\ 
     & 4596.07 & 62.17 ~p~ 1.51 & 1.07(-4) ~p~ 1.92(-5) & 8.03 ~p~ 0.08 \\ 
     & 4661.53 & 61.34 ~p~ 1.25 & 1.07(-4) ~p~ 2.35(-5) & 8.16 ~p~ 0.06 \\ 
     & 4673.84 & 10.44 ~p~ 1.19 & 1.43(-4) ~p~ 5.31(-5) & 7.89 ~p~ 0.30 \\ 
     & 4676.14 & 45.04 ~p~ 1.22 & 7.75(-5) ~p~ 2.30(-5) & 8.10 ~p~ 0.08 \\ 
     &         &              & {\bf 1.25(-4)} ~p~ {\bf 4.86(-5)} & {\bf
   8.10} ~p~ {\bf 0.17} \\ 
Si\,{\sc iii}& 4552.57 & 83.11 ~p~ 1.22 & 6.61(-6) ~p~ 9.81(-7) & 6.82 ~p~ 0.06 \\ 
     & 4567.77 & 62.27 ~p~ 1.10 & 7.33(-6) ~p~ 1.14(-6) & 6.87 ~p~ 0.07 \\ 
     & 4574.57 & 21.00 ~p~ 1.31 & 5.90(-6) ~p~ 1.65(-6) & 6.77 ~p~ 0.12 \\ 
Si\,{\sc iv} & 4116.11 &162.50 ~p~ 1.29 & 4.33(-6) ~p~ 6.08(-7) & 6.64 ~p~ 0.06 \\ 
     & 4631.08 & 42.81 ~p~ 2.00 & 8.41(-6) ~p~ 2.47(-6) & 6.93 ~p~ 0.13 \\ 
     &         &              & {\bf 6.52(-6)} ~p~ {\bf 1.54(-6)} & {\bf
   6.81} ~p~ {\bf 0.10} \\ 
\hline
\end{tabularx}
  \end{table*}

 \begin{table*}
    \renewcommand{\thetable}{A\arabic{table}}
   \renewcommand{\arraystretch}{0.9}
  \centering
   \caption{Comparison of the observed line intensities, band fluxes,
  and flux densities between the {\sc Cloudy} model and the observation. The
  band width for the integrated band flux is as follows;
  0.084, 0.116, 0.087, 0.111, 0.104\,{\micron}
  in $B$, $g^{'}$, $V$, $r^{'}$, and $i^{'}$, respectively.
  0.162, 0.251, 0.260, 0.626, 1.042\,{\micron}
  in $J$, $H$, $Ks$, $W1$, and $W2$, respectively.
0.36, 0.40, 0.60, 1.00, 0.25, 1.00, 2.00, 0.70, 0.40, 0.50, 0.50, 0.50,
  0.50, 0.50, and 1.00\,{\micron} in IRS-1 to -15, respectively.
  \label{T-Cloudy}}
\begin{tabularx}{\textwidth}{@{\extracolsep{\fill}}ccD{.}{.}{-1}D{.}{.}{-1}ccD{.}{.}{-1}D{.}{.}{-1}@{}}
\hline
 Line & $\lambda_{\rm lab.}$ &\multicolumn{1}{c}{$I$({\sc Cloudy})}
 & \multicolumn{1}{c}{$I$(Obs)} & Line & $\lambda_{\rm lab.}$
 & \multicolumn{1}{c}{$I$({\sc Cloudy})} & \multicolumn{1}{c}{$I$(Obs)} \\ 
 &  & \multicolumn{1}{c}{$I$({\hb})=100} &
      \multicolumn{1}{c}{$I$({\hb})=100}
      &  &  & \multicolumn{1}{c}{$I$({\hb})=100} & \multicolumn{1}{c}{$I$({\hb})=100} \\ 
\hline
B24 &   3671\,{\AA} & 0.439 & 0.735 & {\cliii} &   5538\,{\AA} & 0.122 & 0.111 \\ 
B23 &   3674\,{\AA} & 0.487 & 0.909 & {\nii} &   5755\,{\AA} & 1.293 & 1.322 \\ 
B22 &   3676\,{\AA} & 0.545 & 0.916 & {\hei} &   5876\,{\AA} & 1.673 & 1.566 \\ 
B21 &   3679\,{\AA} & 0.613 & 1.046 & {\oi} &   6300\,{\AA} & 0.951 & 0.538 \\ 
B20 &   3683\,{\AA} & 0.696 & 1.111 & {\siii} &   6312\,{\AA} & 0.392 & 0.179 \\ 
B19 &   3687\,{\AA} & 0.798 & 1.388 & {\oi} &   6363\,{\AA} & 0.303 & 0.164 \\ 
B18 &   3692\,{\AA} & 0.924 & 1.465 & {\nii} &   6548\,{\AA} & 35.039 & 37.985 \\ 
B17 &   3697\,{\AA} & 1.081 & 1.605 & B3 &   6563\,{\AA} & 289.424 & 310.085 \\ 
B16 &   3704\,{\AA} & 1.282 & 2.196 & {\nii} &   6584\,{\AA} & 103.400 & 113.836 \\ 
B15 &   3712\,{\AA} & 1.541 & 2.054 & {\hei} &   6678\,{\AA} & 0.467 & 0.480 \\ 
B14 &   3722\,{\AA} & 1.882 & 2.671 & {\sii} &   6716\,{\AA} & 2.688 & 3.884 \\ 
{\oii} &   3726\,{\AA} & 139.511 & 132.411 & {\sii} &   6731\,{\AA} & 4.311 & 5.857 \\ 
{\oii} &   3729\,{\AA} & 76.585 & 91.955 & {\hei} &   7065\,{\AA} & 0.541 & 0.391 \\ 
B13 &   3734\,{\AA} & 2.340 & 2.922 & {\ariii} &   7135\,{\AA} & 0.592 & 0.581 \\ 
B12 &   3750\,{\AA} & 2.966 & 3.836 & {\hei} &   7281\,{\AA} & 0.110 & 0.123 \\ 
B11 &   3771\,{\AA} & 3.849 & 4.991 & {\oii} &   7323\,{\AA} & 6.193 & 4.657 \\ 
B10 &   3798\,{\AA} & 5.130 & 6.224 & {\oii} &   7332\,{\AA} & 4.939 & 3.739 \\ 
{\hei} &   3820\,{\AA} & 0.138 & 0.108 & {\ariii} &   7751\,{\AA} & 0.143 & 0.139 \\ 
B9 &   3835\,{\AA} & 7.068 & 7.876 & P24 &   8334\,{\AA} & 0.145 & 0.218 \\ 
B8 &   3889\,{\AA} & 10.143 & 11.547 & P23 &   8346\,{\AA} & 0.162 & 0.211 \\ 
B7 &   3970\,{\AA} & 15.359 & 16.977 & P22 &   8359\,{\AA} & 0.182 & 0.274 \\ 
{\sii} &   4070\,{\AA} & 1.019 & 0.944 & P21 &   8374\,{\AA} & 0.206 & 0.260 \\ 
B6 &   4102\,{\AA} & 25.012 & 27.051 & P20 &   8392\,{\AA} & 0.235 & 0.271 \\ 
B5 &   4340\,{\AA} & 45.252 & 47.420 & P19 &   8413\,{\AA} & 0.271 & 0.318 \\ 
{\hei} &   4471\,{\AA} & 0.556 & 0.613 & P16 &   8502\,{\AA} & 0.441 & 0.470 \\ 
{\feiii} &   4659\,{\AA} & 0.323 & 0.401 & P13 &   8665\,{\AA} & 0.815 & 0.887 \\ 
{\feiii} &   4702\,{\AA} & 0.108 & 0.112 & P12 &   8750\,{\AA} & 1.036 & 1.079 \\ 
{\feiii} &   4755\,{\AA} & 0.059 & 0.079 & P11 &   8863\,{\AA} & 1.347 & 1.309 \\ 
{\feiii} &   4770\,{\AA} & 0.036 & 0.025 & P10 &   9015\,{\AA} & 1.798 & 1.744 \\ 
{\feiii} &   4881\,{\AA} & 0.106 & 0.130 & {\siii} &   9069\,{\AA} & 4.858 & 2.581 \\ 
{\hei} &   4922\,{\AA} & 0.155 & 0.129 & {\clii} &   9124\,{\AA} & 0.058 & 0.087 \\ 
{\oiii} &   4959\,{\AA} & 1.095 & 1.109 & P8 &   9546\,{\AA} & 3.568 & 3.033 \\ 
{\oiii} &   5007\,{\AA} & 3.296 & 3.326 & {\hi} &  7.48/50\,{\micron} & 3.153 & 3.102 \\ 
{\hei} &   5016\,{\AA} & 0.427 & 0.375 & {\ariii} & 8.99\,{\micron} & 0.618 & 0.757 \\ 
{\NI} &   5198\,{\AA} & 0.036 & 0.056 &  {\hi} &  12.38\,{\micron} & 1.067 & 1.029 \\ 
{\NI} &   5200\,{\AA} & 0.022 & 0.052 & {\neii} & 12.81\,{\micron} & 20.002 & 20.607 \\
{\feiii} &   5271\,{\AA} & 0.187 & 0.172 & {\siii} & 18.71\,{\micron} & 6.684 & 5.263 \\ 
{\cliii} &   5518\,{\AA} & 0.113 & 0.116 & {\siii} & 33.47\,{\micron} & 2.941 & 4.004 \\ 
\hline
 Band & $\lambda_{\rm c}$ & \multicolumn{1}{c}{$I$({\sc Cloudy})} &
      \multicolumn{1}{c}{$I$(Obs)}
      & Band & $\lambda_{\rm c}$ & \multicolumn{1}{c}{$I$({\sc Cloudy})} & \multicolumn{1}{c}{$I$(Obs)} \\ 
 &  & \multicolumn{1}{c}{$I$({\hb})=100} &
      \multicolumn{1}{c}{$I$({\hb})=100}
      &  &  & \multicolumn{1}{c}{$I$({\hb})=100} & \multicolumn{1}{c}{$I$({\hb})=100} \\ 
\hline
$B$ &   0.4297\,{\micron} & 3112.550 & 2569.843 & IRS-04 &  14.50\,{\micron} & 42.844 & 53.079 \\ 
$g^{'}$ &   0.4640\,{\micron} & 3502.697 & 2831.786 & IRS-05 &  15.00\,{\micron} & 11.085 & 13.975 \\ 
$V$ &   0.5394\,{\micron} & 1401.690 & 1280.855 & IRS-06 &  16.50\,{\micron} & 50.613 & 64.398 \\ 
$r^{'}$ &   0.6122\,{\micron} & 2097.757 & 1395.011 & IRS-07 &  22.00\,{\micron} & 145.937 & 177.499 \\ 
$i^{'}$ &   0.7440\,{\micron} & 603.378 & 527.513 & IRS-08 &  23.35\,{\micron} & 53.640 & 64.324 \\ 
$J$ &  1.235\,{\micron} & 168.908 & 153.285 & IRS-09 &  27.00\,{\micron} & 32.776 & 37.026 \\ 
$H$ &  1.662\,{\micron} & 82.062 & 79.281 & IRS-10 &  28.00\,{\micron} & 41.138 & 45.412 \\ 
$Ks$ &  2.159\,{\micron} & 45.506 & 35.859 & IRS-11 &  29.00\,{\micron} & 40.587 & 44.158 \\ 
$W1$ &  3.353\,{\micron} & 38.985 & 36.990 & IRS-12 &  30.00\,{\micron} & 39.898 & 42.550 \\ 
$W2$ &  4.603\,{\micron} & 43.688 & 49.748 & IRS-13 &  31.00\,{\micron} & 38.834 & 40.629 \\ 
IRS-01 &  8.100\,{\micron} & 10.832 & 10.822 & IRS-14 &  32.00\,{\micron} & 37.647 & 38.441 \\ 
IRS-02 &  10.00\,{\micron} & 12.494 & 13.062 & IRS-15 &  34.50\,{\micron} & 73.392 & 64.405 \\ 
IRS-03 &  13.50\,{\micron} & 23.575 & 28.543 &  &  &  &  \\ 
\hline
ID & $\lambda_{\rm c}$ & \multicolumn{1}{c}{$F_{\nu}$({\sc Cloudy})} & \multicolumn{1}{c}{$F_{\nu}$(Model)} & ID &
       $\lambda_{\rm c}$ & \multicolumn{1}{c}{$F_{\nu}$({\sc Cloudy})} &
        \multicolumn{1}{c}{$F_{\nu}$(Model)} \\
 &  & \multicolumn{1}{c}{(Jy)} & \multicolumn{1}{c}{(Jy)} &  &  &
    \multicolumn{1}{c}{(Jy)}
    & \multicolumn{1}{c}{(Jy)} \\
 \hline
MIR-01 &  15.00\,{\micron} & 0.052 & 0.066 & FIR-02 &  90.00\,{\micron} & 0.081 & 0.063 \\ 
MIR-02 &  20.00\,{\micron} & 0.135 & 0.172 & FIR-03 &  100.0\,{\micron} & 0.060 & 0.052 \\ 
MIR-03 &  25.00\,{\micron} & 0.262 & 0.306 & FIR-04 &  120.0\,{\micron} & 0.035 & 0.035 \\ 
MIR-04 &  30.00\,{\micron} & 0.375 & 0.401 & FIR-05 &  140.0\,{\micron} & 0.023 & 0.025 \\ 
FIR-01 &  65.00\,{\micron} & 0.201 & 0.105 &   &  &   &   \\ 
\hline
\end{tabularx}
 \end{table*}

 \clearpage
 
 \begin{figure*}
    \renewcommand{\thetable}{A\arabic{figure}}
 \centering
  \includegraphics[width=\textwidth,clip]{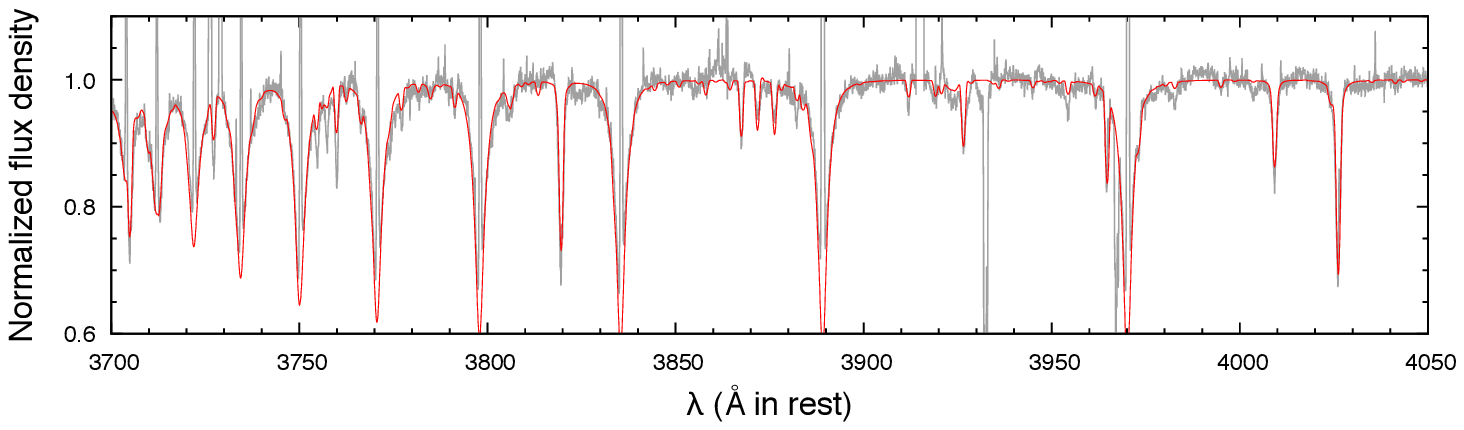}\\
  \includegraphics[width=\textwidth,clip]{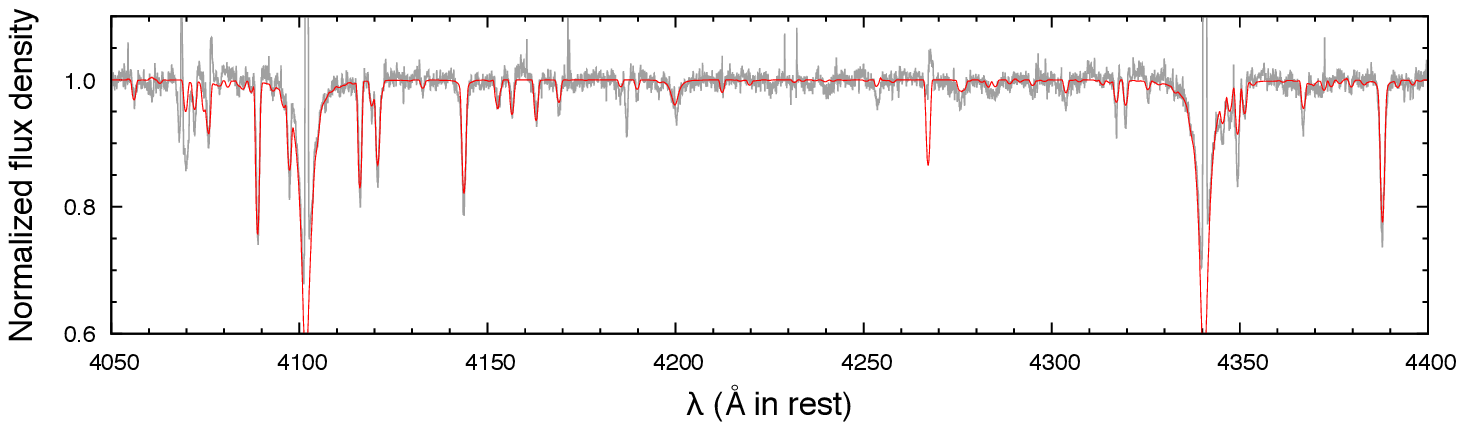}\\
  \includegraphics[width=\textwidth,clip]{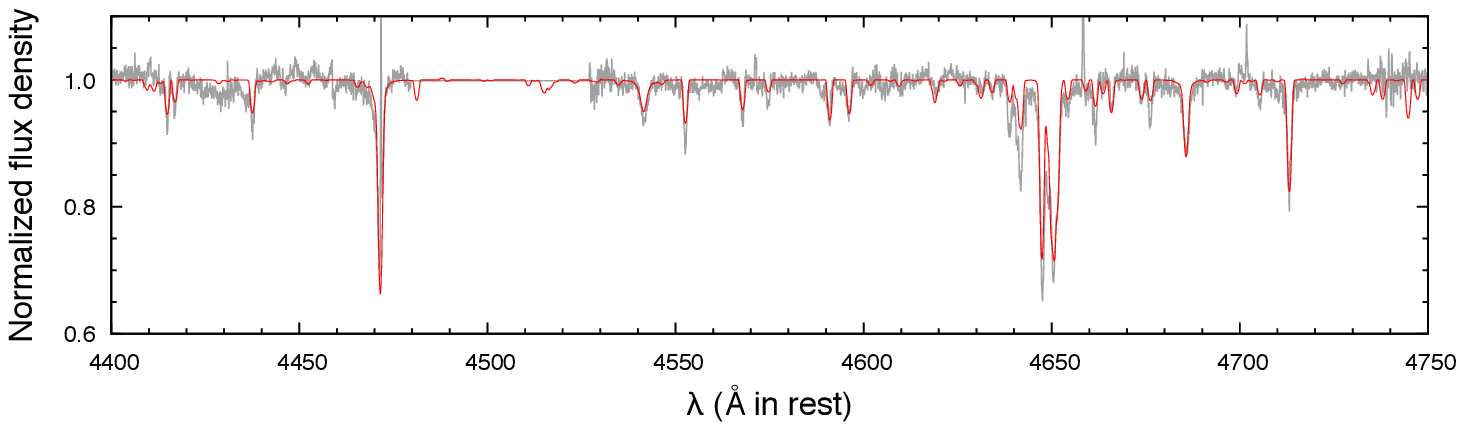}
 \caption{Comparison between the observed HDS (grey line)
      and the {\sc TLUSTY} synthetic spectrum (red line) of SaSt\,2-3.
 The input parameters are listed in Table\,\ref{T-stellar}.
      \label{F-synspecA}
      }
   \end{figure*}

\bsp  
\label{lastpage}
\end{document}